\documentclass[%
 reprint,
nofootinbib,
amsmath,amssymb,
aps,
pra,
]{revtex4-1}
\usepackage{graphicx}
\usepackage{dcolumn}
\usepackage{bm}
\usepackage[makeroom]{cancel} 
\usepackage{ragged2e} 
\usepackage{changepage} 
\usepackage{tabularx} 
\usepackage{booktabs} 
\usepackage{tablefootnote} 
\usepackage{lipsum} 
\usepackage{array} 
\usepackage{todonotes} 

\usepackage{hyperref}
\hypersetup{
	colorlinks   = true, 
	urlcolor     = {blue!80!black}, 
	linkcolor    = {blue!80!black}, 
	citecolor   = {green!50!black} 
}

\usepackage{tikz}
\usetikzlibrary{shapes,decorations}

\usepackage{xcolor}

\usepackage{enumerate}

\usepackage{xcolor} 
\definecolor{ferngreen}{rgb}{0.31, 0.47, 0.26}
\definecolor{forestgreen(web)}{rgb}{0.13, 0.55, 0.13}
\definecolor{green(html/cssgreen)}{rgb}{0.0, 0.5, 0.0}
\definecolor{darkspringgreen}{rgb}{0.09, 0.45, 0.27}
\definecolor{dartmouthgreen}{rgb}{0.05, 0.5, 0.06}

\definecolor{ochre}{rgb}{0.8, 0.47, 0.13}
\definecolor{otterbrown}{rgb}{0.4, 0.26, 0.13}

\definecolor{cerulean}{rgb}{0.0, 0.48, 0.65}
\definecolor{darkcerulean}{rgb}{0.03, 0.27, 0.49}

\definecolor{blueviolet}{rgb}{0.54,0.17,0.89}
\definecolor{amaranth}{rgb}{0.9,0.17,0.31}

\begin{document}

\preprint{APS/123-QED}

\title{A semi-classical analysis of the coupling strength $\gamma$ between motional quantum states of spatially distant trapped ions interacting via a conducting wire}

\author{N. Van Horne}
\affiliation{Florida Atlantic University, USA, 33431}
\affiliation{Centre for Quantum Technologies, National University of Singapore, Singapore 117543}
\email{noahvanhorne@protonmail.com}

\author{M. Mukherjee}
\affiliation{
Centre for Quantum Technologies, National University of Singapore, Singapore 117543
}
\affiliation{
MajuLab, CNRS-UNS-NUS-NTU International Joint Research Unit, UMI 3654, Singapore
}

\date{\today}

\begin{abstract}

	Interfacing ion qubits in separate traps is among the challenges towards scaling up trapped ion quantum computing. This theoretical study focuses on using a conducting wire to couple the motional quantum states of ions in separate planar traps. The approach of interfacing ion traps provides an alternative to coupling distant qubits with lasers or via ion shuttling. We develop a detailed model for the coupling strength $\gamma$, which in turn allows us to calculate the time required to exchange motional quantum states between two ions, $t_{\mathrm{ex.}}$. To maximize the coupling strength $\gamma$, a rule of thumb is developed to relate the distance $d_{\mathrm{eq.}}$ between a trapped ion and a nearby pickup disk, to the optimal radius $r_{\mathrm{opt.}}$ of the disk, $d_{\mathrm{eq.}} / \sqrt{2} \le r_{\mathrm{opt.}} \le d_{\mathrm{eq.}}$. We include a coefficient $\zeta$ which describes three self-capacitances connected in series, relating the capacitances of each part of the coupling system, which corrects an oversight common to several previous works. Our analytical model for the coupling strength is then compared with other existing analytical models in the literature. This requires a standardization of the coupling systems and mapping them to a common set of parameter values. Similarities, differences and limitations of the various models are discussed. 
	This study's focus on the coupling strength $\gamma$ and $t_{\mathrm{ex.}}$ provides a foundational pillar for exchanging quantum information between trapped ions via a wire. 
	In addition, many insights are general and apply broadly to hybrid systems interfacing trapped ions with nearby conductors.

\end{abstract}

\maketitle


	Qubits are the basic unit of information of today's quantum computers. They can be created using a variety of physical systems, much as normal (non-quantum) information bits can be made from small magnets as was once done in floppy disks, or burn marks on CDs, or transistor logic. Among front-running qubit candidates, trapped ions are of particular interest due to their long lifetimes, with demonstrated $T_2$ coherence times over $\sim 50$ seconds at the time of writing (without dynamical decoupling)\cite{2016IonCoherence1s, 2014IonCoherence50sec}. They are suitable for performing quantum gate operations in their own right, and have the potential to complement other technologies such as superconducting qubits, which have faster gate times on the order of tens of nanoseconds \cite{2020SuperCondGateTime50ns,2014supercondQbitCohAndGatetimeSupp,2019ReviewArtic}.\footnote{For context, the shortest gate times for trapped ions are in the range of $1 - 30 ~\mu$s \cite{2018FastGate1point6microsec,2016FastIonGate,2016IonCoherence1s}, barring further development of the strategy outlined in \cite{2017FastIonGate}. The longest coherence times for superconducting qubits are on the order of $1 - 100~\mu$s \cite{2020SuperCondGateTime50ns,2014supercondQbitCohAndGatetimeSupp,2013SupercondQbitCoherence1,rigetti2012superconducting}.} One of the barriers to scaling trapped ion quantum computing is the number of qubits that can be made to interact coherently in a single trapping device \cite{2016ScalingNumberOfIons,2019ReviewArtic,2019LongChainEntangling}. This has motivated research into ways to exchange quantum information between ion qubits in separate traps, for example using ion shuttling \cite{2014Shuttling,2020Shuttling,2020ionShuttling,2020Honeywell}, or photonic interconnects \cite{2019EntanglingLongionChain,2012ion-photonEntanglement,2014PhotonicInterconn}. The other strategy is to couple the motion of a trapped ion to a nearby conducting wire which leads to another ion \cite{heinzen1990quantum,daniilidis2009wiring, zurita2008wiring,marzoli2009experimental,liang2010two,kotler2017hybrid,rica2018double}. 
	For these approaches to work, we introduce two general criteria that must be simultaneously satisfied for any successful transfer of quantum information: 
\begin{adjustwidth}{0em}{0em}
	\begin{enumerate}[1)]			
		\item \label{Crit1} The coherence time of the quantum state should exceed the time it takes to transfer the quantum state: $t_{\mathrm{deco.}} / t_{\mathrm{ex.}} \gg 1$
		
		\item \label{Crit2} The amplitude of the signal carrier (Coulomb interaction, voltage, current, photons, etc.) must exceed the amplitude of the noise which arises throughout the transfer process: Signal / Noise $\gg 1$
		

	\end{enumerate}
\end{adjustwidth}
	These are rigorous statements which elaborate on the DiVincenzo criteria for quantum communication \cite{divincenzo2000physical}. Here, we consider the scheme of a conducting wire which guides the exchange of quantum information between two ions. This method has not yet been successfully implemented. The two criteria above serve as guiding principles which motivate the calculation of four quantities: i) the time $t_{\mathrm{ex.}}$ required to exchange quantum information via a conducting wire, ii) the decoherence time $t_{\mathrm{deco.}}$ of the motional mode of an ion qubit, iii) the classical voltage signal, $V_{\mathrm{sig.}}$, iv) the classical Johnson-Nyquist (thermal) noise voltage $V_{\mathrm{J.N.}}$.

	A limited number of works have studied systems similar to the design developed herein \cite{zurita2008wiring,liang2010two,daniilidis2009wiring,an2022coupling} and calculated exchange times $t_{\mathrm{ex.}}$ for those systems.
	Our work provides an independent in-depth investigation of the coupling strength $\gamma$ and exchange time $t_{\mathrm{ex.}}$. 
	We explicitly discuss surrounding considerations including 
	practical guidelines for optimal dimensions of the coupling device, the current which flows through the device, the role of the resistance of the coupling wire, driving the device on resonance, and how the coupling strength derived here compares with the coupling strength calculated in other works, both quantitatively and qualitatively. This study's focus on the coupling strength $\gamma$ and $t_{\mathrm{ex.}}$ provides a foundational pillar needed to simultaneously meet the two general criteria above. The comprehensive analysis and methodology developed are expected to be useful for a diverse range of experiments aimed at transferring quantum information via a conducting wire using the motional states of trapped ions. Since many of the insights are general, in addition to being applicable to ion-ion coupling systems, results from this work can be applied more broadly to a range of hybrid systems interfacing trapped ions with nearby conductors. 

	The study is organized into four sections. In section \ref{GndedDisk}) we calculate the charge imbalance $Q_{\mathrm{transf.}}$ induced on an isolated, disk-shaped conductor. Section \ref{CouplingCst}) uses the result from section \ref{GndedDisk}) to find the coupling strength $\gamma$ between two charges in separate traps.  In section \ref{LitRev}), a literature review compares the coupling strength calculated here with the coupling strengths calculated in other studies. This requires a standardization of the coupling systems and mapping them to a common set of parameter values. Section \ref{ConclOut}) summarizes our conclusions and outlook.

\section{Induced charge and current in a grounded system}\label{GndedDisk}

	\subsection{Establishing the relevant quantities}\label{pointsofinterest}
	
		\setlength{\parskip}{10pt plus 1pt minus 1pt}
		
			We begin with a brief refresher of electrostatics in conductors. Everywhere within a conductor, the electric field is zero at equilibrium.  This means the potential $V$ within a conductor, which is related to the electric field by $\vec{E}= \vec{\nabla}V$, ~is either a constant, or zero.  However, the charge-density on the surface of a conductor can (and must) vary arbitrarily along the surface. Even if a conductor is grounded, in general the local charge-density on the surface is \textit{not} zero or homogeneous, as charges near the surface must rearrange to ensure that any $\vec{E}$-field outside the conductor is canceled off to zero within the conductor \cite{griffiths}.
			
			\begin{figure}[h!]
				\includegraphics[width=\linewidth]{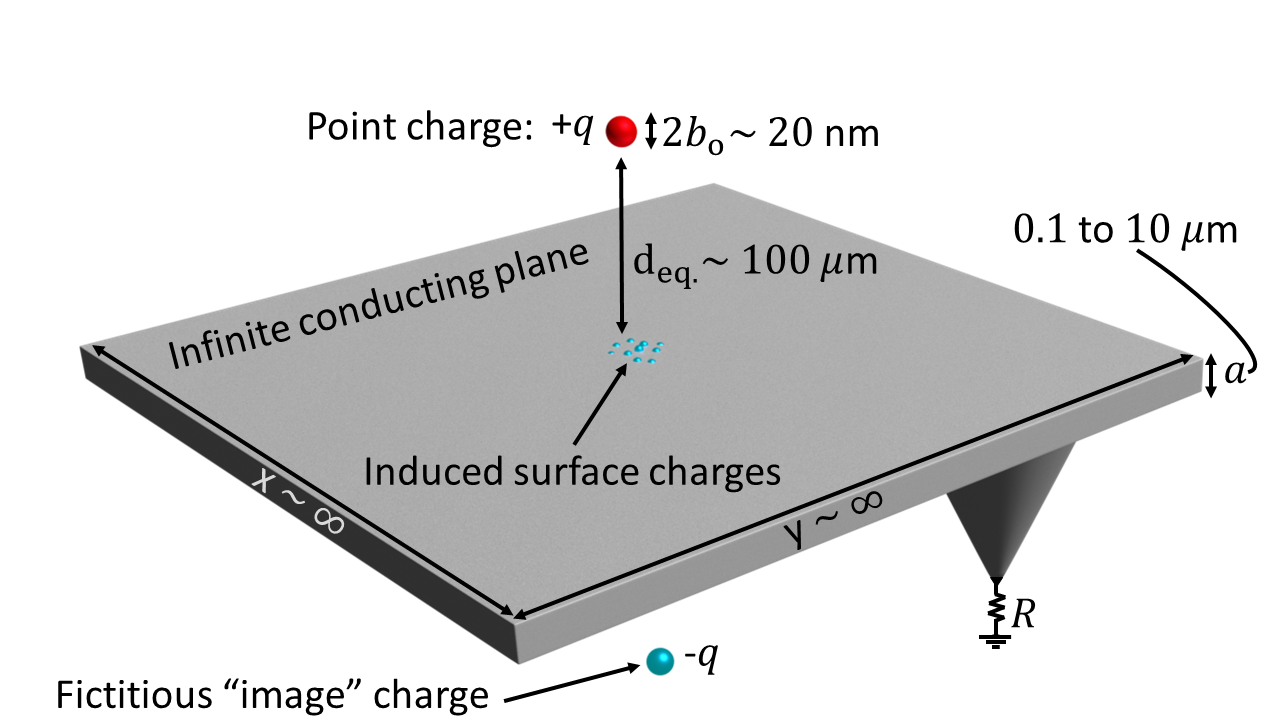}
				\caption{Charged particle floating above a plane. Schematic diagram of a positive charge (red) suspended above an infinite grounded conducting plane. The positive charge causes negative charges (small blue dots) to accumulate on the surface below it. The induced charge is centered around the origin of the coordinate system, ($x = 0, y = 0, z = 0$). It can be shown that the electric field produced above the plane by the induced charges, is identical to the electric field which would be produced by a fictitious "image" charge located at a distance $d_{\mathrm{eq.}}$ below the surface of the plane.}
				\label{fig:Figure1}
			\end{figure}
		
			We start with a simple model, and proceed to fill in details as their relevance becomes apparent.  Let us consider a positive point-charge $+q$ suspended above an infinite grounded conducting plane (Figure \ref{fig:Figure1}). According to the "first uniqueness theorem" of electrostatics, (see Griffiths \cite{griffiths1962introduction}, section 3.2.1), for any given set of boundary conditions for the voltage, along with a known charge-density distribution,  there exists only one function which describes the voltage throughout the volume. This is a slight generalization of the uniqueness of solutions to Laplace's equation, given specified boundary conditions.  The two boundary conditions $V=0$ at the surface of a grounded plane, and $V=0$ far away from the plane, guarantee that a point-charge situated a distance $d$ above the conducting surface, along the $z$ axis, must produce a potential in the space above the plane given by (\cite{griffiths1962introduction}, section 3.1.5):
\begin{eqnarray}
V(x,y,z) = &&\frac{1}{4\pi\epsilon_{\mathrm{o}}}
\Bigg( \frac{q_{\mathrm{c}}}{\sqrt{x^2+y^2+(z-d)^2}} \nonumber\\
&&-\frac{q_{\mathrm{c}}}{\sqrt{x^2+y^2+(z+d)^2}} \Bigg) ~,
\label{Griffvoltage}
\end{eqnarray}		
			where $x$ and $y$ are coordinates in the conducting plane as shown in figure \ref{fig:Figure1}. The induced surface-charge density on a conductor is given by:
\begin{equation}
\sigma = -\epsilon_{\mathrm{o}} \frac{\partial V}{\partial \hat{n}} \label{surfacechargegeneral}
\end{equation}
			where $\hat{n}$ is the perpendicular direction to the surface of the conductor (here, $\hat{n} = \hat{z}$). Thus, for a point charge above a conducting plane, $\sigma$ is given by:
\[
\sigma=-\epsilon_{\mathrm{o}} \frac{\partial V}{\partial z}\bigg|_{z=0}= \frac{-qd}{2\pi(x^2+y^2+d^2)^{3/2}}
\]
			or
\begin{equation}\label{SurfaceCharge}
\sigma = \frac{-qd}{2\pi(r^2+d^2)^{3/2}}
\end{equation}
			in cylindrical coordinates. The induced surface-charge density is negative, which ensures that the field produced by the positively charged particle is canceled off within the conducting plane. If we were to integrate the surface-charge density $\sigma$ over the whole plane, we would find the total induced charge on the surface of the conductor is $-q$. This result is independent of the distance $d$ between the point charge and the surface, which may seem counter-intuitive. However, while the \textit{total} charge induced on the surface is independent of $d$ (and always adds up to $-q$), the charge \textit{distribution} is not.  Specifically, the surface-charge density directly below the particle increases and decreases in magnitude depending on whether the particle is brought closer to or farther away from the surface, respectively. The induced charge goes from "dispersed" to "localized" as the charged particle is brought closer to the surface.  The change in charge-density $\sigma$ for a small change in the vertical distance $\delta d$ between the particle and the surface is given by:
\begin{equation}
\frac{\partial \sigma}{\partial d}=\frac{\partial}{\partial d}\left[Ad(r^2 + d^2)^{-3/2}\right]=A\frac{r^2-2d^2}{(r^2+d^2)^{5/2}} ~, \label{derivsurfacecharge}
\end{equation}
			where $A \equiv \frac{-q}{2\pi}$ is introduced to simplify expression \eqref{derivsurfacecharge}.  It is useful to take a moment to examine this expression. Three cases are of specific interest: $r \ll d, r \gg d$, and  $r = d$.  The first case leads to:
\begin{equation}\label{dgreaterthanr}
A\frac{r^2-2d^2}{(r^2+d^2)^{5/2}} \to \frac{-2A}{d^3} ~.
\end{equation}
			\noindent
			The second case leads to:
\begin{equation}\label{rgreaterthand}
A\frac{r^2-2d^2}{(r^2+d^2)^{5/2}} \to \frac{A}{r^3} ~.
\end{equation}
			The most important point to notice is that the signs of the two expressions, \eqref{dgreaterthanr} and \eqref{rgreaterthand}, are opposite.  This corresponds to the fact that if you are an observer situated on the plane, directly below the suspended particle (case 1, expression \eqref{dgreaterthanr}), and the particle is moved a bit farther away from the surface (i.e. $d$ increases), you will see the negative charge density around you \textit{decrease} (become less negative); however, if instead of being positioned directly below the suspended particle, you are located at some 'far away' place (still on the surface of the plane, case 2, expression \eqref{rgreaterthand}), then as the particle is moved farther away from the surface you will see an \textit{increase} in the negative charge density around you.  This is consistent with our earlier observation that the \textit{total} surface-charge adds up to a constant, $-q$. If the particle is moved closer to the surface, the \textit{increase} in negative charge-density directly below the particle must come at the expense of a \textit{decrease} in negative charge-density at some not-yet-specified 'farther-away' radius. For completeness, we can mention the rather uninteresting case $r=d$.  We find: $A \left(r^2-2d^2\right) / \left((r^2+d^2)^{5/2}\right) \to \left(-A\right) / \left( 2d^3 \right)$.
			
			Given that $\frac{\partial \sigma}{\partial d}$ is positive for some values of $r$ and negative for other values of $r$, a natural question is: "When does $\frac{\partial \sigma}{\partial d}=0$?".
		
		\noindent
			Solving:
\[
A\frac{r^2-2d^2}{(r^2+d^2)^{5/2}}=0
\]
		\noindent
			yields:
\begin{equation}
r=\sqrt{2}d \label{rmincurrent} ~.
\end{equation}
		\noindent
			This is the distance of zero charge-density variation.  If one tries to measure a current at a distance exactly $r=\sqrt{2}d$ from the origin, no current will be measured at all.  Furthermore, if the current-measuring device extends along both sides of this invisible line, it will lose sensitivity because it will simultaneously measure opposite currents, effectively canceling off a portion of the signal.
		
\begin{figure}[h]
	\centering
	\includegraphics[width=\linewidth]{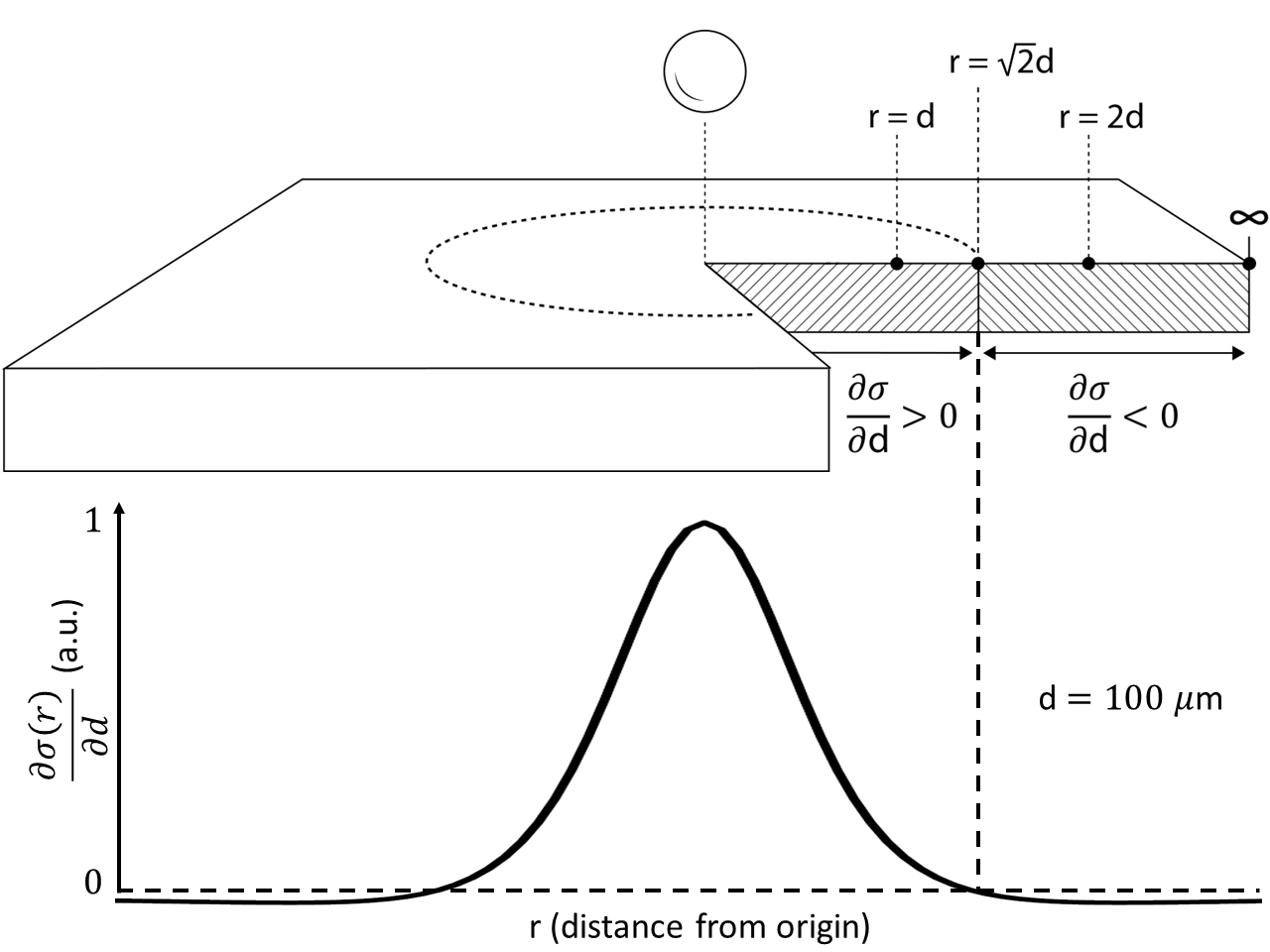}
	\caption{Regions of interest. (Top) The round ball represents a positively charged ion. When the ion moves away from the plane, negative charges induced below the ion in the region $r < \sqrt{2}d$ become dispersed and exit the region as they move towards $r = \infty$. The region $r < \sqrt{2}d$ becomes more positive, meaning $\frac{\partial \sigma}{\partial d} > 0$. In contrast, the region $r > \sqrt{2}d$ gains negative charges from the region $r < \sqrt{2}d$, so it becomes more negative, and $\frac{\partial \sigma}{\partial d} < 0$. (Bottom) A plot of the change in surface-charge density as the ion moves farther from the surface, $\frac{\partial \sigma}{\partial d}$ (in arbitrary units), as a function of distance $r$ from the origin. 
	}
	\label{fig:Figure2}
\end{figure}

			It is useful to know where on the surface the greatest current is, when the particle oscillates, so the current can be measured at that location. In other words, one would like know at what radius $\frac{\partial \sigma}{\partial d}$ is \textit{maximized} as a function of $r$. The problem to solve is: $f(r)=\frac{\partial \sigma}{\partial d}(r)=max$, or:
\begin{equation}
\frac{\partial f(r)}{\partial r} = \frac{\partial}{\partial r}\left[A\frac{r^2-2d^2}{(r^2+d^2)^{5/2}}\right] = 0 ~.
\end{equation}
		\noindent
			This yields two solutions:
\begin{equation}
r=0,
\qquad
\mathrm{and}
\qquad
r=\pm{2d}
\end{equation}
		\noindent
			The first solution at $r=0$ is rather intuitive.  Based on simple reasoning, we might guess that there should also be a second solution somewhere at $r>\sqrt{2}d$. However, that the second maximum is at $r=2d$ is not obvious. (The third solution at $r=-2d$ is not physically meaningful). Of the two solutions, one might predict that the solution at $r=0$ is where the greatest change in surface-charge density occurs. But it is good practice to verify this. We find for the solution at $r=0$:
\begin{equation}
\frac{\partial \sigma}{\partial d}\bigg|_{r=0}=A\frac{-2}{d^3} ~,
\end{equation}
		\noindent
			while for the solution at $r=2d$:
\begin{equation}
\frac{\partial \sigma}{\partial d}\bigg|_{r=2d}=A\frac{2}{5^{5/2}d^3} ~.
\end{equation}
		\noindent
			The two solutions differ in magnitude by a factor of $\frac{1}{5^{5/2}}$, so the change in surface-charge density at $r=2d$ is approximately $~0.02$ times as significant as the change in surface-charge density at $r=0$.
		\noindent
			The discussion so far is summarized in Figure \ref{fig:Figure2}.
		
		\noindent	
			Several relevant quantities discussed so far include:
\begin{enumerate}
	\item The potential $V(x,y,z)$
	\item The surface-charge density $\sigma$
	\item The variation of $\sigma$ with distance $d$ from the \\ surface, $\frac{\partial \sigma}{\partial d}$
	\item The sign and value of $\frac{\partial \sigma}{\partial d}$ at various locations
\end{enumerate}
		
		\noindent
			The significance of these quantities applies to a broad range of experimental configurations.  While specific values may vary from one experimental setup to the next, the main physics will generally remain the same.
		
		\subsection{Induced charge and induced current}
		
			With the general features of the system outlined, we turn to calculating quantitative values relevant to designing an experiment.  First, we look at the total charge which flows into and out of a given region of the conducting plane, during each oscillation of the particle. Specifically, we are interested in the disk-like region defined by the radius $r=\sqrt{2} d$. Measuring the total charge which flows into/out of this region when the particle oscillates gives the maximum electrical current which can be channeled, since adding the change in charge outside of $r=\sqrt{2} d$ would mean adding a contribution of opposite sign.

			\subsubsection{The total charge flow into~/~out of a given region over each oscillation of the particle}\label{ChargeflowQ}
		
			We begin by looking at how the surface-charge density $\sigma$ changes at a single point, $\frac{\partial \sigma}{\partial d} |_{r,\phi}$ , when the particle oscillates.  This expression can then be integrated over a surface area to get the total charge flow into/out of a given region. Referring to Figure 1, let the amplitude of oscillation of a charged particle be denoted by $b_\mathrm{o}$.  In typical ion-trap conditions, for an ion cooled near the ground state the total displacement of the ion, $2b_\mathrm{o}$, is on the order of a few tens of nanometers \cite{daniilidis2013}. The distance $d$ between the ion and the trap electrodes (or conducting surface) is on the order of fifty to two hundred micrometers.  Thus, it is reasonable to assume $d>>2b_\mathrm{o}$, ~and therefore that $\frac{\partial \sigma}{\partial d}$ is approximately constant over $2b_\mathrm{o}$, which is to say $\frac{\partial \sigma}{\partial d}\sim \frac{\partial \sigma}{\partial d}\big|_{d=d_{\mathrm{eq.}}}$
			where $d_{\mathrm{eq.}}$ is the equilibrium distance between the ion and the conducting surface, around which the ion oscillates. Thus, $\frac{\partial \sigma}{\partial d} \sim \mathrm{constant} = A \frac{r^2-2d_{\mathrm{eq.}}^2}{(r^2 + d_{\mathrm{eq.}}^2)^{5/2}}$ ~,
			which is equivalent to the linear relationship:
\begin{equation}\label{deltasigma}
\Delta\sigma = \left( A \frac{r^2-2d_{\mathrm{eq.}}^2}{(r^2 + d_{\mathrm{eq.}}^2)^{5/2}}  \right) \Delta d ~.
\end{equation}
			The variation of the ion's position as it oscillates can be described as sinusoidal, with an amplitude $b(t) = b_\mathrm{o} \sin{\omega t}$. Therefore, integrating equation \eqref{deltasigma} with respect to $d$ gives:
\begin{equation}\label{SurfChrgLinearized}
\sigma = \left( A \frac{r^2-2d_{\mathrm{eq.}}^2}{(r^2 + d_{\mathrm{eq.}}^2)^{5/2}}  \right)b(t) + \sigma_0 ~,
\end{equation}
			where $\sigma_0 = A\frac{d_{\mathrm{eq.}}}{(r^2 +{d_{\mathrm{eq.}}}^2)^{3/2}}$ (equation \eqref{SurfaceCharge} evaluated at $d_{\mathrm{eq.}}$).  Using the expression for $\sigma$ linearized over $d$, which is effectively what has been done to get equation \eqref{SurfChrgLinearized}, it is now possible to evaluate the total charge that flows \textit{into} or \textit{out of} a given region, as the particle oscillates. First, $\sigma$ is integrated over some surface to get the total charge enclosed on that surface.  Second, the result is evaluated once when the particle is at its farthest from the surface, and once again when the particle is at its closest to the surface. Then the difference is taken.
			
			Integrating $\sigma$ over a disk of surface-area extending outwards from $r'=0$ to $r'=r$ (the 'primes' have been added to distinguish between the evaluated and non-evaluated, integrated expression of $\sigma$) gives:
\begin{eqnarray}\label{QtrsfUneval.}
Q=&&\int_{\phi = 0}^{\phi = 2\pi}\int_{r' = 0}^{r' = r} A \Bigg( \frac{r'^2-2d_{\mathrm{eq.}}^2}{(r'^2 + d_{\mathrm{eq.}}^2)^{5/2}}b_\mathrm{o} \sin{\omega t} \nonumber \\ 
&&+ ~\frac{d_{\mathrm{eq.}}}{(r'^2 +{d_{\mathrm{eq.}}}^2)^{3/2}}\Bigg)r'dr'd\phi ~.
\end{eqnarray}
			After the integral is calculated, evaluating the resulting expression when the particle is closest to the surface ($\sin{\omega t}= -1$), and again when it is farthest from the surface ($\sin{\omega t}= +1$), and taking the difference, gives:

%
	
\begin{equation}\label{qtransf}
	Q_{\mathrm{transf.}} = \frac{2qr^2 b_\mathrm{o}}{(r^2 +{d_{\mathrm{eq.}}}^2)^{3/2}}
\end{equation}

			This is the total \textbf{\textit{change}} in charge contained within a disk-like region centered around $r=0$, when an ion above an infinite plane moves from $d_{\mathrm{eq.}} -b_\mathrm{o}$ to $d_{\mathrm{eq.}} +b_\mathrm{o}$. It should be noted that linearizing the surface charge with respect to $d$ to get equation \eqref{SurfChrgLinearized} comes at a cost; in the limit $d \gg r$, the linearized surface-charge density in equation \eqref{SurfChrgLinearized} dies off with $1/d^3$ as the ion moves farther away from the surface. However, in the non-linearized case, equation \eqref{SurfaceCharge} dies off with $1/d^2$. The price of mathematical simplicity is a sacrifice in physical generality---a point which will re-appear as a footnote in section \eqref{VariousSysAndGammas}, and in appendix \ref{AppNonLinGamOpt}.\footnote{Note that since the linearization is only over $d$, in the opposite limiting case when  $r \gg d$, the surface-charge density dies off with $1/r^3$ in both the linearized and non-linearized cases.} We can evaluate expression \eqref{qtransf} for realistic values. For quantum computing applications, a trapped ion is typically cooled to near its motional ground state. For an ion behaving as a quantum harmonic oscillator in its ground state, the amplitude of oscillation (really the standard deviation of the position) around the minimum potential energy is given by:
			\begin{equation}\label{ZeroPtWvFcn}
			b_\mathrm{o} = (\langle 0\vert z^2 \vert 0\rangle)^{\frac{1}{2}} = (\hbar/2m\omega)^{\frac{1}{2}} ~,
			\end{equation}
			where $z$ is the position operator for displacement in the vertical direction, $\hbar = h/\left(2\pi\right)$ is the reduced Plank's constant, $m$ is the mass of the charged particle, and $\omega = 2\pi f$ is the angular frequency along the axis of the harmonic potential. This is defined to be one half the total spread of the zero-point wavefunction. Substituting equation \eqref{ZeroPtWvFcn} into equation \eqref{qtransf} gives 
			\begin{equation}\label{qtransfFreq}
			Q_{\mathrm{transf.}} = \frac{2qr^2}{(r^2 +{d_{\mathrm{eq.}}}^2)^{3/2}}\left( \frac{h}{8\pi^2 mf} \right)^{1/2} ~.
			\end{equation}
			Evaluating equation \eqref{ZeroPtWvFcn} for the mass of a $^{9}\mathrm{Be}^+$ ion $m = 1.5 \times 10^{-26}~$kg, and $f = 5~$MHz gives the spread of the zero-point wavefunction $2b_\mathrm{o} \sim 20~ \times 10^{-9}~$m. Evaluating equation \eqref{qtransfFreq} for $q = -e = 1.6 \times 10^{-19}~ \mathrm{C}~$, $d_{\mathrm{eq.}} = 50 \times 10^{-6}~ \mathrm{m}~$, $r = \sqrt{2}d_{\mathrm{eq.}}$, and $2b_\mathrm{o} \sim 20~ \times 10^{-9}~ \mathrm{m}$, we find: $Q_{\mathrm{transf.}} = 2.6 \times 10^{-23}~ \mathrm{C}$. Note that $Q_{\mathrm{transf.}}$ is maximized for $r = \sqrt{2}d_{\mathrm{eq.}}$, ~as one might expect from equation \eqref{rmincurrent}.
			
			\subsubsection{Induced current}\label{inducedcurrent}
\begin{figure}
	\centering
	\includegraphics[width=\linewidth]{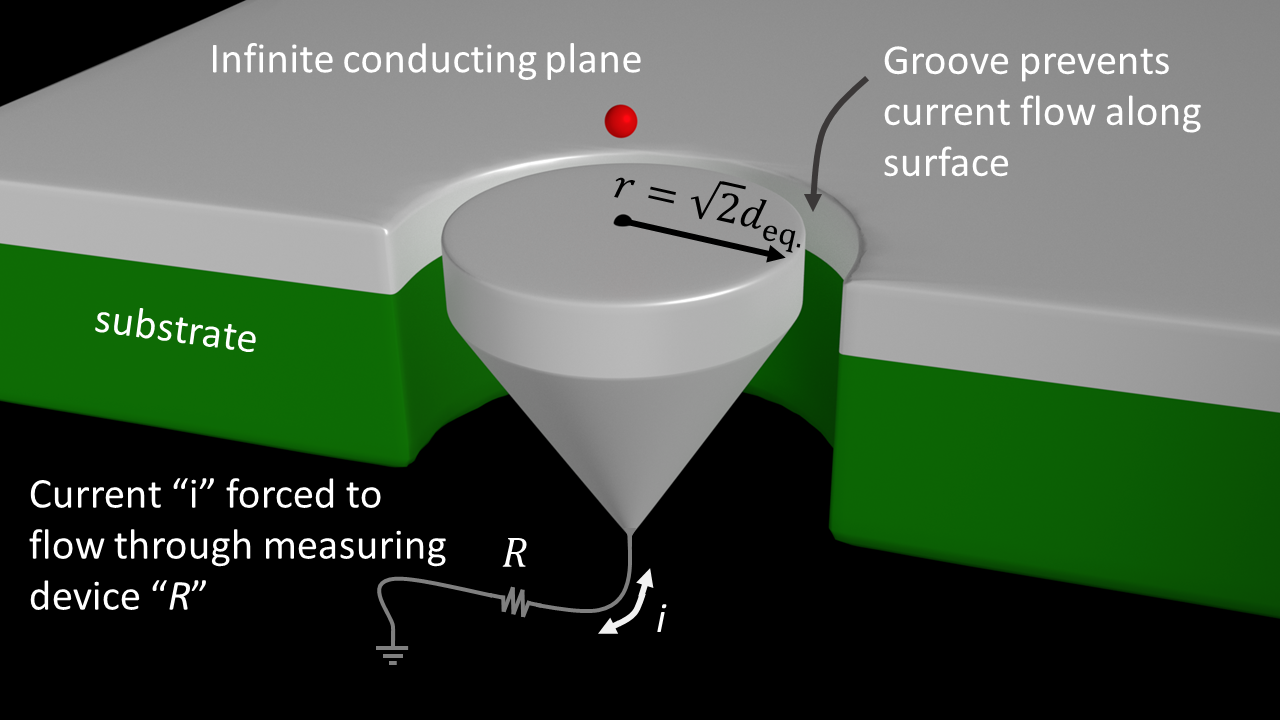}
	\caption{Pickup disk isolated by a groove to channel charge flow. The red circle represents a trapped ion. If the ion oscillates above a continuous infinite conducting plane, the induced charge-imbalance can re-equilibrate by flowing in the in-plane direction. Cutting a groove prevents this, and allows the induced current to be channeled. 
	}
	\label{fig:Figure3}
\end{figure}

			In the previous section \eqref{ChargeflowQ} we calculated the positive charge flow into (really the electron flow out of) a disk-like region centered around $r=0$, when a positive ion moves a small distance away from a grounded conducting plane.  However, this calculation does not specify where the charge flows \textit{from}, or \textit{to}, as it re-distributes to reach the new equilibrium configuration.  For instance, the charge could flow along the surface of the conductor in the in-plane direction, or vertically 'upwards' $/$ 'downwards' through the thickness of the infinite plane.  While this is not relevant when considering equilibrium configurations, the question becomes \textit{essential} if one aims to channel the charge flow to another ion or another type of qubit. For example, if a measuring device is connected to a region of the conducting plane right below the ion (see Figure \ref{fig:Figure3}), what is to prevent the induced currents from flowing sideways along the surface of the plane to re-equilibrate, rather than through the measuring device?  The answer is, nothing. 
		
			Therefore, one must ensure that current is properly channeled.  This can be done by making the relative resistance through the wire to ground, or to another ion, much less than the resistance of any other path to re-equilibration.  In the example of an infinite plane, such a configuration can be implemented by cutting a groove around the disk-shaped region beneath the ion, or otherwise insulating such a disk-shaped region from the rest of the conducting plane (as depicted in Figure \ref{fig:Figure3}).  If a conducting wire is connected from the disk to ground, the \textit{only} way for charge to re-equilibrate is through the wire.  As an alternative to cutting a groove, the rest of the infinite plane could be removed entirely.  Although from an experimental standpoint this may be an improvement, it would also affect the potential between the ion and the disk-shaped conducting surface, meaning all of the equations up until now would have to be revised.  For the sake of developing a model with a potential which is easily tractable using analytical expressions, we will continue to assume an infinite plane with a thin circular groove cut out around a central disk. Practical implications of this assumption are discussed in sections \ref{CalcGamma} and appendix \ref{NeglIndChrg}.	
			With the current properly channeled and using the result from section \eqref{ChargeflowQ}, assuming all of the charge $Q_{\mathrm{transf.}}$ flows to ground in the time of one half-cycle, the time-averaged current throughout each $1/2$-cycle is:
\begin{equation}\label{current}
Q_{\mathrm{transf.}} = I_{\mathrm{av}.}t
\quad 
\Rightarrow
\quad
(2f) \times Q_{\mathrm{transf.}}=I_{\mathrm{av}.} ~,
\end{equation}
			where $f$ is the frequency of oscillation of the ion, typically in the range of $100~$kHz to $10~$MHz. Here, as before, we will assume $f = 5~$MHz. If the trap potential is approximately harmonic, then the ion's motion is approximately sinusoidal.  For a sinusoidal current, the value $I_{\mathrm{av}.}$ is also referred to as $I_{\mathrm{rav}}$, for '\textit{rectified average current}'. Thus \cite{young2007universityAnother}: $I_{\mathrm{av}.}=I_{\mathrm{rav}}=\frac{2}{\pi}I_{\mathrm{max}} \sim 1.3 \times 10^{-16}~\mathrm{A} \label{Iaverage}$, and $I_{\mathrm{max}} \sim 2.0 \times 10^{-16}~\mathrm{A} \label{I_max}$. 
			It should be highlighted that when calculating $I_{\mathrm{av}.}$ we assume all of the charge $Q_{\mathrm{transf.}}$ has time to flow to ground in one half-cycle. This assumption is only valid if the system is designed such that it is true, meaning the resistance $R$ and inductance $L$ of the wire is small enough, for a given capacitance of the disk. To maximize the current, the optimal radius of the disk is the same radius that maximizes $Q_{\mathrm{transf.}}$, $r = \sqrt{2}d_{\mathrm{eq.}}$. This radius can be used to calculate a capacitance for the disk, $C_{\mathrm{d}} = 8 \epsilon_{\mathrm{o}} r$, where $\epsilon_{\mathrm{o}}$ is the permittivity of free space. In section \ref{twoionpotential}, however, we will find that the coupling strength between two ions, ion\#1 and ion\#2, interacting via a conducting wire is maximized for a different radius, $r = d_{\mathrm{eq.}}/\sqrt{2}~$.\footnote{The radius $r$ which maximizes the current is also not the same as the radius which maximizes the energy absorbed by the disk via charge-imbalance, over the course of one half-oscillation. However, the difference between the $r$ values which maximize energy, and current, is minimal.} For the following section we borrow the result from section \ref{twoionpotential} to calculate a bound on the resistance $R$, using the model of a pickup disk attached to ground.

		\subsection{A constraint on the resistance}\label{Rmax}

{\setlength{\parindent}{0cm}
	
		The assumption that all of the charge $Q_{\mathrm{transf.}}$ flows through the wire in the time of one half-oscillation of the ion places an upper bound on the resistance $R$ and inductance $L$ of the wire. Here, we will examine details regarding the upper bound on the resistance $R$. A discussion of the inductance $L$ is left for appendix \ref{OffResAndLneg}, but the main result can be mentioned. We show that the inductance $L$ is intrinsically small, which together with the small capacitance of the coupling system, leads the frequency at which the system is driven, $\omega^{\mathrm{trapped~ion}}_{0}$, to be far below the natural frequency of the coupling system, $\omega^{\mathrm{coupling~system}}_{0}$. In other words, an oscillating trapped ion drives the system in an "electrostatic" fashion in the $\omega^{\mathrm{trapped~ion}}_{0} \ll \omega^{\mathrm{coupling~system}}_{0}$ frequency regime.
		

		Turning to the resistance $R$, we start by considering $Q_{\mathrm{transf.}}$. Since $Q_{\mathrm{transf.}}$ is confined to a disk of finite volume, the mutual repulsion of charges within the disk creates a potential associated with the disk itself. Intuitively, the greater the number of charges confined to a given volume (or surface, which is where the charges end up in a conductor), the greater the potential of that volume or surface. The ratio of the amount of charge confined on an object, to that object's potential relative to ground, is an object's self-capacitance. Figure \eqref{fig:Figure4} (left side) shows a disk where $Q_{\mathrm{transf.}}$ could be confined, resulting in a potential relative to ground, where the net charge-density is taken to be zero. The disk is connected to ground through a resistor, which represents the finite resistance of any real conducting wire.	
	
	\begin{figure}[hbt!]
		\centering
		\includegraphics[width=\linewidth]{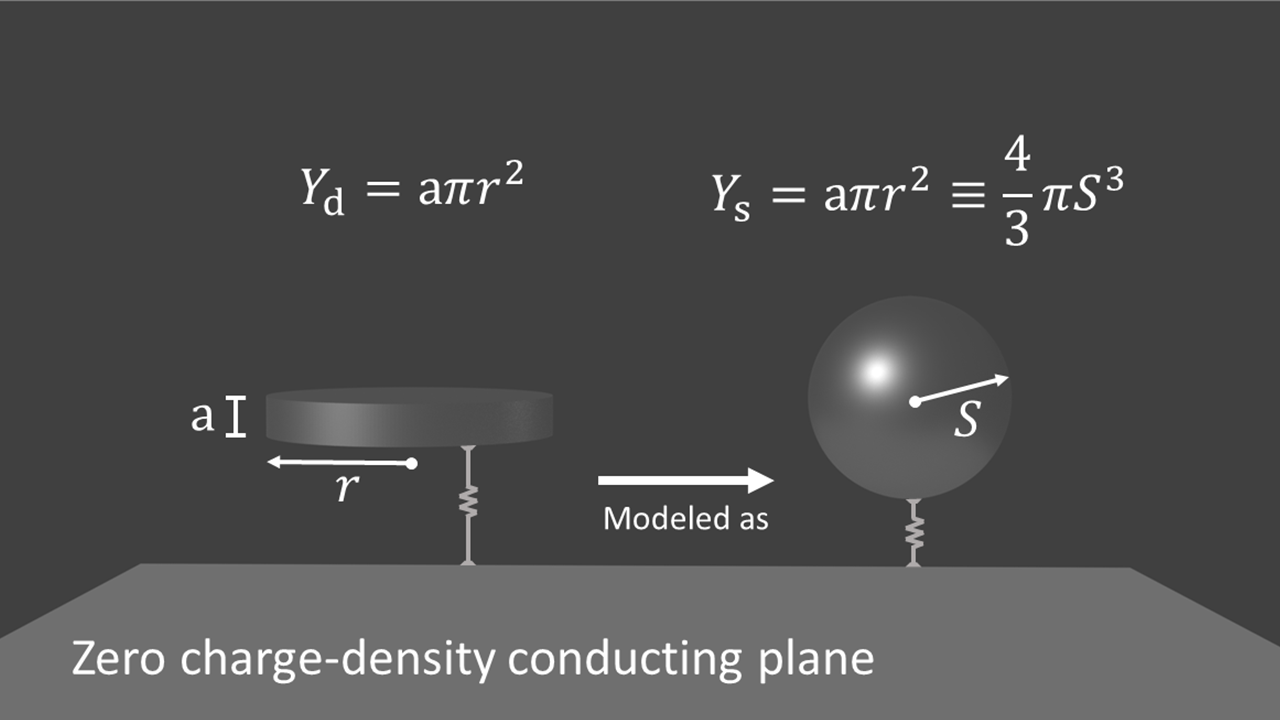}
		\caption{Charge draining from a disk or a sphere. The equations show the volume of a thin disk, $Y_{\mathrm{d}}$, which is imposed as a constraint on the size of a sphere such that $Y_{\mathrm{s}} \equiv Y_{\mathrm{d}}$. Charge placed on a thin disk connected to ground drains off at a certain rate. Charge placed on a sphere with the same volume as the disk drains off more quickly; a sphere minimizes the distance between charges and increases their mutual repulsion.  
		}
		\label{fig:Figure4}
	\end{figure}
	
		To find the upper bound on $R$ we can use one of two approaches. Both approaches use the self-capacitance of the disk, $C_{\mathrm{d}}$, and the fixed time $T/2$ imposed by the period of oscillation of an ion in a trap, to calculate the maximum resistance $R$ which still allows all of the charge to drain to ground. The geometry of an object influences its self-capacitance; hence, the choice of model has a significant impact on the resulting range of acceptable $R$ values. For the two models considered here, a thin disk and a sphere, the self-capacitances are given by $C_{\mathrm{d}} \approx 8 \epsilon_{\mathrm{o}} r$,\footnote{For what qualifies as "thin", we refer to McDonald's result \cite{mcdonald2003capacitance},
		\begin{equation*}
			C_{\mathrm{d}} \approx 4 \pi \epsilon_{\mathrm{o}}\left( \frac{2r}{\pi} \left(1+0.26\sqrt{\frac{\frac{\mathrm{a}}{2}}{r}} \right) \right) ~,
		\end{equation*}
		where $r$ is the radius of the disk and $\mathrm{a}$ is the thickness of the disk. In the limit $\left( \frac{\mathrm{a}}{2} \ll r \right)$ this reduces to the result above.} and $C_{\mathrm{s}} = 4 \pi \epsilon_{\mathrm{o}} S$, respectively. $\epsilon_{\mathrm{o}}$ is the vacuum permittivity, $r$ is the radius of the disk, and $S$ is the radius of the sphere. The first approach to finding the upper bound on $R$ is to recognize that a capacitor connected through a resistor to ground is a low-pass filter. A low-pass filter has a cutoff frequency $\frac{1}{T} = f_{\mathrm{c}} = \frac{1}{2 \pi R C}$. Below this frequency the signal is propagated, so rearranging gives the constraint $R < \frac{1}{2 \pi f_{\mathrm{c}} C}$, where $C$ is the capacitance and $R$ is the resistance through which the capacitor drains to ground. The second approach is to use the capacitor discharge equation $V(t) = V_0 e^{-t/\left( CR \right)}$, where $V(t)$ is the voltage after a time $t$, $V_0$ is the initial voltage, $C$ is the capacitance, and $R$ is the resistance through which the capacitor drains to ground. This equation is valid for both self-capacitance and mutual capacitance. 
	Solving for $R$ gives
	
	\begin{equation}\label{MaxResist}
	R = \frac{t}{C \ln{ \left( V_0 / V_f \right) }} ~,
	\end{equation}	
	
	where we changed the notation $V(t)$ to $V_f$, for "final" voltage. Arbitrarily, we choose to evaluate equation \eqref{MaxResist}, (instead of the low-pass filter equation), for the capacitance $C_{\mathrm{d}}$ of a thin disk. The distance $d_{\mathrm{eq.}} = 50~\mu$m, which determines the optimum radius $r = d_{\mathrm{eq.}} / \sqrt{2}~$, determines the capacitance $C_{\mathrm{d}}~$, and using $f = 5~$MHz, $t=1/\left( 2f \right)$, and letting $V_0 / V_f = 10$ yields $R_{\mathrm{disk}} \sim 17~ \mathrm{M} \Omega$. For comparison, and to illustrate the effect of geometry, we evaluate equation \eqref{MaxResist} for the capacitance $C_{\mathrm{s}}$ of a spherical conductor which has the same volume as the thin disk. The volume of a disk with finite thickness "$\mathrm{a}$", (where $\mathrm{a} \ll r $), is $Y_{\mathrm{d}} = \mathrm{a} \pi r^2$, and the volume of a sphere is $Y_{\mathrm{s}} = \frac{4}{3} \pi S^3$. Imposing $Y_{\mathrm{s}} = Y_{\mathrm{d}}$ gives $S = (\frac{3}{4\pi} \left( \mathrm{a} \pi r^2 \right) )^{1/3}$. Thus, $C_{\mathrm{s}} = 4 \pi \epsilon_{\mathrm{o}} (\frac{3}{4\pi} \left( \mathrm{a} \pi r^2 \right) )^{1/3}$. Assuming $\mathrm{a} = 1~\mu$m, $R_{\mathrm{Sphere}} \sim 40~ \mathrm{M}\Omega$.
	
	The maximum allowable value of $R$ calculated for the thin disk model is less than the maximum resistance allowed by the spherical volume model, which agrees with intuition.  If charges are confined to the surface of a sphere, the average distance between them is minimized compared to if the charges are on the surface of any other shape which has the same volume, and hence on average they experience greater mutual repulsion. 
	In general, the bound on $R$ can be expressed in terms of a function "$g$" such that $R \le g \left( f, d_{\mathrm{eq.}}, \mathrm{a} \right)$, where $f$ is the secular frequency of the trapped ion, $d_{\mathrm{eq.}}$ is its equilibrium distance from the disk, and "$\mathrm{a}$" is the thickness of the thin disk (not to be confused with the wire radius "$a$").
	
	The self-capacitance of an object may change in a dynamical regime if the skin depth at a given frequency approaches the dimension of the object. For a trapped ion oscillating at $5~$MHz, the skin-depth is on the order of $\delta \sim 30~\mu$m \cite{richard}. 
	This is likely greater than or comparable to the thickness of the pickup disk, which \textit{a priori} could be of similar thickness to the trap electrodes, in the range $\mathrm{a} \sim 100~$nm to $\mathrm{a} \sim 10~\mu$m \cite{labaziewicz2008temperature,hite2012100,daniilidis2014surface,sedlacek2018distance}. 
{\setlength{\parindent}{0cm} 
		Unfortunately, the electron configuration within the disk cannot be known in a dynamical regime without dedicated simulations, statistical mechanics calculations, or experiments specific to the thin-disk geometries in question. Hopefully, future studies will develop a fuller understanding of the dynamical behavior of the system.
		
		This section ends with a remark on the interplay between the maximum resistance $R$ of the drainage wire (or measurement device), and the capacitance of the disk. We show above that the wire through which a current is channeled to another ion or a measurement device must not exceed certain upper bounds on $R$. In this description, as in figure \ref{fig:Figure4}, the coupling wire is represented by a wire with finite resistance, but zero capacitance. In reality, the coupling wire also has a finite capacitance. To avoid the wire absorbing charge which flows away from the disk and through the wire, which would diminish the signal, the capacitance and therefore the size of the wire should be as small as possible. Reducing the size of the wire increases its resistance. This trade-off seems potentially problematic. Appendix \ref{AppendixResist} discusses realistic resistances for different types and sizes of wires. Ultimately, in regards to satisfying the two criteria outlined in the Introduction, the time for charges to flow may be less of a constraint on the maximum resistance $R$ than other factors such as increased thermal noise.
	}				

%
}
		
\section{Coupling the motional states of two spatially distant charged particles, using an electrically floating wire}\label{CouplingCst}
			
			Table \ref{tab:DefaultVals} lists parameter values which will be used often in the rest of this study. Where two notations are listed, both notations are used roughly interchangeably.
			
			\begin{table}[ht]
				\caption{Typical parameter values}
				\label{tab:DefaultVals}
				
				\centering
				\begin{ruledtabular}
				\renewcommand{\arraystretch}{1.2}
					\begin{tabular}{lcr}
	
						Description & Variable & Value  \\
						
						\colrule \\
						
						Charge  & $q_{\mathrm{c}}$ & $e = 1.6 \times 10^{-19}~$C  \\
						
						Mass  & $m$ & $m_{_{^{9}\mathrm{Be}^+}} = 1.5 \times 10^{-26}~$kg \\
						
						Dist. btw'n ion \& disk  & $d~$, $d_{\mathrm{eq.}}$ & $50 ~\times 10^{-6}~$m \\
						
						Radius of disk  & $r~$, $r_{\mathrm{disk}}$ & $ d_{\mathrm{eq.}}/\sqrt{2}$  \\
						
						Length of wire  & $l_{\mathrm{w}}~$, $l_{\mathrm{wire}}~$ & $0.01~$m  \\
						
						Wire radius  & $a$ & $10~ \times 10^{-6}~$m  \\
						
						Motional mode freq. rng.  & $\Delta f$ & $500~$Hz  \\
						
						Secular frequency  & $f$ & $5 \times 10^{6}~$Hz  \\
						
						Temperature  & $T$ & $10~$K \\
						
						Resistance  & $R$ & $0.001 ~\Omega$ or $0.0001 ~\Omega$ \\
						
					\end{tabular}
				\end{ruledtabular}
			\end{table}

		\subsection{The physical model: differences and similarities with the "grounded-disk" model}	
			
			In the first part of this study, we considered a disk-like pickup electrode connected to ground via a wire. In what follows, the pickup electrode is instead connected to an electrically-floating wire, which carries the signal to a second 'transmitter' disk placed at some distant location. The surface of the transmitter disk, referred to as disk2, also lies in the plane of the ground-plane, and is positioned below a second trapped ion, ion\#2 (Figure \ref{fig:TwoDisks}).  This new configuration is different from the pick-up disk connected to ground. To give an example, since the system is not grounded, when ion\#1 moves closer to disk1, any charge that moves onto disk1 must come at the expense of charge on disk2.  Despite differences, the setup bears an important calculational similarity to the configuration studied in section \ref{GndedDisk}.  We will take advantage of this and use results from section \ref{GndedDisk} to help characterize the system of two ions coupled by a wire. This section is dedicated to deriving the strength of the coupling interaction between two trapped ions in separate traps. 
			
			Before beginning, we develop an intuitive picture of the mechanics of the system. Suppose ion\#1 moves from its most distant position, to its closest position, relative to disk1. If the disk were initially grounded, an amount of charge $-~Q_{\mathrm{transf.}}$ would be drawn from ground and rise up onto disk1.  However, since the "floating-wire-and-disks" system is not grounded, if $-~Q_{1}$ comes up onto disk1, it must be at the expense of a corresponding $+~Q_{1}$ which develops on disk2, beneath ion\#2 (assuming the intervening wire has zero capacitance).  If ion\#2 is at an equilibrium position $d_{\mathrm{eq.2}}$, now it will be repelled by the charge $+~Q_{1}$ beneath it, until it reaches a new equilibrium position, at a distance farther away from disk2. Thus, before ion\#2 moves up (farther away), it experiences a potential across the distance from $d^i_{\mathrm{eq.2}}$ to $d^{ii}_{\mathrm{eq.2}}$.  In principle, this potential could be calculated---in practice, we calculate the coupling between ion\#1 and ion\#2 directly from the change in the electric field.

			\subsection{Contributions to the potential}\label{twoionpotential}
			
			\begin{figure}[h]
				\centering
				\includegraphics[width=\linewidth]{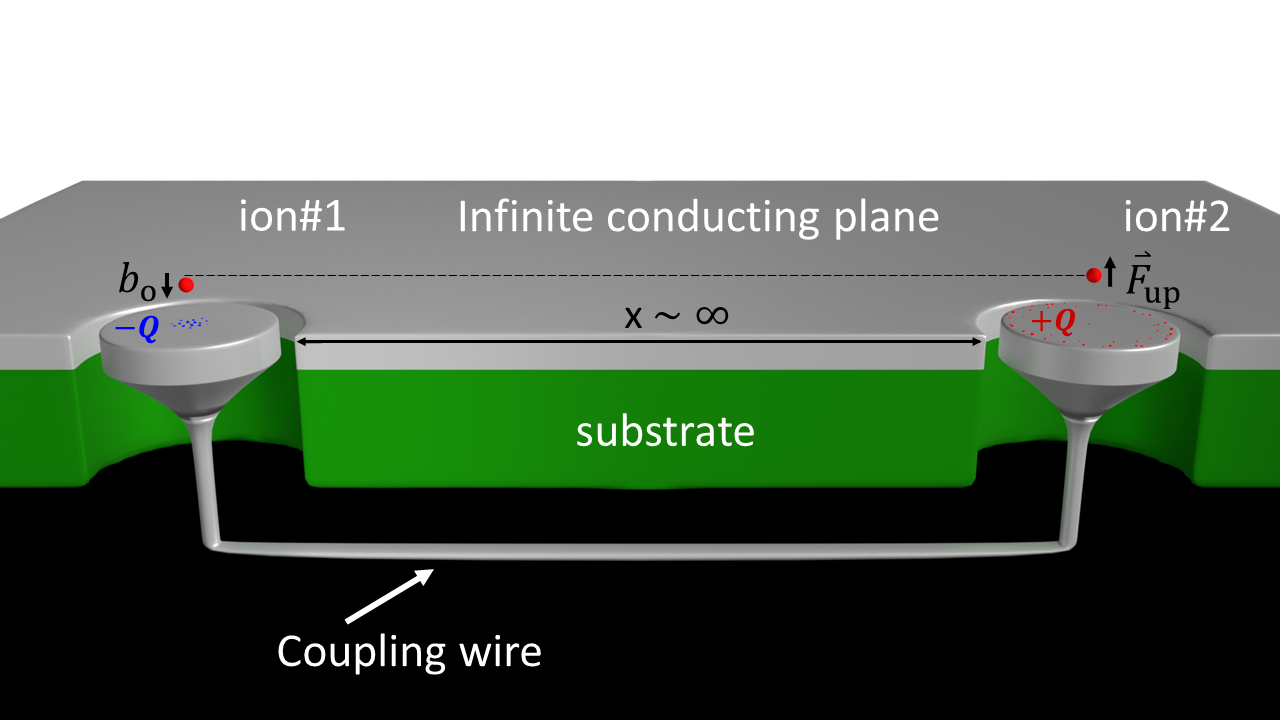}
				\caption{Ion\#1 coupled to ion\#2 using two conducting disks and a connecting wire. The two red dots represent positively-charged ions in separate traps. When ion\#1 on the left moves downwards a distance $b_{\mathrm{o}}$ it induces charge \textcolor{blue}{$-Q$} on disk1. A corresponding positive charge \textcolor{red}{$+Q$} is induced on disk2, which exerts a force $\vec{F}_{\mathrm{up}}$ on ion\#2. The distance $x \sim \infty$ indicates the separation between ion\#1 and ion\#2 (about $0.1~$cm to $1~$cm) is much larger than all other characteristic dimensions of the system, such as the disk radii ($r \sim d_{\mathrm{eq.}} \sim 200~\mu$m).
				}
				\label{fig:TwoDisks}
			\end{figure}
			Due to the symmetry of the setup, we only need to consider one of the two regions in figure \ref{fig:TwoDisks}. To maintain the approximation that the potential is described by two disks surrounded by an infinite plane, the connecting wire is illustrated as going down and running \textit{beneath} the infinite grounded conducting plane. This reduces the chance that the wire itself disturbs the potential. We choose to look at the region above pick-up disk2 (henceforth disk2). There are four main contributions to the potential above disk2. (A possible 5th term describing a fixed bias-voltage is omitted from equation \eqref{fullpotential}):	
			\begin{equation*}
			V_{\infty 2} = 
			V_{\mathrm{ion\#2}} + V^{\mathrm{induced~by~ ion2}}_{\mathrm{disk2~ + ~\infty plane2}}
			\end{equation*}
			\begin{equation}\label{fullpotential}
			~~~~~~~~~~~~~~~~~~~~+ ~V^{\mathrm{induced~ by ~ion1}}_{\mathrm{disk2~  + ~\infty plane2}} + V_{\mathrm{rf-trap}}
			\end{equation}
			Let's consider these contributions one at a time. The first contribution, $V_{\mathrm{ion\#2}}$, is the potential produced by ion\#2 itself. This is given by the expression:
			\begin{equation}\label{IonPotential}
			V_{\mathrm{ion\#2}} =  \frac{1}{4\pi\epsilon_{\mathrm{o}}} \left[\frac{q_{\mathrm{c}}}{\sqrt{r^2_2+(d_{\mathrm{eq.}2}-z_2)^2}}\right] ~,
			\end{equation}
			where the origin of the coordinate system is taken to be the center of disk2, $d_{\mathrm{eq.}2}$ is the equilibrium height of ion\#2, $r_2$ is the radial distance from the origin at which the potential is evaluated, $z_2$ is the  vertical distance from the origin at which the potential is evaluated, and $q_{\mathrm{c}}$ is the charge of ion\#2.
			
			The second contribution, $V^{\mathrm{induced~ by ~ion2}}_{\mathrm{disk2~  + ~\infty ~plane2}}$, or in more compact notation $V^{\mathrm{i2}}_{\mathrm{d\&p2}}$, comes from the surface-charge which is induced by ion\#2 on (disk2  + $\infty$ plane2).  We can identify this contribution by taking the full potential (equation \ref{Griffvoltage}) and subtracting $V_{\mathrm{ion\#2}}$, which gives, after converting to cylindrical coordinates:
			\begin{equation}
			V^{\mathrm{i2}}_{\mathrm{d\&p2}} = \frac{1}{4\pi\epsilon_{\mathrm{o}}} \left[-\frac{q_{\mathrm{c}}}{\sqrt{r^2_2+(z_2+d_{\mathrm{eq.}2})^2}}\right] ~. 
			\end{equation}
			The potential produced by this \textit{induced} charge is of opposite sign from the potential produced by the ion, since the induced charge is of opposite sign from the charge of the ion.
			
			The third contribution is from the surface-charge induced by ion\#1, on (disk2  + $\infty$ plane2).  We call this contribution $V^{\mathrm{induced~ by~ ion1}}_{\mathrm{disk2~  + ~\infty plane2}}$, or more compactly $V^{\mathrm{i1}}_{\mathrm{d\&p2}}$.  To visualize the source of $V^{\mathrm{i1}}_{\mathrm{d\&p2}}$, consider figure \ref{fig:TwoDisks}. Assume that all of the surfaces in the system (disk1, disk2, both planes, and the connecting wire) are initially grounded, at potential $V= 0$.  Additionally, suppose that ion\#1 starts at a distance ${d_{\mathrm{eq.1}}+b_\mathrm{o}}$ above disk1.  From previous calculations for a pick-up disk connected to ground in section \ref{GndedDisk}, when ion\#1 moves from ${d_{\mathrm{eq.1}}+b_\mathrm{o}}$ to ${d_{\mathrm{eq.1}}-b_\mathrm{o}}$, disk\#1 will \textit{try} to draw an amount of charge $-Q_{\mathrm{transf.}}$ upwards onto the surface of disk1.  This would re-establish the equilibrium configuration of (disk1 + $\infty$ plane1).  However, any charge $-Q_1$ drawn onto disk1 must come at the expense of inducing a corresponding charge $+Q_1$ on disk2.  Thus, as ion\#1 oscillates, it induces a periodically oscillating charge on disk 2.  This creates an alternating potential above disk2.
			
			The fourth contribution to $V_{\infty 2}$ is $V_{\mathrm{rf-trap}}$.  This contribution is assumed to be entirely independent of the positions $d_{\mathrm{eq.1}}$ and $d_{\mathrm{eq.}2}$ of the ions.  It is provided by the trap electrodes using an external source, and is well approximated as a harmonic potential, though often with different "spring constants" along different axes of the trap.
			
		\subsection{Calculating the coupling between ion $\#1$ and ion $\#2$}\label{CalcGamma}
		
			The interaction between ion\#1 and ion\#2 can be described by a coupling constant, $\gamma$. The coupling constant $\gamma$ may be thought of as the linear change in force on ion\#2 when the distance between ion\#1 and disk1 changes.  There is also a linear change in the force on ion\#1 when the distance between ion\#2 and disk2 changes, so $\gamma$ contains two terms:
			\begin{equation}\label{gamma}
			\gamma \equiv \frac{1}{2}\left[\frac{\partial}{\partial{b_2}}\left(-\vec{F_1}\right)+\frac{\partial}{\partial{b_1}}\left(-\vec{F_2}\right)\right] ~,
			\end{equation}
			where $b_2$ and $b_1$ represent the instantaneous displacement of ion\#2 and ion\#1 away from their equilibrium positions, as they oscillate. The factor of $\frac{1}{2}$ is to avoid double-counting the energy, as stated in \cite{daniilidis2009wiring}. Equation \eqref{gamma} amounts to averaging the effect of ion\#1 on ion\#2 when ion\#1 moves, with the effect of ion\#2 on ion\#1 when ion\#2 moves, and defining this average to be the coupling strength. It is roughly equivalent to the definition $\gamma \equiv \frac{1}{2}\left[\frac{\partial^2\left(U_1+U_2\right)}{\partial{b_1}\partial{b_2}}\right]$ used in reference \cite{daniilidis2009wiring}, where $U_1$ and $U_2$ are the total energy of ion\#1 and ion\#2 (the total potential they feel, times the charge of the ion $e$). Exact equivalence occurs if $U_1$ has continuous second partial derivatives at all points, which according to Schwarz's theorem guarantees that $\frac{\partial^2 \left(  U_1 \right)}{\partial b_1\partial{b_2}} = \frac{\partial^2 \left(  U_1 \right)}{\partial b_2\partial{b_1}}$. The first term in \eqref{gamma} describes how 	much the force on ion\#1 changes, for a small change in the position of ion\#2. The second term in \eqref{gamma} describes how much the force on ion\#2 changes when the position of ion\#1 changes by a small amount. Another way to think of $\gamma$ is as the 'spring constant' $k$ which expresses the rigidity of the connection between the two ions.  This is illustrated most clearly by re-casting the relationship between the displacement of one ion, and the force felt by the other ion: $\vec{F}_2=k\vec{b}_{1} \Rightarrow \frac{\partial (\vec{F}_2)}{\partial{b_{1}}}= k$.
			
			\noindent
			One might wonder why the partial derivatives are not taken with respect to $d_{\mathrm{eq.1}}$ and $d_{\mathrm{eq.2}}$, and why one should not calculate $\frac{\partial \vec{F}_2}{\partial d_{\mathrm{eq.1}}}$ and $\frac{\partial \vec{F}_1}{\partial d_{\mathrm{eq.2}}}$~, rather than $\frac{\partial \vec{F}_2}{\partial b_{1}}$ and $\frac{\partial \vec{F}_1}{\partial b_{2}}$. Indeed, the total change in force depends on both the change in $d_{\mathrm{eq.1}}$~ \textit{plus} the instantaneous change in $b_1$. However, the contribution from $\frac{\partial \vec{F}_2}{\partial \vec{d}_{\mathrm{eq.1}}}$ is effectively negligible in the overall coupling constant $\gamma_{\mathrm{tot.}} = \gamma_b +\gamma_{d_{\mathrm{eq.}}}$. Over the course of one oscillation of ion\#1 or ion\#2, there should be virtually no change in the average distance of the ion $d_{\mathrm{eq.1}}$ or $d_{\mathrm{eq.2}}$, (for example due to drifts in the trapping potential), but there \textit{will} be a change in the position of the ion due to its short time-scale oscillations. Thus, over a given time $t \sim 1 ~\mu$s~, ~$\Delta d_{\mathrm{eq.1}} \approx \Delta d_{\mathrm{eq.2}} \ll \Delta b_{1} \approx \Delta b_2$. This means during one oscillation, $\vec{F_2}^{\mathrm{tot}}=k_b\vec{\Delta b} + k_{d_{\mathrm{eq.}}}\vec{\Delta d_{\mathrm{eq.}}} \approx k_b\vec{\Delta b}$. What has been referred to as $\gamma$ above should in fact more accurately be referred to as $\gamma_b$.  However, we continue to describe this quantity simply as $\gamma$ for ease of notation, and because there will be no ambiguity moving forward.
			
			Equation \eqref{gamma} shows that for ion\#2 to contribute to $\gamma$, the force $\vec{F}_2$ on ion\#2 must depend on $b_{1}$, the displacement of ion\#1 towards or away from disk\#1. Otherwise, the partial derivative with respect to $b_{1}$ yields zero. Similarly, the force on ion\#1 must depend on $b_{2}$.  Referring to section \eqref{twoionpotential}, among the four contributions to $V_{\infty2}$ (eq. \eqref{fullpotential}) for the potential in region\#2, the first, second, and fourth contributions do not depend on $b_{1}$. Only the third term, $V^{\mathrm{i1}}_{\mathrm{d\&p2}}$, depends on both $b_{2}$ and $b_{1}$. Since $\vec{F_2}$ in equation \eqref{gamma} is the same as writing: "the gradient with respect to $b_{2}$ of $\left( -qV_{\infty2}~ \right)$", any terms in $V_{\infty2}$ which do not depend on $b_{1}$, will correspond to terms in $\vec{F_2}$ which also do not depend on $b_{1}$. Terms $1$, $2$, and $4$ in eq. \eqref{fullpotential} can therefore be ignored for the purpose of calculating $\gamma$:
			\begin{eqnarray}\label{Vinfslashed}
			V_{\infty 2}~ = 
			&&~\cancel{V_{\mathrm{ion\#2}}} + \cancel{V^{\mathrm{induced~ by~ ion2}}_{\mathrm{disk2~ + \infty plane2}}} \nonumber \\
			&& + ~V^{\mathrm{induced~ by~ ion1}}_{\mathrm{disk2~  + ~\infty plane2}} + \cancel{V_{\mathrm{rf-trap}}} ~.
			\end{eqnarray}
			Thus, calculating $\gamma$ reduces to finding $V^{\mathrm{i1}}_{\mathrm{d\&p2}}$. 
			
			To find $V^{\mathrm{i1}}_{\mathrm{d\&p2}}$ we consider the \textit{electric field} produced above disk2. Indeed, the electric field is all that is needed to calculate $\gamma$, because $\gamma$ can be directly calculated from the force $\vec{F}_2=q_{\mathrm{c}} \times \vec{E}_2$.
			The electric field produced by bringing an amount of charge $+~Q_{2}$ onto disk2 is found using the following reasoning. When the charges which make up $+~Q_{2}$ are placed on disk2, they spread out towards the edges of the disk as they try to move as far away from each other as possible.\footnote{A tempting pitfall is to think that when $+~Q_{\mathrm{transf.}}$ goes onto disk2, it spreads out in the same configuration as the surface-charge which is induced below ion\#1, by ion\#1.  However, increasing the net charge on disk2 by moving ion\#1 closer to disk1, (hence drawing electrons off of disk2), is different from increasing the net charge on disk2 by moving ion\#2 away and allowing electrons to flow off of disk2 and re-equilibrate. 
			} Their potential energy is minimized when they form a ring centered around the z-axis. This configuration is reached provided that the charges placed on disk2 are given enough time to redistribute and reach their equilibrium configuration.  For our purposes, the condition is met so long as the charges re-distribute quickly compared to the time-period of one half-oscillation of the ion.  The width of the resulting ring of charge is thin compared to the full radius of the disk, so we consider \textit{all} of the charge to be distributed in a ring of radius $r_{\mathrm{disk}}$. (See proof in appendix \ref{ChargeRing}. The approximation becomes less valid for thinner disks.) Equation \eqref{ringofcharge} describes the electric field along the z-axis, produced by a ring of charge centered around the z-axis and lying in a plane perpendicular to the z-axis \cite{young2007university}:
			\begin{equation}\label{ringofcharge}
			\vec{E}_{\infty2} = \frac{1}{4\pi \epsilon_{\mathrm{o}}}{\frac{Q_2z}{\left(z^2+r^2\right)^{3/2}}}\hat{z} ~.
			\end{equation}
			Here, $z$ is the vertical distance from the plane of the ring, and $r$ is the radius of the ring. When evaluated at the position of ion\#2, and for the radius of disk2, this gives:
			\begin{eqnarray}\label{elecfielddisk2}
			\vec{E}_{\infty2} = 
			&& \frac{1}{4\pi \epsilon_{\mathrm{o}}}{\frac{Q_2d_{\mathrm{eq.2}}}{\left(d_{\mathrm{eq.2}}^2+r_{\mathrm{disk2}}^2\right)^{3/2}}}\hat{z} \nonumber \\
			&& ~~~~~~~~~~~= \frac{1}{4\pi \epsilon_{\mathrm{o}}}{\frac{-Q_1d_{\mathrm{eq.2}}}{\left(d_{\mathrm{eq.2}}^2+r_{\mathrm{disk2}}^2\right)^{3/2}}}\hat{z} ~.
			\end{eqnarray}
			On the right-hand side of expression \eqref{elecfielddisk2},~ $Q_1$ is related to the charge which is induced on disk1 by the displacement of ion\#1; in other words, it is related to $Q_{\mathrm{transf.}}$. The second equality in expression \eqref{elecfielddisk2} represents a strong assumption, that the charge which appears on disk2 due to ion\#1 is exactly the same in magnitude, but opposite in sign from the charge induced on disk\#1 by ion\#1.  This assumption is valid provided the amount of charge which accumulates in the wire and is subsequently given up by the wire during each oscillation of the ion is negligible compared to the total charge which is accumulated and given up by each of the disks.  In other words, it is valid if the capacitance of the wire connecting disk1 and disk2 is negligible compared to the capacitances of disk1 and disk2.  The assumption is lifted below with the introduction of a unitless coefficient $\zeta$, which not only relates $Q_1$ to $Q_{\mathrm{transf.}}$, but also accounts for a finite capacitance of the wire. For the illustrative case above, the second equality can be understood as representing a scenario in which disk1 and disk2 have equal capacitances, and the wire capacitance is zero. In such a situation, the equilibrium state is for each disk to carry an equal amount of the charge-imbalance resulting from the movement of ion\#1 (see appendix \ref{WireCharge_ex1}).
			
			Given a known relationship between $Q_{\mathrm{transf.}}$ and $Q_1$, and a known relationship between $Q_1$ and $Q_2$ (see appendix \ref{WireChargeDist}), the electric field may be rewritten by substituting in $Q_2 = -\zeta Q_{\mathrm{transf.}} = -\zeta \frac{q_{\mathrm{c}} r^2 b_1\left(t\right)}{(r^2 +{d_{\mathrm{eq.}}}^2)^{3/2}}$~. Here, the coefficient $\zeta$ is calculated in appendix \ref{WireCharge_ex3} (equation \eqref{QcQtransf}), and the second equality comes from substituting in the explicit, unevaluated expression for $Q_{\mathrm{transf.}}$ ~(after calculating the integral in \eqref{QtrsfUneval.}, but before equation \eqref{qtransf}). Substituting $Q_2$ into equation \eqref{elecfielddisk2}, the electric field can be rewritten as			
			\begin{eqnarray}\label{E_w/_Qtransf}
			\vec{E}_{\infty2} = 
			&&\frac{1}{4\pi \epsilon_{\mathrm{o}}}{\frac{d_{\mathrm{eq.2}}}{\left(d_{\mathrm{eq.2}}^2+r_{\mathrm{disk2}}^2\right)^{3/2}}} \nonumber \\
			&&
			~~~~~~~~~~~~~~~~~\times \frac{-\zeta q_{\mathrm{c}} r_{\mathrm{disk1}}^2 b_1\left(t\right)}{\left(r_{\mathrm{disk1}}^2 +{d_{\mathrm{eq.1}}}^2\right)^{3/2}}\hat{z} ~.
			\end{eqnarray}
			After multiplying the electric field by $q_{\mathrm{c}}$ we arrive at an expression for the upwards force $\vec{F_2}$ exerted on ion\#2, which depends on $b_{1}$.  From this, and an equivalent expression for the force $\vec{F_1}$ exerted on ion\#1, one can calculate the coupling $\gamma$ (or at any rate, the $\hat{z}$ component of $\gamma$). Dropping the vector notation we have:
			
			\begin{equation*}
			\gamma = \frac{1}{2}\left[\frac{\partial}{\partial{b_{2}}}(-F_1)+\frac{\partial}{\partial{b_{1}}}(-F_2)\right] ~,
			\end{equation*}
			
			therefore,

%
			
			\begin{equation*}
				\gamma = \zeta \frac{1}{8\pi \epsilon_{\mathrm{o}}}{\frac{d_{\mathrm{eq.1}}}{\left(d_{\mathrm{eq.1}}^2+r_{\mathrm{disk1}}^2\right)^{3/2}}}{ \frac{{q_{\mathrm{c}}}^2r_{\mathrm{disk2}}^2}{\left(r_{\mathrm{disk2}}^2 +{d^2_{\mathrm{eq.2}}}\right)^{3/2}}}
			\end{equation*}
			\begin{equation}\label{gammab}
			+ ~\zeta \frac{1}{8\pi \epsilon_{\mathrm{o}}}{\frac{d_{\mathrm{eq.2}}}{\left(d_{\mathrm{eq.2}}^2+r_{\mathrm{disk2}}^2\right)^{3/2}}}{ \frac{ q^2_{\mathrm{c}} r_{\mathrm{disk1}}^2}{\left(r_{\mathrm{disk1}}^2 +{d^2_{\mathrm{eq.1}}}\right)^{3/2}}} ~.
			\end{equation}
			
			\noindent The units of $\gamma$ are Newtons per meter, which corresponds to the force exerted on ion\#2 (ion\#1) for a given displacement of ion\#1 (ion\#2), similar to a standard spring constant. One can verify that this expression results in stronger coupling than the coupling via the direct Coulomb interaction. This is shown in appendix \ref{MinLength}.

			It was mentioned earlier that charges placed on the flat conducting disk will naturally accumulate around the edges, forming a ring.  What was not mentioned is that the positive charges in this ring induce a negative charge-density in the infinite plane around them, thus forming a second, negatively-charged ring.  The second ring partially cancels out the electric field produced by the first ring, and reduces the strength of the coupling constant. We argue in appendix \ref{NeglIndChrg} that this second ring can be ignored, and the calculation of $\gamma$ above is sufficient.

		\subsection{Optimizing the coupling constant $\gamma$}\label{GammaOpt}
			
		The expression for gamma calculated in section \ref{CalcGamma}
		can be optimized as a function of the disk radii to maximize the coupling. Plugging in values for $d_{\mathrm{eq.1}}, d_{\mathrm{eq.2}}, q, \zeta,$ and $\epsilon_{\mathrm{o}}$ gives an expression $\gamma \left( r_{\mathrm{disk1}}, r_{\mathrm{disk2}} \right)$. If one imposes that the coupling system should be symmetrical, $r_{\mathrm{disk1}} = r_{\mathrm{disk2}} = r$, $\zeta$ reduces to $\zeta =\frac{1}{2 + C_b/C_{\mathrm{disk}}}$, where $C_b$ is the wire capacitance. Imposing $d_{\mathrm{eq.1}} = d_{\mathrm{eq.2}} = d$, and substituting in the explicit expression for the disk capacitance, $C_{\mathrm{d}} = 8 \epsilon_{\mathrm{o}} r$ from section \ref{inducedcurrent}, gamma becomes simply $\gamma \left( r \right)$. Now, the optimization of $\gamma$ is an optimization of a function with only one independent variable, $r$, 
\begin{eqnarray}\label{gamma_symm}
&&\gamma =\frac{q^2_{\mathrm{c}}}{4\pi \epsilon_{\mathrm{o}}} \left( \frac{1}{2 + \frac{C_b}{8\epsilon_{\mathrm{o}}r} } \right) \left( \frac{d r^2}{\left(d^2+r^2\right)^{3}} \right)\hat{z} ~.
\end{eqnarray}			
		Unfortunately, maximizing the full expression for $\gamma$ by solving $\frac{\partial}{\partial r}\gamma \left( r \right) = 0$ analytically, is not trivial. However, looking at the portion in the left-most parentheses of equation $\eqref{gamma_symm}$, which corresponds to the expression for $\zeta$~, we see that equation $\eqref{gamma_symm}$ can be simplified in the limit $2 \gg \frac{C_b}{C_{\mathrm{disk}}}$. In this case, $\zeta = \frac{1}{2}$ which is independent of $r$. Optimizing the remaining portion of $\gamma$ one finds

\begin{equation}\label{GammaNL_Opt}
\frac{\partial}{\partial r} \left( \frac{d r^2}{\left(d^2+r^2\right)^{3}} \right) = 0
\qquad
\Rightarrow
\qquad
d = \sqrt{2}r
\end{equation}

		We can also consider the opposite limiting case, where the wire capacitance $C_b$ is much greater than the capacitance of the disks, $2 \ll \frac{C_b}{C_{\mathrm{disk}}}$.
		For a disk capacitance $C_{\mathrm{d}} = 8 \epsilon_{\mathrm{o}} r$ it can be shown that $\gamma$ is optimized for $r_{\mathrm{opt.}} = d_{\mathrm{eq.}}$ (see appendix \ref{C_wireGGC_disk}). Thus, a general rule of thumb is $d_{\mathrm{eq.}} / \sqrt{2} \le r_{\mathrm{opt.}} \le d_{\mathrm{eq.}}$.

\begin{figure}[htbp]
			\centering
			\includegraphics[width=1.0\linewidth]{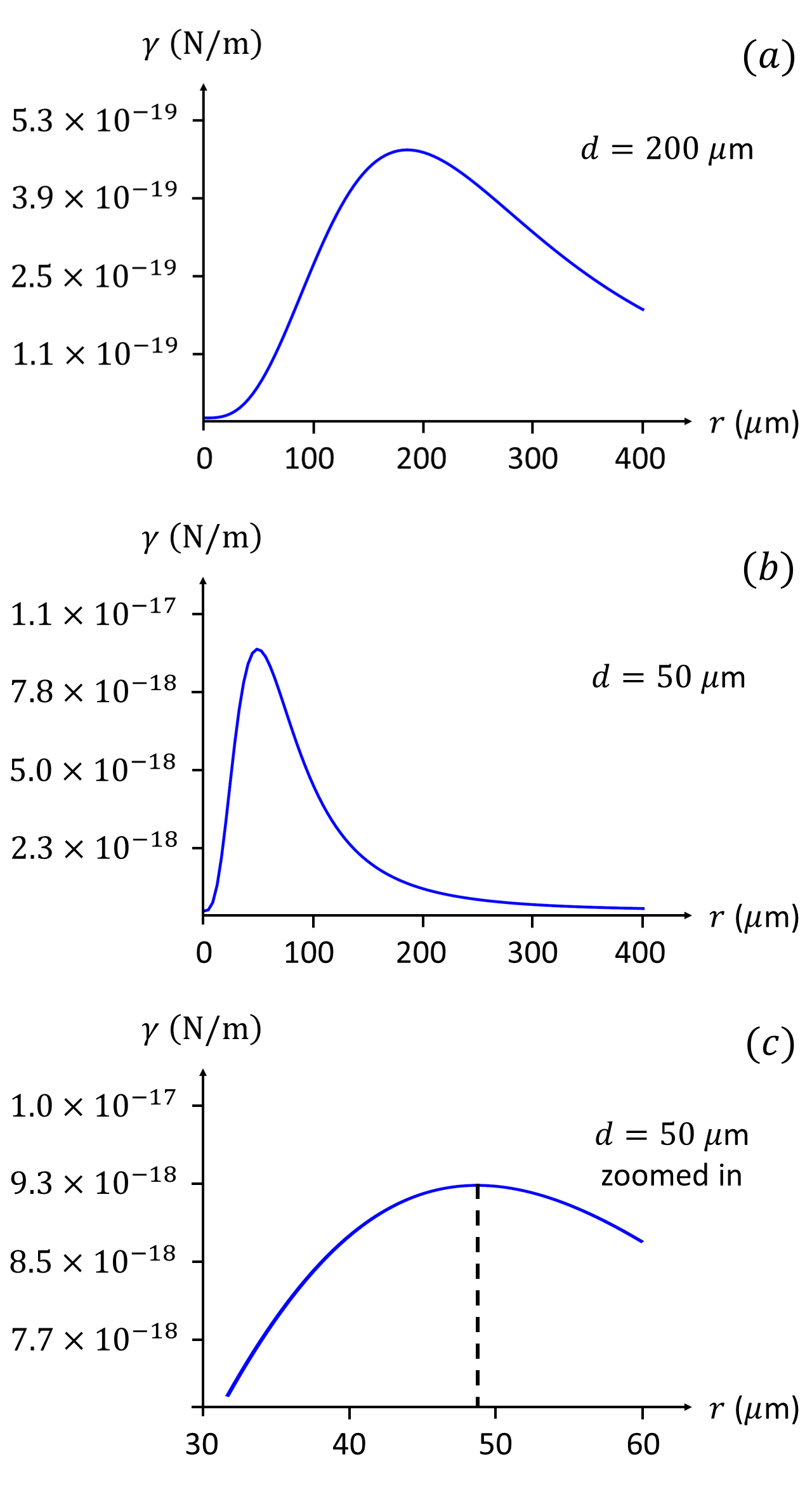}
			\caption{Coupling strength $\gamma$ (N/m) versus $r$ (micro-meters). ($a$) The coupling strength $\gamma$ is plotted as a function of the disk radius $r$, for a distance $d = 200~\mu$m between the two ions and their respective coupling disks. The maximum coupling strength is $\gamma \sim 5 \times 10^{-19}$ N/m. ($b$) The same as in ($a$) but for a distance $d = 50~\mu$m between each ion and its coupling disk. The maximum coupling strength is $\gamma \sim 9 \times 10^{-18}$ N/m. ($c$) the same plot as in ($b$), but zoomed in on the peak.}
			\label{fig:GammaSymm}
		\end{figure}

		A priori, we still do not know whether in general, fully optimizing the expression in equation \eqref{gamma_symm} could yield a significantly improved value of $r_{\mathrm{disk}}$. An exact numerical result for the optimal disk radius $r_{\mathrm{opt.}}$ can be found by plotting $\gamma \left(r\right)$ as a function of $r$, as shown in figure \ref{fig:GammaSymm}.
		
			\noindent To plot $\gamma$, the capacitance $C_b$ is modeled as that of a thin cylindrical wire, $C_b = 2\pi \epsilon_{\mathrm{o}} l_{\mathrm{wire}} / \mathrm{ln} \left( l_{\mathrm{wire}}/a \right)$, where $l_{\mathrm{wire}}$ is the length of the wire and $a$ is its radius. We let $a = 10~\mu$m, and $l_{\mathrm{wire}} = 1~$cm, which gives $C_b = 8.0 \times 10^{-14}~$F. 
			For these parameter values, for two representative distances $d_{\mathrm{eq.}} = 200~\mu$m and $d_{\mathrm{eq.}} = 50~\mu$m, the ratios of wire capacitance to disk capacitance are $C_b / C_{\mathrm{disk}} = 8$, and $C_b / C_{\mathrm{disk}} = 32$, respectively. Both cases fall in the regime $2 \ll \frac{C_b}{C_{\mathrm{disk}}}$, meaning we expect that $r_{\mathrm{opt.}} \sim d_{\mathrm{eq.}}$. This can be verified by looking at the numerical results. Figure \ref{fig:GammaSymm} ($a$) shows the exact numerical result for $d = 200~\mu$m is $r_{\mathrm{opt.}} \approx 180~\mu$m. Figure \ref{fig:GammaSymm} ($b$) and ($c$) show the exact numerical result for $d = 50~\mu$m is $r_{\mathrm{opt.}} \sim 48~\mu$m. We see that for the smaller distance, $d_{\mathrm{eq.}} = 50~\mu$m, the numerical result $r_{\mathrm{opt.}} \approx 48~\mu$m is close to $r = d_{\mathrm{eq.}}$, confirming we are in the $2 \ll \frac{C_b}{C_{\mathrm{disk}}}$ regime. For the larger distance $d_{\mathrm{eq.}} = 200~\mu$m, the analytical result $r_{\mathrm{opt.}} \sim 180~\mu$m is slightly closer to $r_{\mathrm{opt.}} = d_{\mathrm{eq.}} / \sqrt{2} = 140~\mu$m, which is the optimal value for the $2 \gg \frac{C_b}{C_{\mathrm{disk}}}$ regime. Nevertheless, the setup is still closer to the $2 \ll \frac{C_b}{C_{\mathrm{disk}}}$ regime.
	
			Next, we compare numerically optimized results for $\gamma$ with values of $\gamma$ calculated for the $2 \gg \frac{C_b}{C_{\mathrm{disk}}}$ regime, where $r_{\mathrm{opt.}} = d_{\mathrm{eq.}} / \sqrt{2}$, shown in table \ref{tab:Gamma_Opt}. Results are shown for two different values of the distance between the ion and the surface, $d_{\mathrm{eq.}} = 200~\mu$m, and $d_{\mathrm{eq.}} = 50~\mu$m. These two representative distances more or less cover the typical range found in the literature, including the "optimal" value $d_{\mathrm{eq.}} = 132~\mu$m according to one study \cite{bano2026ion}. $\gamma$ is also given for two different wire lengths, $l_{\mathrm{wire}} = 1$~cm, and $l_{\mathrm{wire}} = 10$~cm. The wire radius is $a = 10~\mu$m in all cases.
						
			\begin{table}[h!]
				\caption{Values of $\gamma_{\mathrm{opt.}}$ calculated for $r_{\mathrm{opt.}} = d_{\mathrm{eq.}} / \sqrt{2}$, (the $2 \gg \frac{C_b}{C_{\mathrm{disk}}}$ regime).  }
				\label{tab:Gamma_Opt}
				\begin{ruledtabular}
					\begin{tabular}{lrr}
						\textrm{$\gamma_{\mathrm{opt.}}$ (in $\mathrm{N/m}~, \mathrm{or}~ \frac{\mathrm{kg \cdot m}}{\mathrm{s}^2}$)}&
						\textrm{$d_{\mathrm{eq.}}$}&
						\textrm{$l_{\mathrm{wire}}$} \\
						\colrule \\
						$4.2 \times 10^{-19}$  & $200~\mu$m & 1~ cm \\
						$6.8 \times 10^{-20}$  & $200~\mu$m & 10~ cm \\
						$8.0 \times 10^{-18}$ & $50~\mu$m & 1~ cm \\
						$1.1 \times 10^{-18}$ & $50~\mu$m & 10~ cm \\
					\end{tabular}
				\end{ruledtabular}
			\end{table}
			
			\noindent We focus on the value of $\gamma$ calculated for $d_{\mathrm{eq.}} = 50~\mu$m and $l_{\mathrm{wire}} = 1~$cm, $\gamma = 8.0 \times 10^{-18}~\mathrm{N/m}$. This is slightly lower than the fully optimized value shown in the zoomed-in plot in figure \ref{fig:GammaSymm}, $\gamma \sim 9 \times 10^{-18}~\mathrm{N/m}$. However, the difference between the two values is small despite the fact that the disk radius is small, which increases sensitivity to changes in $r$. This shows that the rule of thumb $d_{\mathrm{eq.}} / \sqrt{2} \le r_{\mathrm{opt.}} \le d_{\mathrm{eq.}}$ optimizes $\gamma$ reasonably well even if one is not certain of the regime in which the setup operates.

		\subsection{Calculating the exchange time $t_{\mathrm{ex.}}$} \label{ExchangeTime-t_ex}	
			
			For the coupling system herein and the other systems in table \ref{tab:GammaVariousSys} (see section \ref{VariousSysAndGammas}), the total energy can be modeled as that of two harmonic oscillators, with an added linear coupling term. The coupling can be described using the analogy of a mass and spring system, where two masses $m_1$ and $m_2$ are attached to two walls by springs with spring constants $k_1$ and $k_2$, and connected together in the middle by a third spring, with spring constant $\gamma$. Taking the displacement of each mass away from its equilibrium position to be $\Delta{x_1}$ and $\Delta{x_2}$, the energy stored in the coupling spring is given by $\frac{1}{2}\gamma \left(\Delta{x_1} - \Delta{x_2} \right)^2$~, which gives rise to a coupling term of the form $\gamma\Delta{x_{1}}\Delta{x_{2}}$. Here, $\Delta{x_1}$ and $\Delta{x_2}$ are measured in such a way that they are both positive for "positive" displacements, towards the right along a number line extending from $0$, at the left-most wall, towards $\infty$, in the direction of the right-most wall (see appendix \ref{WireChargeDist}, figure \ref{fig:2ConnectedSprings}). The full Hamiltonian can then be expressed in the form: $H=\frac{p_1^2}{2m}+\frac{1}{2}m\omega_1^2 \Delta{x^2_1} + \frac{p_2^2}{2m} + \frac{1}{2} m \omega^2_2 \Delta{x^2_2} + \gamma \Delta{x_1} \Delta{x_2}$, where we have made the approximation $\frac{1}{2}k_1 \Delta{x^2_1} + \frac{1}{2}\gamma \Delta{x^2_1} \approx \frac{1}{2}k_1 \Delta{x^2_1}$, and we have defined $\omega_{i=1,2} \equiv \sqrt{k_{i}/{m}}~$. The frequencies $\omega_1$ and $\omega_2$ represent the natural frequencies of the "free" oscillators, due to contributions from all of the trap potentials (for example in equation \eqref{fullpotential} the dominant contributor is $V_{\mathrm{rf-trap}}$), but \textit{not} the contribution from the coupling constant $\gamma$. To describe the system as two coupled quantum harmonic oscillators, $\Delta{x_1}$ must be replaced by $\Delta{\hat{x}_1} = \sqrt{\frac{\hbar}{2}\frac{1}{m_1 \omega_1}}\left(\hat{a}^{\dagger}_1 + \hat{a}_1 \right)$, and the same for $\Delta{x_2}$. Similarly, $p_1$ must be replaced by $\hat{p}_1 = i\sqrt{\frac{\hbar}{2} m_1 \omega_1}\left(\hat{a}^{\dagger}_1 - \hat{a}_1\right)$ and likewise for $p_2$. For this Hamiltonian, it has been shown that under the rotating wave approximation, in the resonance case $\omega_1=\omega_2$, full exchange of motional states can occur after time-evolution \cite{portes2008quantum}. If one ion (ion\#1) is initially in a superposition of states such as $\left( |0\rangle + |n\rangle \right)/ \sqrt{2}$,~ and a second ion (ion\#2) is initially in the ground state $|0\rangle$,~ after a time $t_{\mathrm{ex.}}= \frac{\pi \omega m}{\gamma}$,~ ion\#1 will be in the ground state, while ion\#2 will be in the superposition state $\left( |0\rangle + e^{-i \Theta} |n\rangle \right)/ \sqrt{2}$, where $\Theta = n\pi (m\omega^2/\gamma + 1/2)$. Under the condition $\gamma = n2m\omega^2/(4j-n)$ where $j$ is any integer, the exponential factor multiplying $|n\rangle$ becomes "$1$" and full state exchange occurs \cite{portes2008quantum}. Although the rotating wave approximation is only valid in the limit of small coupling constants ($\gamma/m\omega^2 < 0.1$), \cite{estes1968quantum} for the coupling constants listed in table \ref{tab:GammaVariousSys}, the mass of a beryllium ion, $m_{\mathrm{Be}} = 1.5 \times 10^{-26}~$kg, and typical operating frequencies $f \sim 5~$MHz, in general it is satisfied, since $\frac{1.5 \times 10^{-17} ~\mathrm{N/m}}{\left(1.5 \times 10^{-26}~\mathrm{kg}\right)\left(2\pi \times 5 \times 10^6 ~\mathrm{Hz}\right)^2} = 1 \times 10^{-6}$.	
		
			The criterion $t_{\mathrm{ex.}}= \frac{\pi \omega m}{\gamma}$ provides one half of one of the requirements proposed in the Introduction for achieving successful transfer of quantum information between two particles in separate traps. If the exchange time $t_{\mathrm{ex.}}$ is less than the decoherence time $t_{\mathrm{deco.}}$~, successful quantum information transfer is possible.  However, if $t_{\mathrm{ex.}}$ is greater than or equal to the decoherence time, the time needed to 'send' the information is too great and all of the information will be lost in the process.  This highlights the importance of making $t_{\mathrm{ex.}}$ as small as possible, which can be done by increasing the coupling constant $\gamma$. (\label{Reduce_tex}Other ways to reduce $t_{\mathrm{ex.}}$ include reducing the confinement strength $k_{\mathrm{1}}=\omega^2_{1}m$, or reducing the mass of the particle. Reducing the confinement strength is severely limited by the constraint of increased $1/f^{\mathrm{\tilde{\alpha}}}$ "Anomalous" \cite{leibfried2003quantum, deslauriers2006scaling} heating, discussed in our separate work \cite{van2021coupling}.)
			Substituting equation \eqref{gamma_symm} for $\gamma$ into the expression for $t_{\mathrm{ex.}}$ gives:
		
		\begin{equation*}
			t_{\mathrm{ex.}} = \pi \omega m \qquad \qquad \qquad \qquad ~~~~~~~~~~~~~~~~~~~~~~~~~~~~~~~~~~~~~~~~
		\end{equation*}
		\begin{equation}
		\times \left[ \frac{q^2_{\mathrm{c}}}{4\pi \epsilon_{\mathrm{o}}} \left( \frac{r}{2r + \frac{2 \pi l_{\mathrm{wire}} }{8 \mathrm{ln}\left(l_{\mathrm{wire}}/a\right) } } \right) \left( \frac{d r^2}{\left(d^2+r^2\right)^{3}} \right) \right]^{-1}~.
		\end{equation}	
			
			For the value of $\gamma_{\mathrm{pickup-disk}}$ in table \ref{tab:GammaVariousSys} we can give an associated exchange time $t_{\mathrm{ex.}}= \frac{\pi \omega m}{\gamma}$. We consider a beryllium ion with a mass $m_{\mathrm{Be}} = 1.5 \times 10^{-26}~$kg oscillating at an angular frequency $\omega = 2\pi \times 5~$MHz, and a distance between the ion and the pickup electrode $d_{\mathrm{eq.1}}=d_{\mathrm{eq.2}}=50~\mu$m. This leads to $t^{\mathrm{p.d.}}_{\mathrm{ex.}} = 180~$ms. For a larger distance between the ion and the disk, $d_{\mathrm{eq.1}}=d_{\mathrm{eq.2}} = 200~\mu$m, $t^{\mathrm{p.d.}}_{\mathrm{ex.}} = 3,500~$ms.
			
			This section concludes with a remark on the state exchange time, $t_{\mathrm{ex.}} = \pi \omega m / \gamma$. The expression for $t_{\mathrm{ex.}}$ is explicitly independent of amplitude, or equivalently, of the harmonic oscillator energy of the system.\footnote{If the amplitude of oscillation in a simple-harmonic-oscillator is specified, then so too is its total energy, and vice-versa: $E_{\mathrm{tot}} = \frac{1}{2}kx^2$.} The coupling constant $\gamma$ also does not depend on the amplitude of oscillation of the ion, and this is true for any of the expressions \eqref{gammanoah} - \eqref{gamma_Kotler-Daniilidis}. For a system of two undamped, coupled simple harmonic oscillators, the natural frequency $\omega$ is determined entirely by the masses and spring-constants, and does not depend on the amplitude of the masses' oscillation. Since $t_{\mathrm{ex.}}$ does not depend on the amplitude of oscillation explicitly or implicitly (via $\gamma$), for a fixed trap potential, to a first approximation $t_{\mathrm{ex.}}$ is amplitude-independent. This implies that even for "zero" amplitude of oscillation, a state exchange still occurs. Therefore, if the particle is cooled to its minimum-uncertainty wavepacket, the ground-state, the coupling system should (in theory) still perform its role as an effective mediator of state exchange. The argument above is not strictly accurate because in order for any state exchange to occur, there must exist at least two distinguishable (non-degenerate) states, meaning one of the two ions must have at least one quantum of energy. Additionally, in the introduction of this manuscript we stated that the ratio of the amplitude of the induced signal carrier (voltage or current) to the amplitude of the corresponding noise in the wire should be as large as possible to favor transfer of information. Since the amplitude of the induced voltage or current is proportional to the amplitude of the ion's motion, this relates successful information transfer to the ion's amplitude of oscillation. However, this consideration is added as an independent constraint, and does not change the fact that $t_{\mathrm{ex.}}$ is theoretically independent of the amplitude of oscillation of the charged particles.

		\section{Literature review: comparing results for different systems} \label{LitRev}	
			
			\setlength{\parskip}{15pt plus 1pt minus 1pt}
			
			\subsection{Coupling strength values} \label{VariousSysAndGammas}
			
			The coupling strength derived using the pickup-disk system can be compared with results of other analyses for coupling systems using wire-type architectures. At least four distinct physical implementations have been considered in various studies:
			\begin{enumerate}
				\item Pickup disks linked by a wire (this work), and \cite{zurita2008wiring,marzoli2009experimental,sorensen2004capacitive}.
				
				\item Rectangular pickup electrodes linked by a wire \cite{liang2010two}.
				
				\item A straight suspended wire, with two trapped ions suspended below it \cite{daniilidis2009wiring}.
				
				\item Equivalent-circuit systems \cite{heinzen1990quantum,daniilidis2009wiring, kotler2017hybrid, rica2018double, bohman2018sympathetic}.
			\end{enumerate}
			Table \ref{tab:GammaVariousSys} shows the coupling constant $\gamma$ calculated based on several of these studies. The blue entries rely on an approach used by \cite{zurita2008wiring,marzoli2009experimental,liang2010two} which is slightly different from the one used here, and the brown values rely on another, very different approach outlined in \cite{wineland1975principles}, used in one of two calculations in \cite{daniilidis2009wiring}, and in \cite{heinzen1990quantum, kotler2017hybrid, rica2018double, bohman2018sympathetic, bohman2021sympathetic, an2022coupling}. Most of the studies in the first three implementations above are targeted towards charged particles trapped in rf traps (Paul traps), or Penning traps (in the case of references \cite{zurita2008wiring,marzoli2009experimental}). 
			In \cite{heinzen1990quantum, daniilidis2009wiring, kotler2017hybrid, rica2018double, bohman2018sympathetic, bohman2021sympathetic, an2022coupling} a trapped particle is modeled as a charge placed between two electrodes resembling plates of a parallel-plate capacitor. As the charged particle oscillates, it induces a current within the plates which form part of an equivalent lumped circuit element system that may be linked to another particle, as described in appendix \ref{DerivingIonInductance}. This model, originally described in \cite{wineland1975principles}, has been shown to be inappropriate for calculating coupling constants \cite{van2024improved}. The corresponding results in brown are included in table \ref{tab:GammaVariousSys} for context only.			

			For the four physical implementations above, calculations of the coupling strength in all listed references (except \cite{sorensen2004capacitive}) rely on three general methods, described in the following works. 1) This work, and one calculation in \cite{daniilidis2009wiring}, uses a piecewise description of the system to construct a global model. 2) Reference \cite{zurita2008wiring} formulates the system in terms of Green's functions and uses these to derive results. This is copied in \cite{liang2010two}. 3) Reference \cite{wineland1975principles} introduces an equivalent lumped circuit element model which is widely used (incorrectly) to calculate coupling strengths.
			
\begin{table}[h]
	\caption{Values of $\gamma$ for various systems}
	\label{tab:GammaVariousSys}
	\begin{ruledtabular}
		\begin{tabular}{llr}
			$\gamma$~ (in $\mathrm{N/m}~, \mathrm{or}~ \frac{\mathrm{kg \cdot m}}{\mathrm{s}^2}$)  &
			$d_{\mathrm{eq.}} = 50~ \mu \mathrm{m}$ &
			$d_{\mathrm{eq.}} = 200~ \mu \mathrm{m}$ \\
			\colrule \\
						
			$\gamma_{\mathrm{V.~pickup-disk}}$  & $8.0 \times 10^{-18}$ & $4.2 \times 10^{-19}$ \\
						
			$\gamma_{\mathrm{D.~suspended-wire}}$  & $7.5 \times 10^{-18}$ & $7.9 \times 10^{-19}$ \\
						
			\textcolor{darkcerulean}{
			$\gamma_{\mathrm{Z.~transmission-line}}$} & 
			\textcolor{darkcerulean}{$1.5 \times 10^{-17}$} & 
			\textcolor{darkcerulean}{$8.2 \times 10^{-19}$} \\
						
			\textcolor{darkcerulean}{
			$\gamma_{\mathrm{L.~rectangular-electrodes}}$} & 
			\textcolor{darkcerulean}{$3.0 \times 10^{-19}$} & 
			\textcolor{darkcerulean}{$2.1 \times 10^{-19}$} \\
						
			\textcolor{otterbrown}{
			$\gamma_{\mathrm{W.~mass-spring,~pickup-disk}}$} & \textcolor{otterbrown}{$8.0 \times 10^{-18}$}  & \textcolor{otterbrown}{$8.1 \times 10^{-19}$} \\
						
			\textcolor{otterbrown}{
			$\gamma_{\mathrm{W.~mass-spring,~parallel~plate}}$} & \textcolor{otterbrown}{$3.0 \times 10^{-17}$}  & \textcolor{otterbrown}{$1.6 \times 10^{-18}$} \\
						
			\textcolor{otterbrown}{
			$\gamma_{\mathrm{W.~mass-spring,~suspended~wire}}$} & \textcolor{otterbrown}{$1.1 \times 10^{-17}$}  & \textcolor{otterbrown}{$1.8 \times 10^{-18}$} \\
		\end{tabular}
	\end{ruledtabular}
\end{table}
			
			\noindent Table \ref{tab:GammaVariousSys} shows values of the coupling constant for the pickup-disk system herein as well as the systems in \cite{daniilidis2009wiring, zurita2008wiring, liang2010two, marzoli2009experimental, kotler2017hybrid, rica2018double}. In all cases, we have standardized the coupling setups by attributing them the same total capacitance, and where possible, the same \textit{distribution} of capacitances. To draw meaningful comparisons, values of the coupling strength expressed as "g" in references \cite{kotler2017hybrid, rica2018double}, or as "$\pm \Omega_{12}$" in references \cite{zurita2008wiring, marzoli2009experimental, liang2010two}, are rewritten in terms of $\gamma$ (the same as in reference \cite{daniilidis2009wiring}), using the relationship $\gamma = \left( 2m \omega_\mathrm{o} \right)\mathrm{g} = \pm \left( 2m \omega_\mathrm{o} \right)\Omega_{12}$. This gives a quantity which is independent of operating frequency and mass. As a result, when the phrase "coupling strength" is used here to refer to $\gamma$, it does not mean the same thing as when it is used to describe 'g' or $\Omega_{12}$ in other literature. When the term "coupling strength" is used in relation to 'g' or $\Omega_{12}$, it refers to the rate at which states are exchanged considering relevant properties of a given system including the mass of the particles and their frequencies of oscillation. In this case, g (or $\Omega_{12}$) is in units of s$^{-1}$, and here we refer to this as the Rabi coupling strength. In contrast, in this work when we talk about the "coupling strength" we refer to $\gamma$, meaning the force due to the interaction between two coupled objects, which is expressed in units of N/m. Notice that in section \ref{ExchangeTime-t_ex} we stated that $t_{\mathrm{ex.}} = \frac{\pi \omega m}{\gamma}~$, while $1/g = 1/\Omega_{12} = \frac{2\omega m}{\gamma}$. This indicates that $t_{\mathrm{ex.}}$ is effectively the time of one Rabi oscillation of a two-level system. 
			
			All values of $\gamma$ in table \ref{tab:GammaVariousSys} are evaluated assuming a singly-charged particle at a distance $d_{\mathrm{eq.1}} = d_{\mathrm{eq.2}} = 50 ~\mu$m or $d_{\mathrm{eq.1}} = d_{\mathrm{eq.2}} = 200 ~\mu$m from the coupling system, and for a wire-length of $1$~cm, where applicable. The wire radius for the pickup-disk system is taken to be $a =  10~\mu$m, to maximize the signal-to-noise ratio by minimizing the resistance, as outlined in \cite{van2021coupling} 
			and appendix \ref{AppendixResist}. This leads to a total capacitance for the pickup-disk system $C = \left( 2 \times 8\epsilon_{\mathrm{o}} r_{\mathrm{disk1}} + \frac{2\pi \epsilon_{\mathrm{o}} l_{\mathrm{wire}}}{\mathrm{ln}{\left(l_{\mathrm{wire}}/a \right)}} \right) \sim 8.6 \times 10^{-14}$~F or $1.0 \times 10^{-13}$~F, where the first term in the expression for capacitance is due to the capacitances of the two pickup disks, and the second term is the leading order term for the capacitance of a finite, straight, thin cylindrical wire, where $\epsilon_{\mathrm{o}}$ is the permittivity, $l_{\mathrm{wire}}$ is the length of the wire, and $a$ is the radius of the wire. For the suspended-wire setup, since we impose the requirement that the total capacitance of the coupling system should be the same as for the pickup-disk system, this constraint together with the assumption of a wire of length $l_{\mathrm{wire}} = 1$~cm leads to a wire radius of $a = 15.5 ~\mu$m or $a = 38.5~\mu$m. For the capacitance $C$ in equations \eqref{gammakotler} and \eqref{gamma_Kotler-Noah} (see below), the same total capacitance $8.6 \times 10^{-14}$~F~, or $1.0 \times 10^{-13}$~F is used. To evaluate $\gamma_{\mathrm{L.~rectangular-electrodes}}~$, we assume the surface-area of one rectangular pickup electrode in reference \cite{liang2010two} is the same as the surface-area of a pickup disk, defined by the optimal relationship between radius and equilibrium ion position, $r = d_{\mathrm{eq.}}/\sqrt{2}$. This allows us to define a width $a'$ for the rectangular coupling electrode.

			Table \ref{tab:GammaVariousSys} shows that for a distance $d_{\mathrm{eq.}} = ~50~\mu$m, the coupling constant calculated using the pickup-disk setup falls between the values calculated for the suspended-wire and transmission line studies. 
			The shape of the system influences 
			its function.
			For example, the pickup-disk system is designed to store the bulk of the charge on two disks directly below the trapped particles, while the connecting wire is made as small as possible (within the resistive constraints discussed in \cite{van2021coupling} 
			and appendix \ref{AppendixResist}), to minimize the charge stored on the wire. In contrast, the capacitance of the suspended wire is evenly distributed along its length. Another way to express the dependence on geometry is to observe that if a certain amount of charge $Q$ is placed on an isolated suspended-wire system, it would spread out evenly along the length of the wire. However, if the same amount of charge $Q$ is placed on the pickup-disk system, the majority of the charge ends up on disk1 and disk2, as close as possible to the ions. 

			In table \ref{tab:GammaVariousSys} we see that the values of $\gamma$ for the distances $d_{\mathrm{eq.}} = 50~\mu$m and $d_{\mathrm{eq.}} = 200~\mu$m reveal differences in the distance dependence between the various systems. Comparing the pickup-disk and the suspended-wire systems, multiplying the distance by a factor of 4 reduces the coupling strength of the pickup-disk system by over one order of magnitude. However, for the same change in distance, the coupling strength for the suspended-wire system decreases by less than one order of magnitude. This motivates us to take a closer look at the various expressions for the coupling strength $\gamma$ which are found in the literature.

		\subsection{Comparing different expressions from the literature for $\gamma$} \label{VariousGammas}
				
			It is instructive to consider the expression for the coupling constants derived based on the pickup-disk setup described here, alongside the various systems described above, from \cite{daniilidis2009wiring, zurita2008wiring, liang2010two, kotler2017hybrid}.\footnote{Equation \eqref{gammazurita} is calculated as $2m\omega \Omega_{12}$, where $\Omega_{12}$ is equation (17) in reference \cite{zurita2008wiring}. The function $\alpha \left( \bar{y} \right)$ in reference \cite{zurita2008wiring}, appendix A, is defined as $\alpha \left( \bar{y} \right) \equiv \frac{a^3}{\left( \bar{y}^2 +a^2 \right)^{3/2}}$, where the notation $a$ in reference \cite{zurita2008wiring} is roughly $r$ here, and $\bar{y}$ is $d_{\mathrm{eq.}}$. $\alpha \left( \bar{y} \right)$ should not be confused with $\tilde{\alpha} \left( x \right)$, also in appendix A of reference \cite{zurita2008wiring}.}		
\begin{equation*}
\gamma_{\mathrm{pickup-disk}} = \zeta \frac{1}{8\pi \epsilon_{\mathrm{o}}}{\frac{d_{\mathrm{eq.1}}}{\left(d_{\mathrm{eq.1}}^2+r_{\mathrm{disk1}}^2\right)^{3/2}}}{ \frac{q^2_{\mathrm{c}}r_{\mathrm{disk2}}^2}{\left(r_{\mathrm{disk2}}^2 +{d^2_{\mathrm{eq.2}}}\right)^{3/2}}} \end{equation*}
\begin{equation}\label{gammanoah}
+ ~\zeta \frac{1}{8\pi \epsilon_{\mathrm{o}}}{\frac{d_{\mathrm{eq.2}}}{\left(d_{\mathrm{eq.2}}^2+r_{\mathrm{disk2}}^2\right)^{3/2}}}{ \frac{ q^2_{\mathrm{c}} r_{\mathrm{disk1}}^2}{\left(r_{\mathrm{disk1}}^2 +{d^2_{\mathrm{eq.1}}}\right)^{3/2}}}
\end{equation}
				\bigskip
\begin{eqnarray}\label{gammadaniilidis}
\cite{daniilidis2009wiring}~&&\gamma_{\mathrm{suspended-wire}} =
 \frac{2e^2H^2}{\pi \epsilon_{\mathrm{o}}L}\frac{1}{\mathrm{ln}\left[\left(2H-a\right)/a\right]}~~~~~~~~~~ \nonumber \\
&&~~\times \frac{1}{\left( H^2-h_{0,1}^2\right)\left( H^2-h_{0,2}^2\right)}
\end{eqnarray}
\begin{eqnarray}\label{gammazurita}
\cite{zurita2008wiring}\cite{marzoli2009experimental}~&&\gamma_{\mathrm{trans.-line}} = \frac{e^2 r^4}{\left(d^2_{\mathrm{eq.}} + r^2\right)^3}\frac{1}{2C_\mathrm{o} +C_\mathrm{w}} \nonumber \\ 
&&= \frac{e^2 r^4}{\left(d^2_{\mathrm{eq.}} + r^2\right)^3}\frac{1}{\left( \epsilon_{\mathrm{o}}\frac{4\pi r h}{s} + \pi \epsilon_{\mathrm{o}}L/\mathrm{ln}\left( b/a \right) \right)}\qquad
\end{eqnarray}
\begin{eqnarray}\label{gammaliang}
\cite{liang2010two}~ &&\gamma_{\mathrm{rect.-electrodes}} = \frac{16q^2_{\mathrm{c}}a'^2}{\pi^2\left(4d^2_{\mathrm{eq.}}+a'^2\right)^2}\frac{1}{2C_\mathrm{o} +C_\mathrm{w}}~~~~~~~~ \nonumber \\
&& =\frac{16q^2_{\mathrm{c}}a'^2}{\pi^2\left(4d^2_{\mathrm{eq.}}+a'^2\right)^2}\frac{1}{ 2\epsilon_{\mathrm{o}}\frac{2 l h'}{w'} + \epsilon_{\mathrm{o}}\left(2h/w\right)}
\end{eqnarray}
\begin{eqnarray}\label{gammakotler}
\cite{kotler2017hybrid, rica2018double}~ \gamma^{\mathrm{\vert \vert plate}}_{\mathrm{mass-spring}} = &&
 ~m \omega_0^2 \sqrt{\frac{C_1C_2}{\left(C_1 + C\right)\left(C_2 + C\right)}} \nonumber \\
&& \approx m\omega^2_0 \frac{\sqrt{C_1C_2}}{C} = \frac{\alpha^2 q^2_{\mathrm{c}}}{d^2_{\mathrm{Win.}} C} 
\end{eqnarray}
\begin{equation}\label{gamma_Kotler-Noah}
\gamma^{\mathrm{p.d.}}_{\mathrm{mass-spring}} = \frac{\eta q^2_{\mathrm{c}} r^2}{2d_{\mathrm{eq.}} \left(r^2 + d_{\mathrm{eq.}}^2\right)^{3/2}C}~~~~~
\end{equation}
\begin{equation}\label{gamma_Kotler-Daniilidis}
\cite{daniilidis2009wiring}~\gamma^{\mathrm{s.w.}}_{\mathrm{mass-spring}} = \frac{2 e^2}{\mathrm{ln} \left(\frac {2H-a}{a} \right) \left( H^2 - h^2 \right)C}
\end{equation}
{\setlength{\parindent}{0cm}
			Equation \eqref{gammanoah} corresponds to the present work, equation \eqref{gammadaniilidis} is from reference \cite{daniilidis2009wiring}, equation \eqref{gammazurita} is from reference \cite{zurita2008wiring}, equation \eqref{gammaliang} is from reference \cite{liang2010two}, equation \eqref{gammakotler} is based on reference \cite{kotler2017hybrid} (indirectly reference \cite{wineland1975principles}), and equations \eqref{gamma_Kotler-Noah} and \eqref{gamma_Kotler-Daniilidis} are calculated in appendix \ref{DerivingIonInductance} and reference \cite{daniilidis2009wiring}, respectively, based on the method outlined in reference \cite{wineland1975principles}. In equations \eqref{gammanoah} through \eqref{gamma_Kotler-Daniilidis}, $q_{\mathrm{c}}$ or $e$ are the particle charge and the charge of the electron, respectively, and $d_{\mathrm{eq.1}}$, $d_{\mathrm{eq.2}}$, $d_{\mathrm{eq.}}$ represent distances between the trapped charge and the relevant pickup / transmission electrode. Additionally, $r_{\mathrm{disk1}}$, $r_{\mathrm{disk2}}$, and $r$ represent the radius of the disk-like pickup / transmission electrode (where relevant), and $\epsilon_{\mathrm{o}}$ is the permittivity of free space. In equations \eqref{gammanoah} and \eqref{gamma_Kotler-Noah} the coefficients $\zeta$ and $\eta$, respectively, are crucial efficiency factors derived in appendix \ref{WireChargeDist} which range from $0 \le \zeta, \eta \le 1$, and relate the relative values of the capacitances of the pickup-disks and the capacitance of the connecting wire, the latter of which depends on the length of the wire $l_{\mathrm{wire}}$, and its radius $a$. In equation \eqref{gammadaniilidis}, $L$ is the length of the wire coupling ion\#1 and ion\#2, $H$ is the height of the suspended coupling wire above a ground plane, and $h_{0,1}$ and $h_{0,2}$ are the equilibrium heights of ion\#1 and ion\#2 above the ground plane (so $\left( H-h_{0,1} \right)$ in equation \eqref{gammadaniilidis} is equivalent to $d_{\mathrm{eq.1}}$ in equation \eqref{gammanoah}). Additionally, $a$ is the radius of the coupling wire. In equation \eqref{gammazurita}, $h$ is the thickness of the electrodes, and "the gaps between them (defined by etchings) have width $s$ [\textit{sic}]". Again, $L$ is the length and $a$ is the radius of the wire, and $b$ denotes the separation between the two parallel wires in \cite{zurita2008wiring}, given to be on the order of $500~\mu$m. Here, $C_\mathrm{o}$ denotes the capacitance of one planar Penning trap, and $C_\mathrm{w}$ is the capacitance of the pair of wires connecting two traps. In equation \eqref{gammaliang}, $a'$ is the width of the rectangular coupling electrode (denoted $a$ in \cite{liang2010two}), $l$ is the length of the center pickup electrode, $h'$ is the thickness of the trap electrodes, $w'$ is "the space between trap electrodes [\textit{sic}]", $h$ is the thickness of the transmission line, and $w$ is "the space between the transmission lines [\textit{sic}]". Note that for calculating the coupling strengths in table \ref{tab:GammaVariousSys}, equations \eqref{gammazurita} and \eqref{gammaliang} only depend on the total capacitance $\left(2C_\mathrm{o} + C_\mathrm{w}\right)$, which we imposed to be fixed across all systems. We will come back to this point below. In equation \eqref{gammakotler}, the values $C_1$, $C_2$, and $C$ denote capacitances in an electrical circuit analogous to two masses connected to three springs in the series configuration "$spring_1-m_1-spring_m-m_2-spring_2$". For a system of two coupled ions, the "capacitances" of ion\#1 and ion\#2 ($C_1$ and $C_2$, respectively) are calculated based on the current induced in the coupling wire by the ions' movement, as detailed in reference \cite{wineland1975principles}. The value $C$ denotes the total capacitance of the coupling apparatus, (not including the ions), the same as $\left( 2C_\mathrm{o}+C_\mathrm{w} \right)$ in equations \eqref{gammazurita} and \eqref{gammaliang}. For the system described in references \cite{wineland1975principles, kotler2017hybrid}, and \cite{rica2018double}, $C_1$ and $C_2$ are given by $C_{\mathrm{ion}} = C_1 = C_2 = \frac{\alpha^2 q^2_{\mathrm{c}}}{md^2_{\mathrm{Win.}}\omega^2_0}$, where $\alpha$ is a geometric coefficient set to $\alpha = 1$, $m$ is the mass of the particle (for example a beryllium ion), $q_{\mathrm{c}}$ is the charge of one particle, and $\omega_0$ is the resonant frequency of oscillation of the particle in the trap. The distance $d_{\mathrm{Win.}}$ is the separation between two end-cap electrodes on either side of the charged particle (so $d_{\mathrm{Win.}}/2$ in the notation of references \cite{kotler2017hybrid, wineland1975principles} is $d_{\mathrm{eq.}}$ in the pickup-disk notation). The approximate equality in equation \eqref{gammakotler} comes from the limiting case $C_1 = C_2 \ll C$, justified based on the fact that $C_1 = C_2 \sim 9.8 \times 10^{-19}$~F, while $C \sim 4.1 \times 10^{-14}$ to $8.4 \times 10^{-14}$~F. (To estimate $C$ for this approximation, we do not use the total capacitance $C$ from section \ref{VariousSysAndGammas}, to avoid circular logic. Instead, we assume a cylindrical connecting wire and impose "reasonable" values on the length and radius of the wire, to obtain a rough estimate. The capacitance of the wire is given by $C_{\mathrm{w}} = 2 \pi \epsilon_{\mathrm{o}} l_{\mathrm{wire}}\big/ \mathrm{ln}\left(l_{\mathrm{wire}}/a\right)$, where $l_{\mathrm{wire}}$ is the length of the wire, and $a$ is it's radius. We let $l_{\mathrm{wire}} = 1$~cm, and for $a$ we either use the dimensions proposed in appendix \ref{AppendixResist}, for a lower bound, or the dimensions in reference \cite{daniilidis2009wiring}, for an upper bound, namely $a =  12.6$~nm or $a =  12.5~\mu$m, respectively.) The right-most equality in equation \eqref{gammakotler} comes from substituting in the explicit expressions for $C_1$ and $C_2$ (above). In equations \eqref{gamma_Kotler-Noah} and \eqref{gamma_Kotler-Daniilidis}, the same starting expression is used as in equation \eqref{gammakotler}, as well as the approximation $C_1 = C_2 \ll C$. However, instead of using $C_{\mathrm{ion}} = \frac{\alpha^2 q^2}{md^2_{\mathrm{Win.}}\omega^2_0}$ from references \cite{wineland1975principles, kotler2017hybrid}, we use values we have derived ourselves in appendix \ref{DerivingIonInductance}, or from reference \cite{daniilidis2009wiring}; these are $C_{\mathrm{ion}} = C_{\mathrm{rf}} = \frac{\eta q^2_{\mathrm{c}}    r^2}{2d_{\mathrm{eq.}} m\omega^2 \left(r^2 + d_{\mathrm{eq.}}^2\right)^{3/2}}$~, and $C_{\mathrm{ion}} = \frac{2 e^2}{m \omega^2 ~\mathrm{ln}\left( \frac{2H-a}{a} \right) \left( H^2 - h^2 \right)}$~, respectively.
				
			Expressions \eqref{gammanoah} - \eqref{gamma_Kotler-Daniilidis} can be considered in the limiting case where the distance between the ion and the coupling system becomes large. \textit{Unlike} in table \ref{tab:GammaVariousSys} where optimal ratios for the pickup-disk system are preserved, for these limiting cases the size of the coupling system is kept constant when $d_{\mathrm{eq.}}$ increases. The limiting cases therefore indicate sensitivity to changes in the distance when the coupling apparatus size is fixed, but do not give insights into the sensitivity of a given system to overall scaling. For the pickup-disk system in equation \eqref{gammanoah}, when $d_{\mathrm{eq.}} \gg r_{\mathrm{disk}}$, the coupling reduces to $\gamma_{\mathrm{pickup-disk}} \propto 1/d^5_{\mathrm{eq.}}$. In equation \eqref{gammadaniilidis}, when $H \gg h \gg a$, the coupling reduces to $\gamma_{\mathrm{suspended-wire}} \propto \frac{1}{L~\mathrm{ln} \left( 2H \right)~ H^2}$. In equation \eqref{gammazurita}, the coupling reduces to $\gamma_{\mathrm{trans.-line}} \propto 1/d^6_{\mathrm{eq.}}$. In equation \eqref{gammaliang}, the coupling reduces to $\gamma_{\mathrm{rect.-electrodes}} \propto 1/d^4_{\mathrm{eq.}}$. In equation \eqref{gammakotler}, the form of the expression is the same no matter how far the coupling plates are from the ion, and $\gamma^{\vert \vert \mathrm{plate}}_{\mathrm{mass-spring}} \propto \frac{1}{d^2_{\mathrm{Win.}} C}$. In equations \eqref{gamma_Kotler-Noah} and \eqref{gamma_Kotler-Daniilidis}, the limiting case leads to $\gamma^{\mathrm{p.d.}}_{\mathrm{mass-spring}} \propto 1/d^{4}_{\mathrm{eq.}}$~, and $\gamma^{\mathrm{s.w.}}_{\mathrm{mass-spring}} \propto \frac{1}{\mathrm{ln} \left( 2H \right)~ H^2 ~C}$. 

			Comparing the limiting-cases of the pickup-disk and suspended-wire systems shows that relative to the suspended-wire system, the pickup-disk setup is more sensitive to changes in the distance between the ion and the coupling system. Among the limiting cases for equations \eqref{gammanoah}--\eqref{gamma_Kotler-Daniilidis}, the mass-spring system as modeled in equation \eqref{gammakotler} is the least sensitive to distance, while the transmission line system as modeled in equation \eqref{gammazurita} is the most sensitive.

			We now take a closer look at some of the scaling relationships, limiting cases, and methodologies. The scaling relationship derived here for the pickup-disk system, $\gamma_{\mathrm{pickup-disk}} \propto \frac{1}{d^5_{\mathrm{eq.}}}$~, is unusual. One might expect the coupling strength to die off according to $\frac{1}{d^2_{\mathrm{eq1}}}$~ on one side of the coupling system and $\frac{1}{d^2_{\mathrm{eq2}}}$ on the other side, as ion\#1 and ion\#2 move away from what ultimately become "point-like" pickup-disks of the coupling system. Increasing $d_{\mathrm{eq.1}}$ and $d_{\mathrm{eq.2}}$ simultaneously would then suggest an overall scaling of $\frac{1}{d^2_{\mathrm{eq.}}} \times \frac{1}{d^2_{\mathrm{eq.}}} = \frac{1}{d^4_{\mathrm{eq.}}}$. Expression \eqref{gammanoah} contains two terms, each of which contains two portions; one portion which depends on coordinates 'near' a given ion, for example ion\#1, and another portion which depends on coordinates on the 'far side' of the coupling system, near ion\#2.  While the portion on the far side, due to the ring of charge, dies off with $\frac{1}{d^2_{\mathrm{eq.}}}$ when $d_{\mathrm{eq.}} \gg r$, the portion on the near side attenuates with $\frac{1}{d^3_{\mathrm{eq.}}}$. Thus, $\gamma_{\mathrm{pickup-disk}}$ scales as $\propto \frac{1}{d^5_{\mathrm{eq.}}}$ because the derivative with respect to '$d$' of the charge induced within a circular region on an infinite grounded plane, centered directly below a suspended charge, dies off with $\frac{1}{d^3}$ in the limit $d \gg r$.\footnote{It was initially thought that the $\frac{1}{d^3}$ dependence might be introduced by the linearization of the surface charge density in equation \eqref{SurfChrgLinearized}, since the linearized expression differs from the exact expression \eqref{SurfaceCharge}. Indeed, the non-linearized charge density dies off as $\frac{1}{d^2}$, while the linearized charge density dies off as $\frac{1}{d^3}$. However, starting from the linearized or non-linearized charge density leads to the same result for the coupling strength $\gamma_{\mathrm{pickup-disk}}$ (see appendix \ref{AppNonLinGamOpt}). The limiting case $\gamma_{\mathrm{pickup-disk}} \propto \frac{1}{d^5_{\mathrm{eq.}}}$ is thus accurate.}

			We notice another interesting aspect of the pickup-disk system; in the limiting case $d_{\mathrm{eq.1}} = d_{\mathrm{eq.2}} = 0$~, the coupling strength tends to zero. To understand this, one must recall that the model supposes that charges pushed onto disk2 by ion\#1 spread out to form a ring.  If the distance between ion\#2 and disk2 is reduced so much that ion\#2 goes to the center of the ring, charges in the ring will produce no upwards force in the $z$ direction. The fact that the coupling strength tends to zero when both ions come close to the disks is consistent with the model. 

			The scaling relationship for the suspended-wire setup, $\gamma_{\mathrm{suspended-wire}}$, in the limit $H \gg h \gg a$~ can be roughly understood intuitively. The electric field produced by an infinite straight wire dies off with $\lvert \vec{E} \rvert \propto \frac{1}{2\pi r}$~, where $r$ is the distance away from the wire. Thus, increasing the distance between the wire and ion\#1, and the wire and ion\#2 simultaneously, leads to an attenuation of $\gamma_{\mathrm{suspended-wire}}$ which is $\propto \frac{1}{r} \times \frac{1}{r} = \frac{1}{r^2}$, where we have used the notation $r$ instead of $H$. The logarithm brings a small departure from this, and arises due to the presence of induced charges on the ground plane (see exercises in \cite{grifWireInfPlane}). When the radius of the coupling wire tends to zero, the coupling strength tends to zero. However, in the limit $a=H$, the coupling strength diverges. This may happen because $a \rightarrow H$ necessarily entails $h \rightarrow H$, in which case the distance between the charges and the wire tends to zero. Nevertheless, it is surprising that increasing the wire radius from $55~$nm to $12.5~\mu$m increases the coupling by over a factor of 2. Increasing the wire radius  increases the wire capacitance, which dilutes the signal; one might expect an increase in the wire radius to cause the coupling strength to decrease.

			Perhaps the most striking difference among expressions \eqref{gammanoah} - \eqref{gamma_Kotler-Daniilidis} is that equations \eqref{gammakotler} - \eqref{gamma_Kotler-Daniilidis}, which depend on the \textit{ion capacitance} methodology, are not derived directly from the geometry of the coupling system. Instead, the interaction between the ion and the coupling system is incorporated into "capacitances", which aim to take into account both the interaction between the ion and the trapping field on one hand, and the interaction between the ion and the coupling system on the other. These capacitances are related to the geometry of the system, and contain information about the current induced within the coupling system when the ion moves. However, comparing the mathematical expressions and the values in table \ref{tab:GammaVariousSys} calculated using the \textit{ion capacitance} approach with those calculated from other "in depth" approaches shows that for the same system in general the coupling strength is calculated to be different. For instance, at $d_{\mathrm{eq.}} = 50~ \mu \mathrm{m}$, $\gamma_{\mathrm{D.~suspended-wire}} = 7.5 \times 10^{-18}$ whereas $\gamma_{\mathrm{W.~mass-spring,~suspended~wire}} = 1.1 \times 10^{-17}$. 

			We can take a closer look at the limiting cases for the \textit{ion capacitance} model. The ion capacitance we have derived, and the ion capacitance derived for the suspended wire setup in reference \cite{daniilidis2009wiring}, contain a logarithm which comes from the capacitance of the coupling wire (in $C_{\mathrm{rf}}$ a logarithm is hidden in the coefficient $\eta$). Consequently, when equations \eqref{gamma_Kotler-Daniilidis} and \eqref{gamma_Kotler-Noah} are written out and the capacitance $C$ of the coupling system is modeled as the capacitance of a finite coupling wire, $C = C_{\mathrm{wire}} = 2 \pi \epsilon_{\mathrm{o}} l_{\mathrm{wire}}\big/ \mathrm{ln}\left(l_{\mathrm{wire}}/a\right)$, multiple logarithms appear in each expression, though they originate from the same source. The replicate logarithms are due to the way the induced current is introduced in the calculation of $C_{\mathrm{ion}}$, following the methodology of \cite{wineland1975principles}. Explicitly, in the limit $H \gg h \gg a$ in equation \eqref{gamma_Kotler-Daniilidis}, one logarithm appears from the capacitance of each ion, $C_{1}$, $C_2$. (In reference \cite{daniilidis2009wiring} the ion capacitance is calculated from the ion inductance $L_{i}$, which depends on a parameter $\beta$, which depends on a parameter $\alpha$, which is the potential of an infinite wire above a grounded plane, and contains a logarithm). A second logarithm comes from the expression for $C$, when it is written out. For the pickup-disk system of equation \eqref{gamma_Kotler-Noah}, logarithms appear in the expression for $\eta$ which depends on the capacitance of the connecting wire. The coefficient $\eta$ (or equivalently, $\zeta$) enters in the calculation of $C_{\mathrm{ion}}$ because in the methodology of \cite{wineland1975principles} the position of the ion is expressed in terms of the total charge induced on one of the disks (the integrated current), which depends on all of the capacitances of the coupling system (see appendix \ref{WireChargeDist}). Again, a second logarithm comes when the total capacitance $C$ is written out. These irregularities hint at problems with the equivalent circuit element approach. A detailed investigation in reference \cite{van2024improved} reveals that the equivalent circuit element approach is not appropriate for calculating coupling strengths.

			It was mentioned earlier that equations \eqref{gammazurita} and \eqref{gammaliang} only depend on the total capacitance $\left(2C_\mathrm{o} + C_\mathrm{w}\right)$. If this were true, one could ignore the detailed expressions describing how the capacitances are distributed in the system. However, the way the capacitances are distributed cannot be ignored. It is shown in appendix \ref{WireChargeDist} that the \textit{ratio} of the pickup-electrode capacitance to the total capacitance determines the coupling strength. We can show that if equation \eqref{gammazurita} is rewritten explicitly in terms of the ratio of the pickup-electrode capacitance to the total capacitance, the result closely resembles our expression for the coupling strength in the case of a symmetric coupling system, equation \eqref{gamma_symm}. For convenience, equation \eqref{gamma_symm} (with the coefficient $\zeta$ rearranged and expressed in terms of the disk capacitance $C_{a}$) and \eqref{gammazurita} are given here.
			
			\begin{equation}\label{gamma_simp}			
			\gamma =\frac{q^2_{\mathrm{c}}   }{4\pi \epsilon_{\mathrm{o}}} \left( \frac{d r^2}{\left(d^2+r^2\right)^{3}} \right) \left( \frac{C_{a}}{2 C_{a} + C_b } \right) ~,
			\end{equation}
			
			\begin{equation}\label{gammaZur2}
			~\gamma_{\mathrm{trans.-line}} = \frac{e^2 r^4}{\left(d^2_{\mathrm{eq.}} + r^2\right)^3}\frac{1}{2C_\mathrm{o} + C_\mathrm{w}} ~.
			\end{equation}
						
			To draw a comparison, we let the wire capacitance $C_\mathrm{w}$ in equation \eqref{gammaZur2}, from reference \cite{zurita2008wiring}, be the same as the wire capacitance $C_\mathrm{b}$ here, and the Penning trap capacitance $C_\mathrm{o}$ in equation \eqref{gammaZur2} be the same as $C_a$ here. This means $C_\mathrm{o} = C_a = 8 \epsilon_{\mathrm{o}} r$, or $r = C_a / \left( 8 \epsilon_{\mathrm{o}} \right)$. Factoring out a single factor of $r$ from equation \eqref{gammaZur2} and substituting ($r$ in terms of $C_a$) into that factor, and letting $e = q_{\mathrm{c}}$, equation \eqref{gammaZur2} becomes
			\begin{equation}\label{Same-same}
			~\gamma_{\mathrm{trans.-line}} = \frac{q^2_{\mathrm{c}}   }{ 8 \epsilon_{\mathrm{o}}}
			\frac{r^3}{\left(d^2_{\mathrm{eq.}} + r^2\right)^3}\frac{C_a}{2C_a + C_b} ~.
			\end{equation}
			Equations \eqref{gamma_simp} and \eqref{Same-same} are nearly identical, with the difference that equation \eqref{gamma_simp} contains a factor of $d / \pi$, whereas equation \eqref{Same-same} has a factor of $r/2$.

			\begin{section}{Conclusions and outlook}\label{ConclOut}
 
			We have proposed a system for coupling the motional quantum states of spatially distant ions via a conducting wire, and calculated its associated coupling strength $\gamma$. This allows us to calculate the time $t_{\mathrm{ex.}}$ required to exchange quantum information between ion qubits in separate surface traps using a conducting wire. Our calculation of the coupling strength $\gamma$ addresses an oversight existing in the majority of previous works \cite{heinzen1990quantum, daniilidis2009wiring, kotler2017hybrid, rica2018double, bohman2018sympathetic, bohman2021sympathetic, an2022coupling} by correctly incorporating an efficiency factor $\zeta$ which accounts for the self-capacitances of the three main elements of the coupling system. We find that optimization of the design herein leads to a rule of thumb which relates the distance $d_{\mathrm{eq.}}$ between a trapped ion and a nearby pickup disk, to the optimal radius $r_{\mathrm{opt.}}$ of the disk. When the pickup disks' capacitance is much larger than the capacitance of the connecting wire, the optimal relationship is $r_{\mathrm{opt.}} = d_{\mathrm{eq.}}$, while in the opposite limit the optimal relationship is $r_{\mathrm{opt.}} = d_{\mathrm{eq.}} / \sqrt{2}$. In general, $d_{\mathrm{eq.}} / \sqrt{2} \le r_{\mathrm{opt.}} \le d_{\mathrm{eq.}}$. In addition to optimizing the coupling system itself, we have analyzed several key factors including the current which flows through the device, the role of the resistance of the coupling wire, and the possibility of driving the system on resonance.
			
			Although we have described several ways to increase the coupling strength, there are ways it could be further increased. For instance, one could use an optimized design as described in \cite{van2020making}, where the pickup discs are brought out of the trapping plane and in closer proximity to the ions; this would reduce state exchange times $t_{\mathrm{ex.}}$. Overall, our results are aligned with the very limited number of other compelling results calculated in the literature. We note that many existing results are based on one, deeply flawed \cite{van2024improved} approach  introduced in \cite{wineland1975principles}. This motivated our quantitative and qualitative review of coupling strengths calculated in the literature.  

			This work focuses on results which are immediately applicable for a planar coupling system, due to the potential for planar trapped ion systems to be scaled for quantum computing applications. However, several of our results may also be applicable to other hybrid systems involving interactions between ions and nearby conductors. One example is a hybrid system coupling ion qubits to solid-state qubits, such as superconducting qubits.			
			We hope the analysis and methodology developed herein will be generally useful and provide a foundation for a rich set of future studies and experiments. 
			
			
			\end{section}	

}
\begin{equation*}
\end{equation*}

\begin{acknowledgments}
			We thank Angelina B. Frank for making figures \ref{fig:Figure2} and \ref{fig:2ConnectedSprings}, and Marco Fellous-Asiani for verifying the calculation of $Q_{\mathrm{transf.}}$.  
\end{acknowledgments}

\bibliography{References}

\renewcommand{\theHsection}{A\arabic{section}}

\begin{appendix}

\section{Charge in an outer ring of a conducting disk; justification of the 'ring of charge' approximation}
\label{ChargeRing}


			We claim that when charges are placed on a conducting disk, most of the charges end up in a "ring" around the edges of the disk. To justify this, one must demonstrate that at equilibrium, most of the charge is located in an annular segment of the disk with a width $\xi \ll$ than the radius of the disk. In this appendix the radius of the disk will be called "$a$", following the notation in \cite{mcdonald2003capacitance}. We define "$\xi \ll a$" to mean $\xi \le \frac{a}{4}$, and "most of the charge" to mean that at least $\frac{3}{4} \times Q$ should be within the annular segment covered by $\xi$, where $Q$ is the total charge on the disk.	Following McDonald \cite{mcdonald2002conducting}, and Sir Thompson (Lord Kelvin) \cite{thomson1884papers}, the surface charge density of a circular disk over most of its surface area (the parts near the edges of the disk notwithstanding), is given by $\sigma_{\mathrm{circular~ disk}}(r) = \frac{Q'}{4\pi a \sqrt{a^2 - r^2}}$, where $a$ is the radius of the disk, $r$ is the distance away from the center of the disk at which the surface charge density is evaluated, and $Q'$ is a parameter to be determined, (which would be the total charge $Q$ of the two flat surfaces combined, in the limit that the height $2b$ of the disk is $2b = 0$). A more accurate description of the surface charge density near a corner is given by: $\sigma_s = \frac{K}{\sqrt[\leftroot{0}\uproot{0}3]{s}}$, where $K$ is a constant and $s$ is distance away from the corner \cite{jackson1999classical} \cite{ mcdonald2003capacitance}. In the thin disk system, the intersection between the top (or bottom) face of the disk, and the side of the disk, produces a corner which extends as a ring around the center of the disk. The variable "$s$" describes moving inwards a distance "$s$" from the $90^{\mathrm{o}}$-corner edge of the ring, where $s \le b$, and $2b$ is the height of the disk. The same expression $\sigma_s = \frac{K}{\sqrt[\leftroot{0}\uproot{0}3]{s}}$ can be used to describe the charge density on the side of the disk, moving away from the $90^{\mathrm{o}}$-corners, towards the center of the rectangular strip which wraps around the circumference of the disk. To ensure consistency there must be continuity of charge density at every point on the top (and bottom) surface of the disk. Therefore, somewhere on the disk $\sigma_{\mathrm{circular~ disk}}(r) = \sigma_s(s)$. We can postulate that this occurs a distance $b$ inwards from the edge of the disk, and then allow the free parameter $Q'$ to vary in such a way as to satisfy the constraint that the total charge on the disk must sum up to $Q$.
Continuity of charge density imposes: $K 
\big / \sqrt[\leftroot{0}\uproot{0}3]{b} = Q' \big/ 4\pi a \sqrt{a^2 - (a-b)^2}$, or 
\begin{equation}\label{BigK}
K = \frac{Q'b^{1/3}}{4 \pi a \sqrt{a^2 - (a-b)^2}}
\end{equation}
			The total charge contained in the region "near the corners" of the disk can be found by integrating $\sigma_s$ a distance $b$ (in other words, half the disk-height) from the top-corner towards the center of the rectangular strip, from the bottom-corner towards the center of the rectangular strip, and also inwards forming an 'outer ring' of charge on the top and bottom faces of the disk. This yields $Q_s = 4\int_0^b \! \frac{K}{\sqrt[3]{s}}2\pi a \ ds = 
			\frac{3b}{\sqrt{a^2 - {(a-b)}^2}}Q'$, where we have substituted expression \eqref{BigK} in for $K$. The total charge on the rest of the disk, over two 'inner small disks' on the top and bottom faces, may be found by integrating $\sigma_{\mathrm{circular~ disk}}$ from $r=0$ to $r=\left(a-b\right)$. This yields $Q_g = 2\int_0^{a-b} \frac{Q'}{4 \pi a \sqrt{a^2 - {r}^2}} 2 \pi r dr = \left(1-\frac{\sqrt{a^2 - {(a-b)}^2}}{a}\right)Q'$. The sum of these two expressions gives the total charge on the disk, $Q = Q_g + Q_s$, which provides a relationship between $Q$ and $Q'$:
\begin{equation}\label{Q'andQ}
Q'= \frac{Q}{\frac{3b}{\sqrt{a^2 - {(a-b)}^2}} +1 -\frac{\sqrt{a^2 - {(a-b)}^2}}{a}}
\end{equation}
			In the limit $a \gg b$, expression \eqref{Q'andQ} reduces to $Q' \approx Q~$, which can be understood intuitively. When the disk is thin compared to its radius, only a small percent of its charge resides on the thin edge perpendicular to the circular faces of the disk.

{\setlength{\parindent}{0cm}
			We want to find the charge contained in the outer-ring region a distance $\xi \le \frac{a}{4}$ away from the corner. In some cases, if the width $2b$ of the conducting disk is large compared to the radius $a$ of the disk, the distance $b$ inwards from the side of the disk may be greater than $\frac{a}{4}$. In this case, it makes sense to only use the expression for the charge density "near a corner" to calculate the charge in the outer-ring $\xi \le \frac{a}{4}$. Experimentally, the typical thickness of surface-trap electrodes ranges from $2b = 100~$nm to $2b = 10~\mu$m \cite{labaziewicz2008temperature,hite2012100,daniilidis2014surface,sedlacek2018distance}, and for what follows we consider a thickness up to $2b = 35~\mu$m. From section \ref{GammaOpt}, for the limiting case where the disk capacitance is much greater than the wire capacitance, the optimal relationship between the radius of the disk and the equilibrium height of the suspended ion is $r \sim d_{\mathrm{eq.}} / \sqrt{2}$ (here, the notation "$r$" refers to the standard notation used throughout the rest of the article, outside of this appendix). Throughout this article we have used the two example values $d=50~\mu$m and $d=200~\mu$m, which respectively imply disk radii of $a = \frac{1}{\sqrt{2}}50~\mu \mathrm{m} = 35~\mu$m, and $a = \frac{1}{\sqrt{2}}200 ~\mu \mathrm{m} \sim 140~\mu$m. We can calculate the charge in the outer ring for the two extreme aspect ratios: ($2b = 35~\mu$m, $a = 35~\mu$m), and ($2b = 100~$nm, $a = 140~\mu$m).
	
			In the first case, the thickness $\xi = a / 4$ of the outer ring is equal to $b/2$. Since here $a / 4$ is less than $b$, we only use the expression for the charge density "near a corner" to calculate the charge in the outer-ring $\xi \le \frac{a}{4}$. We must calculate the charge on the side of the disk, as well as inwards a distance $b/2$ towards the center of the disk. The charge on the side of the disk is $Q_s / 2$. The charge inwards a distance $b/2$ towards the center of the disk, for the top and bottom faces is $2\int_0^{b/2} \! \frac{K}{\sqrt[3]{s}}2\pi a \ ds = \frac{3bQ'}{2^{5/3}\sqrt{a^2 - (a-b)^2}}$. Adding these together, the total charge in the region $\xi \le \frac{a}{4}$ is
			\begin{equation}
			Q^{\xi \le \frac{a}{4}}_{\mathrm{Edges}} = \frac{3b\left( 2^{2/3}+1 \right)}{2^{5/3}\sqrt{a^2-(a-b)^2}}Q'
			\end{equation}
			Plugging in the relationship between $Q'$ and $Q$ from \eqref{Q'andQ}, and evaluating for $b=17.5 ~\mu$m, $a=35 ~\mu$m, yields $Q^{\xi \le \frac{a}{4}}_{\mathrm{Edges}} \sim 0.76Q$. Hence, the approximation of a ring of charge is justified.
	
			Next, we look at the second case, for a thinner disk. Now, $b = \xi = 0.05~\mu$m is significantly less than $\frac{a}{4}=35~\mu$m, so the amount of charge contained in the thin sliver defined by $b$ is small compared to the total charge on the disk. To find the charge in the region $\xi \le \frac{a}{4}=35~\mu$m we cannot only consider the "edges" of the disk. Indeed, since $b/a$ is small and the ratio of charge in the region $s<b~$, to the charge in the rest of the disk, is small, one could consider using only the expression for the charge density on the circular surface $\sigma_{\mathrm{circular~ disk}}(r) = \frac{Q'}{4\pi a \sqrt{a^2 - r^2}}$. In any case, this expression should be valid in the region $r < \frac{3}{4}a$. To avoid worrying about what happens near the edges of the disk, since $b < a/4$, we can use a short-cut and calculate the charge at a distance $\xi \ge \frac{a}{4}$ away from the edges, or in other words, the charge in the central part of the disk. We use the same integral as used to calculate $Q_g$ (above equation \eqref{Q'andQ}), but integrate from $r=0$ to $r=\frac{3}{4}a$. Thus, $Q^{\mathrm{central}} = 2\int_0^{3a/4} \frac{Q'}{4 \pi a \sqrt{a^2 - {r}^2}} 2 \pi r dr = \left(1-\sqrt{\frac{7}{16}}\right)Q'$. Plugging in the relationship between $Q'$ and $Q$ from \eqref{Q'andQ}, and evaluating for $b=50 ~$nm, $a=140 ~\mu$m, yields $Q^{\mathrm{central}} = 0.33 Q$. The charge in the outer annular region, $Q^{\xi \le \frac{a}{4}}_{\mathrm{Edges}}~$, is $Q$ minus the charge in the central area, in other words $Q^{\xi \le \frac{a}{4}}_{\mathrm{Edges}} = Q - 0.33 Q = 0.67 Q$. In this case our criterion for "most of the charge" ($75 \%$) being in the annular segment covered by $\xi \le \frac{a}{4}$ is not satisfied, but it is not too far off. If we calculate $Q^{\mathrm{central}}$ for the region from $r=0$ to $r=a/2$, we find $Q^{\mathrm{central}} = \left(1 - \sqrt{3/4}\right) Q' = 0.13~Q'$ $\approx 0.13~Q~$, which means over $86\%$ of the charge is contained in the region $r \ge a/2$. Thus, although the approximation that charge distributes in a 'ring' becomes less accurate for thinner disks, the approximation is still reasonably well justified. 
	
			The analysis above can be supplemented with an additional observation. We showed that more than $50~\%$ of the charge lies in the $r \ge a/2$ region, but this region also accounts for more than $50~\%$ of the surface-area of the disk. Our calculations do not inform us about the charge density throughout the disk. For this we recall the expression for the charge density, which in the limit of a thin disk ($Q' \approx Q~$) is $\sigma_{\mathrm{circular~ disk}}(r) = \frac{Q}{4\pi a \sqrt{a^2 - r^2}}~$  \cite{mcdonald2002conducting,thomson1884papers}~. This expression depends on the radial distance $r$ from the center of the disk, and increases continuously as $r$ approaches $a$. It guarantees the charge density is not, for instance, homogeneous across the disk. Although we do not know \textit{a priori} whether representing the surface-charge distribution on a conducting disk as a ring is more accurate than representing it as a homogeneous distribution, we can say by looking at the charge density that the ring of charge model is both plausible and consistent with known results.
	
			To put numbers to the theoretical results above, we calculate an average charge density for the two regions, ($r=3a/4$ to $r=a$), and ($r=a/2$ to $r=a$), using the expression for a thin disk. The area of the first region is $\pi a^2 - \pi \left(\frac{3}{4}a\right)^2 = \pi a^2 \left(\frac{7}{16}\right)$. The charge density in this region is then $\rho^{\xi \le \frac{a}{4}} = \frac{16}{7} \times 0.67~Q / \left(\pi a^2 \right)$. The area of the second region is $\pi a^2 \left( \frac{12}{16} \right)$. The charge density in this region is $\rho^{\xi \le \frac{a}{2}} = \frac{16}{12} \times 0.86~Q / \left(\pi a^2 \right)$. Since $\frac{16}{7} \times 0.67 = 1.53$, while $\frac{16}{12} \times 0.86 = 1.15$, the charge density nearer to the edge of the disk is greater than the charge density closer to the center of the disk.

}

	\section[The distribution of charge on connected conducting \\ self-capacitors]{The distribution of charge on connected conducting self-capacitors}\label{WireChargeDist}

		\subsection{Two connected metallic self-capacitors}\label{WireCharge_ex1}

\begin{figure}[h]
	\centering
	\includegraphics[width=\linewidth]{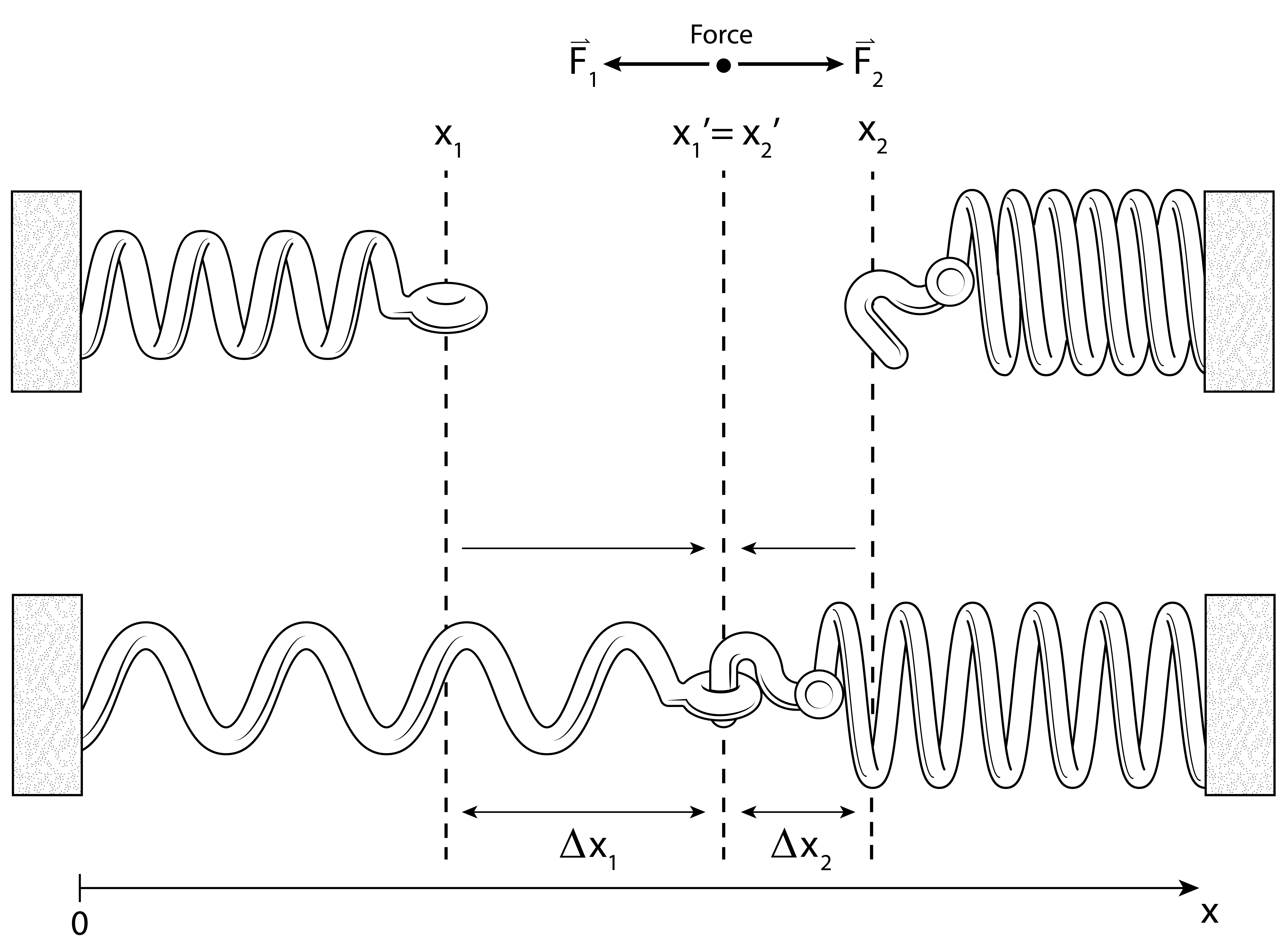}
	\caption{Two connected springs. (Top) Two springs attached to walls are disconnected and in a relaxed position. (Bottom) The two springs are brought together, attached, and then allowed to relax to the equilibrium state shown. The notations $\vec{F_1}$ and $\Delta \vec{x}_1$ in the figure correspond to $\vec{F_a}$ and $\Delta \vec{x}_a$ in the text, and similarly for $\vec{F_2}$.}
	\label{fig:2ConnectedSprings}
\end{figure}

			To help guide intuition, a system of two capacitors connected by a wire with zero capacitance can be mapped to a system of two springs anchored to two walls on either side, and joined together in the middle. We imagine that the springs are initially disconnected from one another, and then pulled together, attached, and left to equilibrate. Therefore, at equilibrium both springs are somewhat stretched.  The conversion of quantities from a system of springs to a system of electrical components is as follows:	
\begin{equation}\vec{F} = \pm k \Delta \vec{x}
\qquad
\longleftrightarrow
\qquad
\vec{V} = \pm \left(\frac{1}{C}\right)\vec{Q}
\end{equation}	
\noindent	
			Specifically, $\vec{F}\leftrightarrow \vec{V}$, $k\leftrightarrow\frac{1}{C}$, and $\Delta\vec{x}\leftrightarrow \vec{Q}$. Here, $\vec{V} \mathrm{and}~ \vec{Q}$ are described as vectors to denote that they can be positive or negative along a one-dimensional number line. The $\pm$ denotes that the direction of the force depends on which side of the spring one displaces, while keeping in mind that $\Delta\vec{x}$ is always measured as a \textit{positive} number (see figure \ref{fig:2ConnectedSprings}). The leftward force due to the leftmost spring is $\vec{F}_a = -(k_a\Delta \vec{x}_a)$, while the right-ward force due to the rightmost spring is $\vec{F}_c = k_c\Delta \vec{x}_c$, where $\Delta \vec{x}_a$ and $\Delta \vec{x}_c$ are both positive numbers measured along the same $x$-axis. At equilibrium the leftward force and the rightward force must have the same magnitude and opposite direction, so: $\vec{F_a} = -\vec{F_c}$. Hence, $k_a \Delta \vec{x_a} =  k_c \Delta\vec{x_c}$, which translates to $\frac{1}{C_a}\vec{Q}_a = \frac{1}{C_c}\vec{Q}_c$.  Thus, $C_a = C_c$ implies $\vec{Q}_a = \vec{Q}_c$.

		\subsection{Three connected metallic self-capacitors}\label{WireCharge_ex2}

			The same reasoning is followed as in section \ref{WireCharge_ex1}, but now with an additional third spring placed between spring $a$ and spring $c$, called spring $b$. This leads to the following two equations:			
\begin{equation}\label{ForceEq1}
k_a \Delta \vec{x_a} =  k_{\mathrm{eq_1}} \left( \Delta \vec{x_b} + \Delta \vec{x_c} \right)
\end{equation}
\begin{equation}\label{ForceEq2}
k_c \Delta \vec{x_c} =  k_{\mathrm{eq_2}} \left( \Delta \vec{x_a} + \Delta \vec{x_b} \right)
\end{equation}
{\setlength{\parindent}{0cm}			
			The notation $k_{\mathrm{eq_1}}$ refers to the equivalent spring constant of the two springs $k_b$ and $k_c$ in series, and $\left( \Delta\vec{x_b} + \Delta\vec{x_c} \right)$ is the total distance by which this equivalent spring is stretched.\footnote{The subscript in $k_{\mathrm{eq_1}}$ stands for "equivalent", not to be confused with the subscript in $d_{\mathrm{eq.}}$ in other parts of the manuscript, which stands for "equilibrium".} Solving \eqref{ForceEq2} for $\Delta\vec{x_b}$ gives $\Delta\vec{x_b} = \frac{k_c\Delta\vec{x_c}}{k_{\mathrm{eq2}}} - \Delta\vec{x_a}$, which can be inserted into equation \eqref{ForceEq1}. Re-arranging \eqref{ForceEq1} then leads to:		
	\begin{equation}
	\Delta \vec{x_c} = \Delta \vec{x_a}\left(1+\frac{k_a}{k_{\mathrm{eq1}}}\right)\left(\frac{1}{\frac{k_c}{k_{\mathrm{eq2}}}+1}\right) ~.
	\end{equation}		
	Writing out $k_{\mathrm{eq1}} = \frac{k_b k_c}{k_b + k_c}$, $k_{\mathrm{eq2}} = \frac{k_a k_b}{k_a + k_b}$ and substituting $\frac{1}{C_i}$ with $i=a,b,c$ for $k_{i}$, and $Q_i$ for $\Delta \vec{x_{i}}$ gives (after simplification):		
	\begin{equation}\label{QcQa}
	Q_c = Q_a\left(1+\frac{C_c+C_b}{C_a}\right)\left(\frac{1}{1+\frac{C_b+C_a}{C_c}}\right)
	\end{equation}
			which further reduces to $Q_c = Q_a\frac{C_c}{C_a}$. If $C_a=C_c$, then $Q_a = Q_c$, the same as was concluded in section \ref{WireCharge_ex1}. Similarly, a relationship can be derived between $\Delta\vec{x_c}$ and $\Delta\vec{x_b}$,		
	\begin{equation}
	\Delta \vec{x_c} = \Delta \vec{x_b}\left(1+\frac{k_a}{k_{\mathrm{eq1}}}\right)\left(\frac{1}{\left(\frac{k_a}{k_{\mathrm{eq1}}}\right)\frac{k_c}{k_{\mathrm{eq2}}}-1}\right) ~.
	\end{equation}		
	This leads to the relationship between $Q_c$ and $Q_b$:			
	\begin{eqnarray}\label{QcQb}
	Q_c = ~&&Q_b \left( 1+\frac{C_c+C_b}{C_a} \right) \nonumber \\
	&& ~~~~~~\times ~\left( \frac{1}{\left( 1+\frac{C_b}{C_c}) \right) \left( 1+\frac{C_b}{C_a} \right) -1} \right) ~,
	\end{eqnarray}
			which further reduces to $Q_c = Q_b \frac{C_c}{C_b}$.
}

		\subsection{Expressing the charge $Q_c$ in terms of $Q_{\mathrm{transf.}}$}\label{WireCharge_ex3}

{\setlength{\parindent}{0cm}			
			Section \ref{WireCharge_ex2} provides expressions for $Q_c$ in terms of $Q_a$, and for $Q_c$ in terms of $Q_b$.  However, the final intent is to find the charge deficit (or surplus) that results on disk2, when ion\#1 moves closer (or farther away from) disk1. What is known is the amount of charge $Q_{\mathrm{transf.}}$ which \textit{would} flow onto disk1, if it were connected to ground.  This provides the total energy brought to disk1 by the movement of ion\#1,~ $E = \frac{Q^2_{\mathrm{transf.}}}{2 \left( C_{a} + C_{b} + C_{c} \right) }$,~ where $C_a = C_{\mathrm{disk1}}$, $C_b = C_{\mathrm{wire}}$, and $C_c = C_{\mathrm{disk2}}$. What remains unknown is how this energy will distribute among $C_a$, $C_b$, and $C_c$, when the system adjusts to the new potential. (Presumably this adjustment occurs adiabatically, since the resistance $R_{\mathrm{wire}}$ and inductance $L_{\mathrm{wire}}$ is low enough for the charge to flow between the capacitors much more quickly than the period of one oscillation. See subsection \ref{Rmax}, and appendices \ref{AppendixResist} and \ref{OffResAndLneg}). Specifically, the aim is to determine the amount of charge which ends up on $C_c$, which determines the strength of the resulting force on ion\#2 when ion\#1 moves.  From the results of section \ref{WireCharge_ex2}, it is now possible to express $Q_c$ in terms of $Q_{\mathrm{transf.}}$. With $C_{\mathrm{tot.}} \equiv \left( C_{a} + C_{b} + C_{c} \right)~$,	
	\begin{equation}\label{energy3cap}
	E = \frac{Q^2_{\mathrm{transf.}}}{2C_{\mathrm{tot.}}} = \frac{Q^2_{a}}{2C_{a}} + \frac{Q^2_{b}}{2C_{b}} + \frac{Q^2_{c}}{2C_{c}} ~.
	\end{equation}	
			Using \eqref{QcQa} and \eqref{QcQb},~ and the notation $\Lambda$ and $\Upsilon$ for the portions which depend on capacitance in \eqref{QcQa} and \eqref{QcQb}, respectively (so that, e.g. $Q_c = Q_a \Lambda$) this becomes
	\begin{equation*}
		\frac{Q^2_{\mathrm{transf.}}}{2C_{\mathrm{tot.}}} = \frac{Q^2_c}{2}\left( \frac{1}{C_a \left(\Lambda^2\right)} + \frac{1}{C_b \left(\Upsilon^2\right)} + \frac{1}{C_c}\right) ~,
	\end{equation*}	
			which re-arranges to	
	\begin{equation}\label{QcQtransf}
	Q_c = \frac{1}{\sqrt{\left(\frac{1}{C_a \left(\Lambda^2\right)} + \frac{1}{C_b \left(\Upsilon^2\right)} + \frac{1}{C_c}\right)C_{\mathrm{tot.}}}}Q_{\mathrm{transf.}} ~,
	\end{equation}	
			and further simplifies to: \\ $Q_c = \left( C_c~/~\left( C_a + C_b + C_c \right) \right) Q_{\mathrm{transf.}}$. This can be summarized succinctly as $Q_c = \zeta Q_{\mathrm{transf.}}$. In the limiting case $C_a = C_c$ and $C_b = 0$, this reduces to $Q_c = \frac{1}{2}Q_{\mathrm{transf.}}$. Therefore, equation \eqref{QcQtransf} taken together with equation \eqref{QcQa} shows that in this limiting case, the charge deficit (or surplus) on $C_a$ is equal to the charge deficit (or surplus) on $C_c$ and the amount of charge deficit (surplus) on each is $Q_a = Q_c = \frac{1}{2}Q_{\mathrm{transf}}$. Note that the \textit{absolute} amount of charge on the pickup-disks-and-wire system never changes; the system is never in contact with an external source (or sink) of charges. If it is assumed that initially the net charge is zero, then it remains permanently zero.
}

\subsection{Expressing the charge $Q_a$ in terms of $Q_{\mathrm{transf.}}$}\label{WireCharge_ex4}

			It is also possible to express the charge induced on disk1, $Q_a$, in terms of $Q_{\mathrm{transf.}}$. We start with the expression for the total energy induced on the coupling system, as in equation \eqref{energy3cap}:
\begin{equation*}
	E = \frac{Q^2_{\mathrm{transf.}}}{2C_{\mathrm{tot.}}} = \frac{Q^2_{a}}{2C_{a}} + \frac{Q^2_{b}}{2C_{b}} + \frac{Q^2_{c}}{2C_{c}} ~.
\end{equation*}
			This can be rewritten by substituting the relationships derived in section \ref{WireCharge_ex2}, $Q_b = Q_c \frac{C_b}{C_c}$ and $Q_c = Q_a\frac{C_c}{C_a}$, as
\begin{equation*}
	\frac{Q^2_{\mathrm{transf.}}}{2C_{\mathrm{tot.}}} = \frac{Q^2_{a}}{2C_{a}} + \frac{ \left( Q_c \frac{C_b}{C_c} \right)^2}{2C_{b}} + \frac{\left( Q_a\frac{C_c}{C_a} \right)^2 }{2C_{c}} ~.
\end{equation*}
			Substituting in $Q_c = Q_a\frac{C_c}{C_a}$ again, and simplifying, this becomes
\begin{equation*}
	\frac{Q^2_{\mathrm{transf.}}}{2C_{\mathrm{tot.}}} = \frac{Q^2_{a}}{2C_{a}} + \frac{ Q^2_a C_b }{2 C^2_a} + \frac{ Q^2_a C_c }{2 C^2_a} ~.
\end{equation*}
			Multiplying the first term by $C_a/C_a$ and then pulling out the factor of $Q^2_a / 2 C^2_a$ which is common to all three terms leads to
\begin{equation*}
	\frac{Q^2_{\mathrm{transf.}}}{2C_{\mathrm{tot.}}} = 
	\frac{Q^2_a}{2 C^2_a}\left( C_{a} + C_b + C_c \right) ~.
\end{equation*}
			Rearranging gives
\begin{equation*}
	\frac{Q^2_{\mathrm{transf.}}C^2_a}{\left( C_{a} + C_b + C_c \right)^2} = Q^2_a ~.
\end{equation*}
			Finally, after taking the square root this becomes
\begin{equation}\label{CoeffEta}
Q_a = \left( \frac{C_a}{ C_{a} + C_b + C_c } \right) Q_{\mathrm{transf.}} = \eta~ Q_{\mathrm{transf.}} ~,
\end{equation}
			where we have defined the coefficient $\eta \equiv C_a / \left( C_a + C_b + C_c \right)$. Comparing the coefficient $\eta$ from equation \eqref{CoeffEta} with the coefficient $\zeta$ defined below equation \eqref{QcQtransf}, we see that $\eta$ and $\zeta$ are identical in the case of a symmetrical coupling system where $C_a = C_c$.

			\subsection{Charge distribution on a system of $N$ non-interacting conductors.}
			
			Sections \ref{WireCharge_ex3} and \ref{WireCharge_ex4} show a trend which suggests that in general, for a given total charge imbalance $Q_{\mathrm{transf.}}$ on a system of $N$ "non-interacting" (but connected) conductors, the equilibrium charge imbalance $Q_{i}$ on the $i$th conductor is determined by the ratio $\left( \frac{C_i}{ C_{\mathrm{tot.}}} \right)$ according to $Q_{i} = \left( \frac{C_i}{ C_{\mathrm{tot.}}} \right)Q_{\mathrm{transf.}}$. Here, $C_i$ is the self capacitance of the $i$th conductor and $C_{\mathrm{tot.}} \equiv \sum_{i=1}^{i=N} C_i$ is the sum of self capacitances of all $N$ conductors. The phrase "non-interacting" means that while charge is allowed to equilibrate between the conductors, any electric fields and associated mutual capacitances are neglected. The statement above can be proven as follows.
				
			For any isolated conductor, the self-capacitance $C_i$ is defined as the ratio of charge to potential: $C_i = \frac{Q_i}{V}$. Rearranging, the charge on conductor $i$ is: $Q_i = C_i V$. This holds for each of the $N$ conductors since they all share the same potential $V$ (they are connected, so they are a single, equipotential system). Adding up the charge imbalance on all conductors gives: $Q_{\mathrm{transf.}} = \sum_{i=1}^{N} Q_i = \sum_{i=1}^{N} (C_i V) = V \sum_{i=1}^{N} C_i = V C_{\mathrm{tot.}}$, or $V = \frac{Q_{\mathrm{transf.}}}{C_{\mathrm{tot.}}}$. Substituting $V$ back into the expression for $Q_i$ above, $Q_i = C_i V$, yields:
				
			\begin{equation}
			Q_{i} = \left( \frac{C_i}{ C_{\mathrm{tot.}}} \right)Q_{\mathrm{transf.}}~,
			\end{equation}
				
			\noindent which concludes the proof. Note that in more general situations where self-capacitance does not dominate and there is strong mutual coupling between the capacitors, the full capacitance matrix of the system must be used. Additionally, this derivation only holds for an ideal isolated system with no external fields, and at static equilibrium.

		\section[Neglecting the induced ring of charge in the ground plane]{Neglecting the ring of induced charge in the ground plane}\label{NeglIndChrg}

			In section \ref{CalcGamma} we observed that the movement of ion\#1 induces a ring of charge on disk2, and this ring of charge induces a second ring of opposite charge in the infinite conducting plane surrounding disk2. The purpose of this appendix 
			is to argue that this second ring can be neglected when calculating the force on ion\#2 due to the movement of ion\#1. Hence, the second ring of induced charge can be neglected in the calculation of the coupling strength $\gamma$.
			
			Suppose a tube is drawn around the ring of charge induced on the disk (called the "first" or "smaller-radius" ring), forming a torus. By Gauss's law the total electric field flux through this tubular surface is equivalent to the total net charge contained inside of it.  If one small segment of the tube is considered, it can be approximated as a cylindrical surface surrounding a cylindrical rod.  Let "$a$" be the radius of this cylindrical surface (the distance from the core of the tube, to the surface of the tube). Let $\theta$ be the angle which wraps around the core of the tube.  Because we are looking at a small "cylindrical" segment, we can make the approximation that the outward flux away from the core of the ring is homogeneous in all directions. (This assumption is valid if the gap between the disk and the infinite plane is large compared to the height and radius of the disk. If the gap is small, induced charges outside of the disk can cause charges near the edge of the disk to distribute unevenly.) Since the angle $\theta$ through which flux lands on the second ring is small compared to $2\pi$, the fraction of the total charge on the first ring which is induced on the outer conducting plane must necessarily be small.  If it were considered, it would result in a slight reduction in the coupling between ion\#1 and ion\#2.

			Note that mathematically, even the slightest tilt of the infinite plane in which disk2 is etched will result in the full charge $-Q_{\mathrm{ring}}$ being induced on the infinite plane. However, in an infinite plane which is only slightly tilted, the majority of the induced charges are induced at a location in the plane which is far away from the inner ring, and from ion\#2. The contribution of these charges to the force on ion\#2 is negligible.

	\section{Optimizing the coupling strength $\gamma$ in the limit of large wire capacitance}\label{C_wireGGC_disk}

			The problem to solve is $\frac{\partial}{\partial r} \gamma \left(r\right) = 0$ in the limit $C_{\mathrm{b}} \gg 2C_{\mathrm{disk}}$. Explicitly,
\begin{equation*}
	\frac{\partial}{\partial r} \left[
	\frac{q^2_{\mathrm{c}}   }{4\pi \epsilon_{\mathrm{o}}} \left( \frac{r}{2r + \frac{C_b}{8\epsilon_{\mathrm{o}}} } \right) \left( \frac{d r^2}{\left(d^2+r^2\right)^{3}} \right)
	\right] = 0 ~.
\end{equation*}
			In the limit $C_{\mathrm{b}} \gg 2 \times 8\epsilon_{\mathrm{o}} r$, this becomes
\begin{equation*}
	\frac{\partial}{\partial r} \left( \frac{d r^3}{\left(d^2+r^2\right)^{3}} \right) = 0 ~.
\end{equation*}
			This reduces to
\begin{equation*}
	\left( d^2 + r^2 \right)^4 - 2r^2 \left( d^2 + r^2 \right)^3 = 0 ~.
\end{equation*}
\begin{equation*}
	\Rightarrow  \quad  d = r
\end{equation*}

\section{Resistance of the coupling wire}\label{AppendixResist}

			First, we consider a normal (non superconducting) cylindrical wire. The alternating-current (AC) resistance of a cylindrical wire is \cite{richard}	
\begin{equation}\label{Skin_resist}
R = \frac{l}{\sigma \left(2\pi a \delta - \pi \delta^2 \right)} ~,
\end{equation}		
			where $\sigma$ is the conductivity of the material, $a$ is the wire radius, $\delta$ is the skin-depth of the current in the wire, and $l$ is the wire length. The skin depth is given by $\delta = 1 \big/ \sqrt{\pi f \mu \sigma}$~, where $f$ is the frequency of interest, $\mu$ is the permeability of the material (approximately $\mu_0$), and $\sigma$ is the conductivity of the material \cite{richard}. A typical skin-depth for copper at $5~$MHz is $\delta \sim 30~\mu$m, which is larger than the radius $a$ of the proposed wire. For a moment, suppose the converse is true, $a \gg \delta$, and consider expression \eqref{Skin_resist} as the wire radius $a$ shrinks and approaches the skin-depth $\delta$. In this case, equation \eqref{Skin_resist} tends to $R = \frac{l}{\sigma \left(2\pi \delta^2 - \pi \delta^2 \right)} = \frac{l}{\sigma \pi \delta^2}$.  In other words, the expression regains the form of the resistance of a cylindrical wire for a DC current, which is proportional to the inverse of the cross-sectional area \cite{richard}. Given that in our case the skin depth $\delta$ is $> a$, we use the expression $R = \frac{l}{\sigma \pi \delta^2}$ to calculate the resistance. Taking $a \sim 10 \times 10^{-6}$~m, the conductivity of copper at room temperature ($20~^{\mathrm{o}}\mathrm{C}$), $\sigma = 6.0 \times 10^{7} ~\Omega ^{-1} \cdot \mathrm{m}^{-1}$, and a $0.01$~m long wire gives			
\begin{equation}\label{WireResist}
R_{\mathrm{wire}} = \frac{l}{\sigma \pi a^2} \sim 0.53~ \Omega ~.
\end{equation}
			At lower temperatures the conductivity of copper increases substantially, and at $30~$K the conductivity is $\sigma = 1.2 \times 10^{10} ~\Omega ^{-1} \cdot \mathrm{m}^{-1}$ (calculated as 1 over the total resistivity $\rho = 0.0083 \times 10^{-8}~\Omega \cdot~$m given in reference \cite{matula1979electrical}). At this temperature the total resistance decreases to $R_{\mathrm{wire}} \sim ~ 0.0027 ~\Omega~$.

			A smaller wire decreases the wire capacitance, which increases the coefficient $\zeta$ and therefore the coupling strength $\gamma$, and reduces the exchange time $t_{\mathrm{ex.}}$. However, it also increases resistance and associated noise, decreasing the signal-to-noise ratio. If the signal-to-noise ratio is less of a constraint than the decoherence to exchange time ratio (see \cite{van2021coupling}), 
			the wire radius could be significantly decreased. Presently, wires can be made with dimensions below $50~ \mathrm{nm}~ \times ~10 ~\mathrm{nm}$, \cite{bezryadin2000quantum, lau2001quantum} which leads to a cross-sectional area equivalent to a cylindrical wire of radius $a \sim 12.6 ~\mathrm{nm}$, or roughly $\sim 10~$nm. For such dimensions, calculating the resistance using equation \ref{WireResist} gives only a lower bound. As the surface-area to volume ratio of the wire increases with decreasing diameter, the preponderance of edge effects leads to a decrease in the effective conductivity. Additionally, at small scales on the order of $25$~nm the energy of electrons can become noticeably quantized due to their confinement within a narrow potential well \cite{Nanowire}. In this regime the conductance is calculated in multiples of the conductance quantum, $G = 2e^2/h~$ \cite{ConductQuantum}, where $e$ is the charge of the electron and $h$ is Planck's constant. The total conductance is given by combining the available conductance channels. 

			In section \ref{Rmax} it was calculated that the maximum resistance which would still allow for all of the charge-imbalance to flow away from a \textit{grounded} disk1 in the time of one half-cycle of the ion, assuming a frequency of oscillation of $5~$MHz, is on the order of $\sim 10~\mathrm{M}\Omega$. The value $R \sim 0.5 ~\Omega$ calculated with equation \eqref{WireResist} for a wire with radius $a = 10~ \mu$m is far below this threshold, meaning charges should fully drain onto and away from pickup-disk1, as ion\#1 moves closer to (farther from) it. However, if the wire radius is reduced to the order of $\sim 10~$nm, the resistance of a non superconducting wire increases dramatically to $R \sim 330 ~\mathrm{k}\Omega$. Moreover, in the coupling setup, pickup-disk1 is not connected to ground, but instead to a wire with finite capacitance, meaning the actual time needed to drain pickup-disk1 is greater than what is calculated in section \ref{Rmax}. Therefore, the values for the resistances calculated in section \ref{Rmax} are over-estimates. Taken together, these factors suggest that for wires with radii on the order of $\sim 10~$nm, increased resistance could prevent full charge equilibration.

			Next, we consider a superconducting coupling wire and experimental figures of merit. A study in reference \cite{lau2001quantum} concludes that the resistance of superconducting nanowires is driven by quantum phase slips, over a free energy barrier which is proportional to the cross-sectional area of the wire. This suggests that the cross-sectional area remains an important parameter even in superconducting nanowires, and increasing the diameter of the wire may reduce its resistance. In reference \cite{bezryadin2000quantum}, at a temperature $T < 4~$K, the resistance of a $165~$nm long, $6~\mathrm{nm~thick}  \times 21~\mathrm{nm~wide}$ superconducting nanowire is $R_{\mathrm{wire}} \lesssim 10~\Omega$ (down from a normal resistance of $R_N = 4.53~\mathrm{k}\Omega$ for the same wire above its critical temperature). This is equivalent to the cross-sectional area of a cylindrical wire of radius $a \sim 6.3~$nm. Taking $R_{\mathrm{wire}} = 10~\Omega$ and scaling this up to $1~$cm assuming the resistance of the wire increase linearly with length
			gives $R_{\mathrm{wire}} \sim 606~\mathrm{k}\Omega$. Reference \cite{boogaard2004resistance} makes measurements on a slightly fatter wire. They find that at a temperature $T = 0.6~$K, a $3~\mu$m long, $100~\mathrm{nm~thick} \times 250~\mathrm{nm~wide}$ superconducting wire connected to \textit{non}-superconducting, conducting leads, has a resistance of $R_{\mathrm{wire}} \approx 0.2~\Omega$. Scaling the length to $l = 0.01~$m gives $R_{\mathrm{wire}} \approx 6.6~\Omega$. The cross-sectional area of this wire is equivalent to a cylindrical wire of radius $a \sim 90~$nm. In the latter case, the resistance $R_{\mathrm{wire}} \approx 0.2~\Omega$ is believed to be largely due to the non-superconducting nature of the leads which connect to either end of the superconducting wire \cite{boogaard2004resistance}.  If the interfaces between the two non-superconducting leads and the superconducting wire are responsible for the majority of the resistance, increasing the length of the superconducting wire should increase the wire resistance less than the linear extrapolation used above. Additionally, the resistance of a fully superconducting wire with the same dimensions would be less. A theoretical minimum resistance is given by the spreading resistance due to reservoirs at either end of the superconducting wire (such as our pickup disks), which in reference \cite{boogaard2004resistance} is expected to be $~11~\mathrm{m}\Omega$. We note that in \cite{boogaard2004resistance} the wire itself is suspended, and the reservoirs on either side of the superconducting wire are made of a normal conductor.

			Using somewhat fatter wires, for a coiled $50~\mu$m diameter NbTi wire $\approx~37~$ meters long, in $2009$ Ulmer et al. achieved a Q-factor of Q~$\approx 0.04 \times 10^6$ for a superconducting resonator operating at $f = 1.6~$MHz, $T = 3.9~$K, with a residual equivalent series resistance $R_{\mathrm{res}} = 0.58~\Omega$ \cite{ulmer2009quality}. Scaling the length down to $1~$cm and scaling the resistance correspondingly leads to $R_{\mathrm{res}} = 0.00016~\Omega$. We obtain another estimate from reference \cite{erickson2014frequency}, which studies a wire with $Q \sim 1 \times 10^7$. The relationship between Q-factor and resistance is $Q = 1/\left(\omega R C \right)$, where $\omega = 2 \pi f$ and $C$ is the wire capacitance. This can be rearranged to calculate the resistance $R$. For a $2~\mu$m wide, $25~$cm long wire with $Q \sim 1 \times 10^7$ at $f = 60~$MHz and $T=3~$K, $R \sim 0.00024~\Omega$, which scales to $R \sim 0.00001~\Omega$ for a $1~$cm wire.

	\section{Calculating $\gamma$ from a non-linearized $Q^{\mathrm{N.L.}}_{\mathrm{transf.}}$}\label{AppNonLinGamOpt}

			Starting with the result from reference \cite{griffiths1962introduction} which gives the charge enclosed in a circular region centered around $r=0$,
\begin{equation}\label{FullInducedCharge}
Q = \frac{qd}{\sqrt{r^2+d^2}} - q ~,
\end{equation}
			one can look at the \textit{derivative} of the electric field above disk2 with respect to the distance '$d_1$' between ion\#1 and disk1 (which is what is needed in the end), and realize that it depends on the \textit{derivative} of the induced charge on disk1 with respect to $d_1$, or in other words, the derivative of $Q$. Combining this observation with the results from section \ref{CalcGamma} and appendix \ref{WireChargeDist} gives
\begin{eqnarray*}
	\frac{\partial \vec{E}^{N.L.}_{\infty2}}{\partial d_1}~
	=
	&&~\frac{1}{4\pi \epsilon_{\mathrm{o}}}\frac{-\zeta d_{2}}{\left(d^2_{2}+r_{\mathrm{disk2}}^2\right)^{3/2}}
	\frac{\partial Q_1}{\partial d_1}
	\hat{z} \\
	&& = \frac{q_{\mathrm{c}}}{4\pi \epsilon_{\mathrm{o}}}\frac{-\zeta d_{2}}{\left(d^2_{2}+r_{\mathrm{disk2}}^2\right)^{3/2}}\frac{r^2_{\mathrm{disk1}}}{\left(r^2_{\mathrm{disk1}} + d^2_1\right)^{3/2}} \hat{z}
\end{eqnarray*}
			Note that since $\zeta$ does not depend on $d_1$, it is factored out of the derivative, as is the portion of the electric field which depends only on '$d_2$', related to the "ring of charge" model. Multiplying by the charge of the trapped particle, $q_{\mathrm{c}}$, and using the definition of the coupling constant leads to 
\setlength{\parskip}{15pt plus 1pt minus 1pt}
\begin{equation*}
	\gamma^{\mathrm{N.L.}} =
	\frac{1}{2}\bigg[ \frac{\partial \big(-q_{\mathrm{c}} \vec{E}^{N.L.}_{\infty1} \big)}{\partial d_2} + \frac{\partial \big(-q_{\mathrm{c}} \vec{E}^{N.L.}_{\infty2} \big)}{\partial d_1} \bigg] 
\end{equation*}
\begin{equation*}
	=
	\zeta\frac{1}{8\pi \epsilon_{\mathrm{o}}}\frac{d_{1}}{\left(d^2_{1}+r_{\mathrm{disk1}}^2\right)^{3/2}}\frac{q^2_{\mathrm{c}}   r^2_{\mathrm{disk2}}}{\left(r^2_{\mathrm{disk2}} + d^2_2\right)^{3/2}} \hat{z} 
\end{equation*}
\begin{equation}\label{GammaNL}
+ ~\zeta\frac{1}{8\pi \epsilon_{\mathrm{o}}}\frac{d_{2}}{\left(d^2_{2}+r_{\mathrm{disk2}}^2\right)^{3/2}}\frac{q^2_{\mathrm{c}}   r^2_{\mathrm{disk1}}}{\left(r^2_{\mathrm{disk1}} + d^2_1\right)^{3/2}} \hat{z} ~,
\end{equation}

\setlength{\parskip}{15pt plus 1pt minus 1pt}
{\setlength{\parindent}{0cm}
			which is the same result derived in section \ref{CalcGamma}.
}

\section{Using the equivalent circuit element method to calculate $C_{\mathrm{hyb.}}$}\label{DerivingIonInductance}

			Here, we calculate an "equivalent capacitance" for an ion in a harmonic trap following the method in \cite{wineland1975principles}. The method in reference \cite{wineland1975principles} has been shown to be inappropriate for calculating coupling constants \cite{van2024improved}, but we include the calculation here for the sake of the comparisons in sections \ref{VariousSysAndGammas} and \ref{VariousGammas}. A difference with the calculation in \cite{wineland1975principles} is that we do not use the expression $i ~\overset{\leftarrow}{=}~ ev_z/d$~, which describes the current $i$ induced to two parallel plates separated by a distance $d$, when a charge $e$ moves with velocity $v_z$ perpendicularly to the plates \cite{shockley1938currents,sirkis1966currents}. This expression is derived for the current induced within two fixed-voltage plates, but the coupling system in the present manuscript is electrically floating and does not remain at a fixed voltage. Instead, we use an expression better suited to the geometry and capacitances of a realistic coupling system. By a "realistic" coupling system, we mean one in which the coefficient $\eta$ (see appendix \ref{WireCharge_ex4}) is not zero. The calculation is specific to a model where a charged particle is suspended above one of two pickup-disk electrodes, which are connected to each other by a conducting wire. We begin by considering the sum of the forces acting on a trapped charge:
{\setlength{\parindent}{0cm}
	\begin{equation}\label{ForcesOnIonExplic}	m\frac{d}{dt}\left(\frac{dz}{dt}\right) = -kz + eE_{\mathrm{temp}}~,
	\end{equation}
			where $\vec{E}_{\mathrm{temp}}$ is the electric field due to temporarily induced charges. Next, we write $\vec{E}_{\mathrm{temp}}$ in terms of its corresponding potential $V_{\mathrm{temp}}$, by starting with the potential produced by a singly charged particle above an infinite grounded plane, $V_{\mathrm{tot}} = \frac{1}{4\pi \epsilon_{\mathrm{o}}}
	\Bigg( \frac{e}{\sqrt{r^2+(z-d_{\mathrm{eq}})^2}} \nonumber
	-\frac{e}{\sqrt{r^2+(z+d_{\mathrm{eq}})^2}} \Bigg)$, and subtracting the potential produced by the particle itself, $V_{\mathrm{particle}} =  \frac{1}{4\pi \epsilon_{\mathrm{o}}} \left(\frac{e}{\sqrt{r^2+(d_{\mathrm{eq}}-z)^2}}\right)$, where the variable "$r$" represents the radial distance away from the origin, which is located directly below the suspended charge. "$d_{\mathrm{eq}}$" represents the distance between the charge and the plane, and as a point of clarification, in equation \eqref{ForcesOnIonExplic} the notation $z$ refers to displacement away from the minimum of the harmonic potential, whereas in the expression for $V_{\mathrm{tot}}$ the notation $z$ refers to the distance away from the plane at which the potential is evaluated. Calculating the corresponding electric field,
	\begin{eqnarray}\label{E-induced}
	E_z~\overset{\leftrightarrow}{=}~ 
	&&\frac{-\partial{V}}{\partial{z}} ~\overset{\leftarrow}{=}~ \frac{-e}{4\pi \mathrm{\epsilon_o}}\left(\frac{z+d_{\mathrm{eq}}}{(r^2 +(z+d_{\mathrm{eq}})^2)^{3/2}}\right) .~~~~
	\end{eqnarray}
			To keep track of independent and dependent variables, the single arrow over the rightmost equals sign in equation \eqref{E-induced} and the expression for the current $i ~\overset{\leftarrow}{=}~ ev_z/d$ (above), $~\overset{\leftarrow}{=}~$, points from the independent variable (the charge) to the dependent variable (the potential gradient in equation \eqref{E-induced}, or the induced current in equation \eqref{CurrentAndVelocity2}). The double-arrow over the leftmost equals sign indicates that electric field and potential gradient both "act" on each other; changing one automatically implies changing the other, indeed, they are different ways of expressing the same thing. We use causal equality ("arrow") notation \cite{van2024improved} to help avoid or identify mistakes in our derivation. Since equation \eqref{E-induced} gives the field due to the full static induced charge but we are only interested in the field due to the temporarily induced charges, we replace $e$ by $Q_{\mathrm{temp}}$~, so $E_z$ becomes $E_{\mathrm{temp}}$. We are interested in the field which is produced by the temporary charge, specifically at the position of the charged particle, which is to say at $z=d_{\mathrm{eq}}$, $r=0$.  Thus:
	\begin{equation}\label{E_tempQ_temp}
	E_{\mathrm{temp}}\bigg|_{z=d_{\mathrm{\mathrm{eq}}},~r=0} ~\overset{\leftarrow}{=}~  \frac{-Q_{\mathrm{temp}}}{4\pi\mathrm{\epsilon_o}}\left(\frac{1}{4d_{\mathrm{eq}}^{2}}\right)~.
	\end{equation}
			If the potential $\left( V_{\mathrm{tot}}-V_{\mathrm{particle}} \right)$, which is used to calculate equation \eqref{E-induced}, is evaluated at $z=d_{\mathrm{eq}}$, $r=0$, for only the charge $Q_{\mathrm{temp}}$, we find:
	\begin{equation}\label{V_tempQ_temp}
	V_{\mathrm{temp}}\bigg|_{z=d_{\mathrm{eq}},~r=0} ~\overset{\leftarrow}{=}~ \frac{-Q_{\mathrm{temp}}}{4\pi\mathrm{\epsilon_o}}\left(\frac{1}{2d_{\mathrm{eq}}}\right)~.
	\end{equation}
			Thus, we can rewrite $E_{\mathrm{temp}}$ in terms of $V_{\mathrm{temp}}$ as:
	\begin{equation*}
			E_{\mathrm{temp}} ~\overset{\leftrightarrow}{=}~ V_{\mathrm{temp}}\left(\frac{1}{2d_{\mathrm{eq}}}\right)\hat{z}~.
	\end{equation*}
			Here, we have related the two causal equations \eqref{E_tempQ_temp} and \eqref{V_tempQ_temp}.
			The equation of motion for the charged particle, \eqref{ForcesOnIonExplic}, can therefore be written as:
	\begin{equation}
	m\frac{d}{dt}\left(\frac{dz}{dt}\right) \overset{\leftarrow}{=} -kz + \frac{eV_{\mathrm{temp}}}{2d_{\mathrm{eq}}}~.
	\end{equation}
			Expressing the displacement "$z$" in terms of an integral and using the relationship for a mechanical harmonic oscillator $\omega = \sqrt{\frac{k}{m}}$~,
	\begin{equation}\label{WriteIntegral}
	m\frac{d}{dt}\left(\frac{dz}{dt}\right) + m\omega^2 \int{\frac{dz'}{dt'}dt'} = \frac{eV_{\mathrm{temp}}}{2d_{\mathrm{eq}}}~.
	\end{equation}
			Now, we can draw a relationship between the quantity $dz'/dt'$ which denotes the velocity of the particle, and the current $i$ which the movement of the particle induces in the pickup-disk circuit. We start with the total charge $Q(t)$ that a single charged particle induces in a disk-like region, similar to what was calculated in section \ref{ChargeflowQ} (see below). In the expression for $Q(t)$, $\eta$ is a coefficient which accounts for the finite capacitance of the various parts of the coupling system, and $b\left(t\right)$ is the instantaneous displacement of the charged particle away from its equilibrium position
	\begin{equation*}
		Q\left(t\right) ~\overset{\leftarrow}{=}~ \eta e\left[\frac{r^2 b\left(t\right)}{(r^2 + d_{\mathrm{eq}}^2)^{3/2}} + \frac{d_{\mathrm{eq}}}{{(r^2 +{d_{\mathrm{eq}}}^2)}^{1/2}} - 1 \right]~.
	\end{equation*}
			Hence, the current through the pickup-disk system is related to the change in charge induced by the motion of the charged particle by:
	\begin{eqnarray}\label{CurrentAndVelocity2}
	i \equiv \frac{dQ}{dt} ~\overset{\leftarrow}{=}~ \eta \frac{er^2}{(r^2 + d_{\mathrm{eq}}^2)^{3/2}}&&\frac{d b\left(t\right)}{dt} \nonumber \\
	&& = \eta \frac{er^2}{(r^2 + d_{\mathrm{eq}}^2)^{3/2}} \frac{dz}{dt} ~~~
	\end{eqnarray}
			where we have neglected the impedance due to the resistance of the coupling wire $R_{\mathrm{wire}}$ and the inductance of the wire $L_{\mathrm{wire}}$. Neglecting $R_{\mathrm{wire}}$ is valid if $R_{\mathrm{wire}}$ is small enough that the induced charge re-equilibrates continuously as the trapped particle moves, or in other words in a "low frequency" approximation. The same applies for neglecting the inductance. These two conditions are satisfied for a range of possible experimental implementations (see appendices \ref{AppendixResist} and \ref{OffResAndLneg}). Plugging the relationship between $i$ and $dz/dt$ back into the equation of motion \eqref{WriteIntegral} gives
	\begin{eqnarray}\label{EqnMotionLionCrf}
	\frac{m}{\eta er^2}&&\left(r^2 + d_{\mathrm{eq}}^2\right)^{3/2}\frac{d \left(i\right)}{dt} \nonumber \\
	&&+ ~\frac{m\omega^2}{\eta er^2}\left(r^2 + d_{\mathrm{eq}}^2\right)^{3/2} \int{i dt} = \frac{eV_{\mathrm{flow}}}{2d_{\mathrm{eq}}}~.
	\end{eqnarray}
			Recalling that $\int{idt} \equiv Q$~, and $L\frac{d(i)}{dt}+\frac{Q}{C} = V$, the effective inductance and capacitance are defined as:
	\begin{equation}\label{LionCrf2}
	L_{\mathrm{hyb.}} \equiv \frac{2d_{\mathrm{eq}}m}{\eta e^2 r^2}\left(r^2 + d_{\mathrm{eq}}^2\right)^{3/2}~,
	\end{equation}
			and
	\begin{equation}\label{C_Temp_wrong} 
	C_{\mathrm{hyb.}} \equiv \frac{\eta e^2 r^2}{2d_{\mathrm{eq}} m\omega^2 \left(r^2 + d_{\mathrm{eq}}^2\right)^{3/2}}~.
	\end{equation}
			This case example contains the necessary correction on the induced current. However, the derivation still contains errors associated with defining the inductive and capacitive elements in equations \eqref{LionCrf2} and \eqref{C_Temp_wrong}, as explained in \cite{van2024improved}. One way to understand the issue is that using equation \eqref{C_Temp_wrong} one does not know how much the charged particle will be displaced as a result of adding a given amount of charge to one of the pickup disks, and using equation \eqref{LionCrf2} one does not know how much a change in current within the coupling system will cause the trapped charge to accelerate. Expressions \eqref{LionCrf2} and \eqref{C_Temp_wrong} are not equivalent to circuit elements, and using them to reason using equivalent circuits does not provide successful predictions. Nevertheless, they are included for the completeness of the comparison in section \ref{VariousGammas}.
}

\section{Proof that trapped ions drive the coupling system in the $\omega^{\mathrm{trapped~ion}}_{0} \ll \omega^{\mathrm{coupling~system}}_{0}$ frequency regime}\label{OffResAndLneg}

			The coupling system is comprised of three components: two disks, and a connecting wire. Each of these has a self capacitance, electrical resistance, and inductance. The coupling system is therefore an L-R-C circuit. It is useful to know whether the frequency at which the trapped ions drive the coupling system is close to its resonant frequency. Driving the coupling system on resonance would not increase the coupling strength $\gamma$. However, it can ensure that the ratio of energy stored in the system $E^{\mathrm{stored}}$ to power dissipated through losses in the wire and disks in each cycle of oscillation $P^{\mathrm{dis.}}$, is maximized \cite{jackson2004novel}. This is equivalent to the ratio of energy transferred from one part of the oscillator to the other in one half-oscillation $E^{\mathrm{trf.}}$, to energy dissipated in one half-oscillation $0.5P^{\mathrm{dis.}}$, being greatest.\footnote{The reason maximizing $E^{\mathrm{stored}}/P^{\mathrm{dis.}}$ is equivalent to maximizing $E^{\mathrm{trf.}}/P^{\mathrm{dis.}}$ is the following. The total energy stored in the coupling system is limited by the fact that more energy flowing, $E^{\mathrm{trf.}}$, corresponds to a larger electrical current. The power dissipated increases with the current squared ($P^{\mathrm{dis.}}=IV=I^2R$), even though the power input remains constant, for example $1$ phonon per half-cycle. Minimizing the power dissipated \textit{per unit of power input} means that for a constant power input, the total stored energy is maximized. Therefore, maximizing the stored energy, which is what happens when a system is driven on resonance, is equivalent to minimizing the \textit{energy lost per unit of energy input}.} In turn, we conjecture that this implies the ratio of quantum information transmitted between the ions, to quantum information lost due to noise, is greatest, i.e. $I^{\mathrm{nfo.}}_{\mathrm{tansmitted}} \big/ I^{\mathrm{nfo.}}_{\mathrm{lost}} = max$.  
			Thus, driving the coupling system on resonance could increase the signal to noise ratio in the coupling system, which is one of the governing criteria outlined in the Introduction.

			Before estimating the resonant frequency of the coupling system, we comment on the dynamics of the system as a whole. The system can be treated as two independent ions oscillating at their natural frequencies, driving the coupling system. 
			This is justified as follows. The overall system can be represented as three masses connected to $4$ springs, where ion\#1 and ion\#2 are two masses on either side of a $3$rd mass. The $3$rd mass represents the impedance of the coupling system (the sum of the impedances of the two disks and the wire). When considering the dynamics of the system, we compare the strength of the interaction between an ion and the trapping potential, $k_{\mathrm{trap}} = m \left( \omega^{\mathrm{trapped~ion}}_{0} \right)^2$, with the strength of the interaction between an ion and the coupling system, on the order of $\gamma$. The coupling strength between the ions and the central mass is significantly weaker than the interaction between the ions and the trap potential, $\gamma \ll k_{\mathrm{trap}}$, which makes it reasonable to neglect the effect of the central mass (the coupling system) when considering the dynamics of the ions. The system is well approximated as two independent ions, each connected to a wall by it's own spring.\footnote{Another case can be considered where the coupling $\gamma$ is strong, on the order of $k_{\mathrm{trap}}$, but the central "mass" is neglected on the basis of its low inductance (justified below). In this limit, the system reduces to two masses with springs on either side, coupled together by a central spring. Such a system, with a coupling spring constant $k_m$, has two normal modes: $\omega^i_{\mathrm{coupled}} = \omega_0$, and $\omega^{ii}_{\mathrm{coupled}} = \left( \omega^2_0 + 2\omega^2_m \right)^{1/2}$, where $\omega_m = \sqrt{k_m/m}$ would be the natural frequency of one of the two masses if it were connected only to the coupling spring. We see that in the limit $k_m \ll k_{\mathrm{trap}}~$ the two modes become equal, $\omega^{ii}_{\mathrm{coupled}} \approx \omega^{i}_{\mathrm{coupled}} = \omega_0~$, and the ions oscillate at their individual resonant frequencies $\omega^{\mathrm{trapped~ion}}_{0}$. We can suppose that the coupling strength is just strong enough to bring the ions to move with one of the two possible relative phases. Specifically, we assume here that the ions move in phase (with a relative phase of zero), and their oscillation is at the frequency of the first normal mode, $\omega^i_{\mathrm{coupled}} = \omega_0$.} The ions drive the coupling system at their own natural frequency $\omega^{\mathrm{trapped~ion}}_{0}$, and changes to the coupling system do not significantly alter these dynamics.


			We now apply constraints to estimate the resonant frequency of the coupling system. The total impedance of disk1+wire+disk2, is given by the impedance of an L-R-C series circuit: $Z=\sqrt{R^2+{(X_L-X_C)}^2}$. The impedance is minimized for $X_L = X_C$, which is to say ${\omega_0}L_{\mathrm{wire}} = \frac{1}{{\omega_0}C_{\mathrm{tot.}}}$, from which the resonant frequency is ${\omega_0} =\sqrt{ 1 / \left( L_{\mathrm{wire}} C_{\mathrm{tot.}} \right) }~$. The total capacitance is $C_{\mathrm{tot.}} = \left( C_{\mathrm{disk1}} + C_{\mathrm{wire}} + C_{\mathrm{disk2}} \right)$. We can estimate these individual capacitances, and hence $C_{\mathrm{tot.}}$. First, we estimate the disk capacitances, $C_{\mathrm{disk1}} = C_{\mathrm{disk2}} \equiv C_{\mathrm{d}}$. Once the coupling strength $\gamma$ is optimized as a function of $r_{\mathrm{disk1}}$, $r_{\mathrm{disk2}}$ (see section \ref{GammaOpt}), the radii of disk1 and disk2 are fixed by the distance between the ion and the disk, $r = d_{\mathrm{eq.}}/\sqrt{2}$. Assuming the thicknesses "$\mathrm{a}$" (non-italicized) of disk1 and disk2 are much smaller than their radii (the limit $r \gg \mathrm{a}$), fixing the disks' radii also fixes their capacitances, $C_{\mathrm{d}} \approx 8 \epsilon_{\mathrm{o}} r$. Thus, $d_{\mathrm{eq.}}$ determines the disk capacitance $C_{\mathrm{d}}$. 
			The capacitance of a cylindrical wire is $C_\mathrm{w} = 2\pi \epsilon_{\mathrm{o}} l_{\mathrm{w}} / \mathrm{ln} \left( l_{\mathrm{w}}/a \right)$, where $l_{\mathrm{w}}$ is the length of the wire and "$a$" (italicized) is the radius of the wire. The length of the wire is bounded in appendix \ref{MinLength} by $l_{\mathrm{w}} \ge 0.5~$mm, so below we will use $l_{\mathrm{w}} = 0.01~$m as in table \ref{tab:DefaultVals}. The capacitance $C_\mathrm{w}$ does not depend strongly on the choice of wire radius $a$, since the dependence is logarithmic. The wire radius is chosen to be large to minimize resistance and associated Johnson-Nyquist noise, but small compared to $l_{\mathrm{w}}$ to ensure the validity of the model. 


			We claimed above that the resonant frequency of each ion does not depend on the coupling system,  $\omega^{\mathrm{trapped~ion}}_{\mathrm{coupled}} \approx \omega^{\mathrm{trapped~ion}}_{0}$. The choice of frequency of the ions is chosen to simultaneously satisfy the two criteria $t_{\mathrm{deco.}} / t_{\mathrm{ex.}} \ge 10$ and $V_{\mathrm{sig.}} / V_{\mathrm{J.N.}} \ge 10$ identified in the Introduction of this manuscript. This means the frequency $\omega_{0}$ at which the coupling system is driven by the ions is fixed. Since both the resonant frequency and the total capacitance of the coupling system are fixed, if the coupling system's resonant frequency is to match the resonant frequency of the ions, its inductance must be tailored to satisfy $L_{\mathrm{wire}} = 1/\left(\left(\omega^{\mathrm{trapped~ion}}_0\right)^2 C_{\mathrm{tot.}} \right)$. Expressing the total capacitance $C_{\mathrm{tot.}}$ in terms of the parameters of the coupling system using $C_{\mathrm{d}} = 8\epsilon_{\mathrm{o}} r$ and $C_{\mathrm{w}} = 2 \pi \epsilon_{\mathrm{o}} l / \mathrm{ln}\left(l/a\right)$, gives
\begin{equation*}
	L_{\mathrm{wire}} = \frac{1}{\omega^2_0 \left( 2 \times 8\epsilon_{\mathrm{o}} r_{\mathrm{disk}} + \frac{2\pi \epsilon_{\mathrm{o}} l}{\mathrm{ln}{\left(l/a \right)}} \right)}\qquad \qquad \qquad
\end{equation*}
\begin{equation}\label{LgInduct}
\qquad \qquad  \qquad \qquad \qquad = 0.01 ~\mathrm{H}~.
\end{equation}
			Equation \eqref{LgInduct} is evaluated for $\omega_0 = 2 \pi \times 5 \times 10^6~$Hz, $r_{\mathrm{disk}} = 35~\mu$m, $l = 0.01~$m, and $a = 10 \times 10^{-6}~$m. Due to the low frequency $\omega_{0}$ at which the ions oscillate and the small capacitance $C_{\mathrm{tot.}}$ of the coupling system, the inductance $L_{\mathrm{wire}}$ needed for the resonance of the coupling system to match the resonance of the ions, is exceedingly large. It is not possible to produce such a large inductance, $L_{\mathrm{wire}} = 0.01 ~\mathrm{H}$, using the coupling wire. The inductance of a typical $1~$cm long $10~\mu$m radius straight coupling wire is $\lesssim 1 ~\mu$H. This implies the coupling system cannot be designed to be on resonance with the ions. Since the actual inductance of the coupling system is much smaller than the inductance needed for resonance with the ions, the resonant frequency of the coupling system is greater than the resonant frequency of the ions. In other words, the frequency at which the ions drive the coupling system is less than the coupling system's intrinsic resonant frequency. In this far-from-resonance regime, the charge distribution within the coupling system redistributes continuously as the ions oscillate, provided the resistance $R$ is low enough, which is verified in section \ref{Rmax} and appendix \ref{AppendixResist}. Thus, we conclude that trapped ions drive the coupling system in a roughly electrostatic fashion, in the $\omega^{\mathrm{trapped~ion}}_{0} \ll \omega^{\mathrm{coupling~system}}_{0}$ frequency regime. This conclusion might be revisited for trapped electrons, which have a much smaller mass than atoms and a correspondingly higher oscillation frequency. 

			To summarize, for coupling electrodes with radii on the order of $\sim 50 ~\mu$m and ions oscillating at a frequency of $\sim 5~$MHz, engineering the coupling system to be an $LC$ resonator with a natural frequency on the order of $\sim 5~$MHz (the frequency the trapped ions), requires a coupling wire with an inductance $L_{\mathrm{wire}} \sim 0.1~ \mathrm{H}$. This is too large to achieve using a typical coupling wire with length $l_{\mathrm{wire}} \sim 1~$cm, and radius $a \sim 10~\mu$m. 

	\section{Minimum length of the coupling wire}\label{MinLength}

			It is good practice to verify that the coupling system proposed herein provides an improvement over direct coupling via the Coulomb interaction. One way to prove this is to make the coupling wire long enough that any coupling observed is dominated by the coupling system. In this regime, turning the coupling system off or on for example by altering its resistance, can be used to monitor the coupling system's effect.

			We start with the expression for the Coulomb force between two charged particles:
\begin{equation}
F = \frac{q_1q_2}{4 \pi l^2_{\mathrm{Coul.}} \epsilon_{\mathrm{o}}},
\end{equation}
			where $q_1$ is the charge of one ion, $q_2$ is the charge of another ion, $l_{\mathrm{Coul.}}$ is the distance separating $q_1$ and $q_2$, and $\epsilon_{\mathrm{o}}$ is the vacuum permittivity. From this, the coupling strength is
\begin{equation}
\gamma_{\mathrm{Coul.}} \equiv -\frac{\partial F}{\partial l_{\mathrm{Coul.}}} = -\frac{\partial}{\partial l_{\mathrm{Coul.}}} \left[ \frac{q_1q_2}{4 \pi \epsilon_{\mathrm{o}}} l^{-2}_{\mathrm{Coul.}} \right] ~~~~~~~~~~~~~~~~~\nonumber
\end{equation}
\begin{equation}
~~~~~~~~~~~~~~~~~~~~~~~~~~ = \frac{q_1q_2}{2 \pi \epsilon_{\mathrm{o}} l^3_{\mathrm{Coul.}}} ~.
\end{equation}
			(Reference \cite{brown2011coupled} finds the same result, for two ions of identical mass in identical trap potentials).

			To decide on an appropriate length for the coupling wire, we specify a target ratio $\gamma / \gamma_{\mathrm{Coul.}} \ge 10$, where $\gamma$ denotes coupling via the conducting wire. Using equation \eqref{gamma_symm} for $\gamma$, the inequality to solve is

\begin{equation*}
	\frac{q^2_{\mathrm{c}}}{4\pi \epsilon_{\mathrm{o}}} \left( \frac{r}{2r + \frac{C_b}{8\epsilon_{\mathrm{o}}} } \right) \left( \frac{d r^2}{\left(d^2+r^2\right)^{3}} \right)~~~~~~~~~~~~~~~~~~~~~
\end{equation*}
\begin{equation}
~~~~~~~~~~~~~~~~~~\ge 10 \times \frac{q_1q_2}{2 \pi \epsilon_{\mathrm{o}} l^3_{\mathrm{Coul.}}} ~.
\end{equation}

			Letting $r = d/\sqrt{2}$ (see section \ref{GammaOpt}), this rearranges to

\begin{equation}
l_{\mathrm{Coul.}} \ge \left[ 10 \times \left( \frac{27 \sqrt{2} d^2}{2} \left[ \frac{2d}{\sqrt{2}} + \frac{C_{\mathrm{b}}}{8\epsilon_{\mathrm{o}}}
\right] \right) \right]^{1/3} ~.
\end{equation}

			Substituting in the capacitance of a finite wire, $C_\mathrm{b} = 2 \pi \epsilon_{\mathrm{o}} l_{\mathrm{wire}} / \mathrm{ln} \left( l_{\mathrm{wire}}/a \right)$, where $l_{\mathrm{wire}}$ is the length of the coupling wire and $a$ is its radius gives

\begin{equation*}
	l_{\mathrm{Coul.}} \ge \left[ 10 \times \left( \frac{27 \sqrt{2} d^2}{2} \left[ \frac{2d}{\sqrt{2}} + \frac{2 \pi l_{\mathrm{wire}}}{8 \mathrm{ln}\left(\frac{l_{\mathrm{wire}}}{a}\right)}
	\right] \right) \right]^{1/3} ~.
\end{equation*}

\begin{figure}[h]
			\centering
			\includegraphics[width=\linewidth]{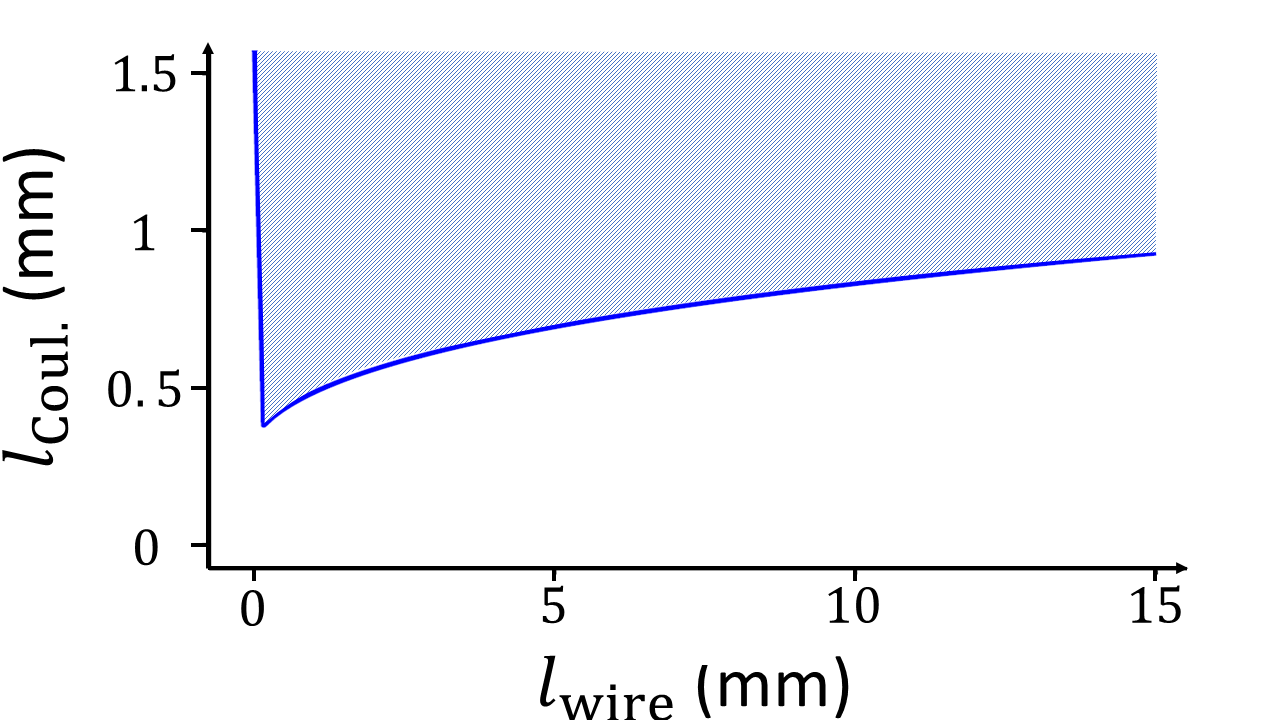}
			\caption{Wire lengths $l_{\mathrm{wire}}$ and ion separations $l_{\mathrm{Coul.}}$ which satisfy $\gamma / \gamma_{\mathrm{Coul.}} \ge 10$. The shaded region shows combinations of $l_{\mathrm{Coul.}}$ and $l_{\mathrm{wire}}$ that satisfy the inequality $\gamma \ge 10 \times \gamma_{\mathrm{Coul.}}$}
			\label{fig:l_CoulAsfcnOf-l_wire}
		\end{figure}

			Plotting this expression as a function of $l_{\mathrm{wire}}$ for $d = 50 \times 10^{-6}~$m and the wire radius $a = 10 \times 10^{-6}~$m gives the curve in figure \ref{fig:l_CoulAsfcnOf-l_wire}.
		
			Figure \ref{fig:l_CoulAsfcnOf-l_wire} can be used as follows. First, pick a value for $l_{\mathrm{wire}}$, for example $10~$mm. Then draw a vertical line upwards until it intersects with the curve. All values of $l_{\mathrm{Coul.}}$ in the shaded region above the point of intersection satisfy $\gamma \ge 10 \times \gamma_{\mathrm{Coul.}}$. To give an example, suppose $d = 50 \times 10^{-6}~$m and   $l_{\mathrm{Coul.}} \ge 0.5~$mm so that decoherence occurs more quickly than state exchange via the direct Coulomb interaction. For instance, we may choose $l_{\mathrm{Coul.}} = 1~$mm. Next, suppose we want to use a wire of length $l_{\mathrm{wire}} = 10~$mm. Looking at figure \ref{fig:l_CoulAsfcnOf-l_wire} we see that if $l_{\mathrm{Coul.}}$ is greater than about $0.8~$mm, the condition $\gamma \ge 10 \times \gamma_{\mathrm{Coul.}}$ is satisfied. Due to the physical layout of the coupling system, if $l_{\mathrm{wire}} = 10~$mm the actual value of $l_{\mathrm{Coul.}}$ will also be about $l_{\mathrm{Coul.}} \approx 10~$mm, which means the criterion $\gamma \ge 10 \times \gamma_{\mathrm{Coul.}}$ will be satisfied. In this scenario where $l_{\mathrm{Coul.}} \ge 0.8~$mm and $l_{\mathrm{Coul.}} \ge 0.5~$mm, if state exchange is observed between two ions, one can be confident that the state exchange is mediated by the coupling system rather than the direct Coulomb interaction.

			With the coupling system described herein, the distance $l_{\mathrm{Coul.}}$ will in general be approximately equal to $l_{\mathrm{wire}}$. Following the procedure above, the cutoff on $l_{\mathrm{Coul.}}$ is nearly always smaller than $l_{\mathrm{wire}}$. This indicates that for most choices of $l_{\mathrm{wire}}$ the criterion $\gamma \ge 10 \times \gamma_{\mathrm{Coul.}}$ is automatically satisfied. The one notable exception is when $l_{\mathrm{wire}}$ approaches the wire radius $a$ and the bound on $l_{\mathrm{Coul.}}$ diverges as shown in figure \ref{fig:l_CoulAsfcnOf-l_wire}. This unrealistic artifact comes from a limitation in the expression for the capacitance of a finite cylindrical wire, which diverges when the wire radius $a$ approaches the wire length $l_{\mathrm{wire}}$. It is irrelevant provided $l_{\mathrm{wire}} \gg a$.

\end{appendix}

\end{document}